\documentclass[english,11pt,reqno]{article}
\usepackage{latexsym, amsmath, amssymb,amsthm, natbib,verbatim, xy}
\xyoption{all}
\pdfoutput=1 

\usepackage[dvipsnames]{xcolor}
\usepackage{tgpagella}
\usepackage{mathpazo}
\usepackage{float}
\usepackage[labelfont=bf]{caption}
\usepackage[left=1in,top=1in,right=1in,bottom=1in]{geometry}
\usepackage{setspace,color}
\usepackage{graphicx}
\usepackage{enumerate}
\usepackage[multiple]{footmisc}
\usepackage{hyperref}
\usepackage{fancyhdr}
\usepackage{subfig}
\usepackage{rotating}
\usepackage{soul}
\usepackage{etoolbox}
\patchcmd{\section}{\scshape}{\bfseries}{}{}

\newcommand{\fig}[1]{{\color{black} #1}}

\newtheorem{theorem}{Theorem}
\newtheorem{corollary}{Corollary}
\newtheorem{definition}{Definition}
\newtheorem{lemma}{Lemma}
\newtheorem{proposition}{Proposition}

\theoremstyle{definition}

\newtheorem{example}{Example}

\def\cala{\mathcal{A}} 
  
\def\calx{\mathcal{X}}

\def\calc{\mathcal{C}} 
\def\cals{\mathcal{S}}

\def\calq{\mathcal{Q}}
\def\calx{\mathcal{X}} 
 
\def\calp{\mathcal{P}} 

\def\i{\mathsf{i}}
\def\b{\mathsf{b}}
\def\t{\mathsf{t}}

\renewenvironment{quote}{%
   \list{}{%
     \leftmargin1.5cm   
     \rightmargin\leftmargin
   }
   \item\relax
}
{\endlist}

\begin{document}

\title{Redesigning the US Army's Branching Process:\\ A Case Study in Minimalist Market Design\thanks{This paper replaces 
and subsumes \cite{greenberg/pathak/sonmez:21}. 
All opinions expressed in this manuscript are those of the authors and do not represent the opinions of the 
United States Military Academy (USMA), United States Cadet Command, the United States Army, or the Department of Defense. 
We are grateful for excellent research assistance from Kate Bradley and Robert Upton. Eryn Heying provided superb help with research administration. 
Scott Kominers and seminar audiences at the 2021 NBER Market Design Workshop, 2021 Winter School at Delhi School of Economics, 
Cambridge, Hamburg, Iowa,  NYU-Abu Dhabi and Wharton provided helpful feedback. 
The Army's Office of Economic and Manpower Analysis provided administrative branching data for
this project to Kyle Greenberg as part of a restricted use agreement with USMA and MIT that specifies that data can only be stored, accessed, and analyzed within USMA's information system.  Any parties interested in accessing this data must make a direct application to USMA.  Pathak acknowledges support from the National Science Foundation for this project. Pathak is a co-founder of Avela.}}
\author{Kyle Greenberg \and Parag A. Pathak  \and Tayfun S\"{o}nmez\thanks{Greenberg: Department of Social Sciences, United States Military Academy at West Point, email: kyle.greenberg@westpoint.edu.
Pathak: Department of Economics, MIT and NBER, email: ppathak@mit.edu, S\"{o}nmez: Department of Economics, Boston College, email: sonmezt@bc.edu.}}

\date{March 2023}

\maketitle

\begin{abstract}

We present the proof-of-concept for minimalist market design \citep{sonmez:23} as an effective methodology
to enhance an institution based on the desiderata of stakeholders with minimal interference.  
Four objectives – respecting merit, increasing retention, aligning talent, and enhancing trust – guided reforms to
US Army's centralized branching process of cadets to military specialties
since 2006. USMA’s mechanism for the Class of 2020 exacerbated challenges
implementing these objectives. Formulating the Army's desiderata as rigorous axioms, we analyze their
implications. Under our minimalist approach to institution redesign, the Army’s objectives uniquely identify a branching mechanism. 
Our design is now adopted at USMA and ROTC.
    
\end{abstract}

\thispagestyle{empty}

\newpage

\section{Introduction}

Consider an economic, political, or social institution that is deployed to fulfill a number of objectives.  
Typically it has many components, each serving its own purposes, and interacting with each other in various ways. 
For example, in the context of an auction design that involves equity considerations for minority-owned businesses, 
a component may be used to collect private information from the participants, 
a second component may be  used to process this information, 
a third component may be used to determine the pricing of various outcomes, and 
a fourth component may be used to ensure a fair outcome. 
Now suppose that the institution fails in some of its objectives. Maybe some of its components are broken, or maybe there is 
an issue with the interface between various components. 
How can a design economist be helpful in addressing these failures? 

Before formulating a potential answer to this question, let us imagine how experts in other areas would respond to similar challenges. 
How would a surgeon address an analogous failure on a human body or a mechanic on a broken car?
These experts would first identify the root cause of the problem, whether it has to do with a component itself or an interface between 
various components and directly address the failure. 
For example, a  surgeon would remove diseased tissue or organs, repair body systems, or replace diseased organs with transplants. 
Similarly, a mechanic would directly repair or replace the worn part of the broken car. 

\textit{Minimalist market design\/} \citep{sonmez:23} is a paradigm under which a design economist operates in a similar way. 
In this paradigm, the first task is to identify the primary objectives of the system operators or other stakeholders in designing
the institution. In many cases, various components of an institution,  interfaces between its components, or 
its mission have evolved over time. The history of an institution is often instructive in 
identifying the primary objectives of various stakeholders in designing and deploying the institution. 
The second task under minimalist market design is to find out whether the current institution satisfies these
primary objectives or not. If it doesn't, then there is potential for policy impact with a compelling alternative design.  
To materialize this potential into a successful redesign, the root causes of the failures should be identified. 
That is, akin to a surgeon or a mechanic, 
a design economist following this minimalist approach identifies which components or interfaces are responsible for the failure. 
As the third task, the failures of the current institution are addressed by interfering only with its flawed components
and interfaces, as if the surgeon performs a ``minimally invasive'' procedure. Hence the adjective ``minimalist'' is the
signature feature of this design paradigm. 

Drawing on our more than a decade-long integrated research and policy effort on the US Army's process for matching
cadets to military specialties known as branches (henceforth referred to as the ``branching'' process), the first contribution of this paper is to present the first  direct application and subsequent proof-of-concept 
of minimalist market design as a whole.\footnote{\cite{sonmez/switzer:13} marks the inception of this direct application.  
In this paper, we report how it finally reached successful completion consistent with the initial prescription in \cite{sonmez/switzer:13}.} 
Although this institute design paradigm evolved through our experiences from our earlier research and policy efforts in school choice and kidney exchange, the
US Army's branching process  is the first application where the minimalist market design paradigm was deliberately and systematically followed 
at all stages of the research and policy program, and eventually succeeded in changing an 
important institution.\footnote{\cite{sonmez:23} presents how the framework developed from
research and policy efforts to reform student assignment systems  \citep{balinski/sonmez:99, abdulkadiroglu/sonmez:03, 
aprs:05} along with establishing kidney exchange systems  \citep{roth/sonmez/unver:04, roth/sonmez/unver:05}, 
and describes how it has recently proven useful in other settings following its successful deployment
for the US Army's branching process in Fall 2020. These settings include the design of pandemic rationing schemes for
scarce medical resources \citep{pathak/sonmez/unver/yenmez:20}  and the design of Indian affirmative action systems \citep{sonyen22}.
}
   
\subsection{The Making of a Partnership between the US Army and Market Designers}

A military specialty in the US Army, known as a \textit{branch\/}, is an important factor in the career progression of cadets.  
Each year, the US Army assigns thousands of graduating cadets from the United States Military Academy (USMA) at West Point and the Reserve Officer Training Corps (ROTC) to a branch through centralized mechanisms.  
These \textit{branching mechanisms\/} at West Point and ROTC determine the branch assignments for 70 percent of newly commissioned Army officers \citep{dodpoprep2018}. Prior to 2006, positions at each branch were assigned purely based on a performance ranking called \textit{Order of Merit List (OML)\/}. 
Thus, the original mission of the branching system was to allocate the positions in a way that reflects the hierarchical structure of the Army.

In 2006, the US Army created an incentive scheme 
within its branching systems with the goal of increasing officer retention \citep{colarusso/lyle/wardynski:10}.  Under this incentive scheme, 
known as the \textit{BRADSO program\/}, cadets receive heightened priority for a fraction of a branch's positions  (henceforth ``flexible-price'' positions)
if they express a willingness to extend the length of their service commitment.\footnote{ADSO is short for Active Duty Service Obligation.  BRADSO stands for Branch of Choice Active Duty Service Obligation. BRADSO slots are 25\% of the total branch allocations at USMA from the Class of 2006 through 2020 and 35\% for the Class of 2021, and either 50\% or 60\% of total branch allocations at ROTC depending on the graduating class. USMA and ROTC cadets receive branches through separate centralized branching systems.}  
Under the Army terminology, a cadet who volunteers for this incentive scheme at a given branch $b$ is said to \textit{BRADSO\/} 
for branch $b$. 
The  USMA leadership accordingly embedded the BRADSO incentive scheme into a new branching mechanism we call \textit{USMA-2006\/}. 
The adjustment of the mechanism reflected the changing mission of the branching system which now included retention.

Although the Army's branching process is a natural application of the celebrated \textit{matching with contracts\/} model by  \cite{hatfield/milgrom:05}, 
it remains outside the scope of the original theory in this paper. 
USMA-2006 was designed at a time when the matching with contracts model was still being developed. 
Consequently, the connection between the Army's practical problem and the original theory had some missing pieces. 
These pieces were later completed by \cite{hatfield/kojima:10}, albeit in an abstract framework.\footnote{Further elaboration is provided by \cite{echenique:12}, \cite{schlegel:15}, and \cite{jagadeesan:19}.} 
The connection between abstract theory and the Army's practical problem was subsequently discovered in \cite{sonmez/switzer:13}. 
In addition, \cite{sonmez/switzer:13} also proposed an alternative mechanism for the Army
by embedding the BRADSO incentive scheme directly within the \textit{cumulative offer mechanism\/} by \cite{hatfield/milgrom:05}. 

While this proposal had desirable theoretical properties, it required a more complex strategy space  in which
cadets have to rank branches and \textit{contractual terms\/} (also referred to as \textit{prices\/})  jointly.  
Under the USMA-2006 mechanism, cadets only rank branches and separately indicate their willingness to BRADSO for 
any branch.  The Army considered the existing strategy space more manageable than a more complex alternative
and kept the USMA-2006 mechanism in the intervening years.   

In 2012, the US Army introduced the Talent-Based Branching program to develop a ``talent market'' where additional information 
about each cadet influences the priority a cadet receives at a branch \citep{colarusso2016starting}.  
This program allowed branches and cadets to better align their interests and fit for one another.  Under Talent-Based Branching, 
branches rated cadets into one of three tiers: high, medium, and low. These cadet ratings were originally a pilot initiative, 
but for the Class of 2020, the US Army decided to use them to adjust the underlying OML-based prioritization, 
constructing priorities at each branch first by the tier and then by the OML within the tier.  

The desire to use the branching system to improve  talent alignment created a new objective for the system, thus changing its
mission yet another time.
Since the decision to integrate cadet ratings into the branching mechanism took place under an abbreviated timeline, the US Army maintained the same strategy
space for the mechanism as in previous years and devised the USMA-2020 mechanism to accommodate heterogeneous branch priorities.  
In their design, the Army created two less-than-ideal theoretical possibilities in the USMA-2020 mechanism. 
First, a cadet could be charged BRADSO under the USMA-2020 mechanism even if she does not need heightened priority to receive a position at that branch.
While this was also possible under USMA-2006, it was nearly four times as common under USMA-2020. 
 Second, under USMA-2020, a cadet's willingness to BRADSO for a branch could improve her priorities even for the base-price positions.
Surveys of cadets designed by the military coauthor of this paper showed that these aspects potentially undermined trust in the branching system, 
and led the Army to reconsider a refinement of the cumulative offer mechanism, despite its more complex strategy space. 
At that stage, the Army established a partnership with the two civilian coauthors of this paper after
nearly a decade since  \cite{sonmez/switzer:13} was first brought to their attention.

As  the second main contribution of our paper, we report  the design and successful deployment of a new branching system for the Class of 2021, 
the \textit{multi-price cumulative offer mechanism\/},  a refinement of the  cumulative offer mechanism  \citep{hatfield/milgrom:05}
that uses a specific choice rule for each branch that reflects the Army's objectives of retention and talent alignment.  
In our main formal result, Theorem \ref{cosm}, we show that the Army's objectives, when formulated through five axioms, 
uniquely give rise to the \textit{multi-price cumulative offer mechanism\/}.  
Thus, in our setting, the cumulative offer process and a
specific choice rule emerge from foundational axioms,  even though branches are not assumed to be endowed with choice rules in our model. 
Therefore, the foundations for two major components of the branching system, the allocation procedure and the choice rules that feed into the procedure, 
are established jointly in our main formal result. As we further elaborate in Section \ref{sec:significance}, this is a departure from earlier 
literature where the foundations for these two parts are established separately. 

The axioms which characterize the multi-price cumulative offer mechanism are as follows:

\textit{Individual rationality\/}: No cadet should be assigned an unacceptable branch-price pair. 

\textit{Non-wastefulness\/}:  No position at a branch can be left idle while there is a cadet who is unassigned,  
unless she would rather remain unassigned than receive the position at the cheapest possible price. 

\textit{No priority reversal\/}: No cadet $i$ should prefer the branch-price package $(b,t)$ of another cadet $j$ to her own assignment, even though
she had a higher baseline priority for branch $b$. 

\textit{Enforcement of the price responsiveness policy\/}: No cadet $i$ should prefer a branch-price package $(b,t)$ to her own assignment while there is another cadet $j$ who received a position at branch $b$ at a different price $t' \not= t$, 
even though 
\begin{enumerate}
\item cadet $i$ has higher claim for a flexible-price position at branch $b$ at price $t$ than cadet $j$ has at price $t'$, and 
\item it is feasible to award the
 flexible-price position at branch $b$ to cadet $i$ at price $t$ instead of cadet $j$ at price $t'$.  
\end{enumerate}

\textit{Strategy-proofness\/}: No cadet ever benefits from misrepresenting her preferences over branch-price pairs. 

Of these axioms, only \textit{enforcement of the price responsiveness policy\/} is novel to our model and analysis. 
This condition perfectly formulates a key objective for the Army and is therefore critical for the broader mission of the branching system. 
Moreover,  while our formal analysis is motivated by the Army's branching application, it can be directly applied for any
extension of a priority-based indivisible goods allocation model (such as the \textit{school choice\/} model by \cite{abdulkadiroglu/sonmez:03}), 
where priorities of individuals can be increased with a costly action (such as paying higher tuition) at a fraction of the units of allocated goods.\footnote{It is already well-established that any such extension can be modeled as a special case of the matching with contracts model.   However, the novel insight we offer is that the underlying choice rules for institutions can also be endogenously obtained 
in such extensions as a direct implication of natural axioms for these extensions.    These choice rules have been exogenously given in earlier literature. 
This distinction is the sense in which our analysis unifies the design of two major components of a priority-based allocation system, 
i.e., the design of its allocation procedure for given choice rules and the design of the choice rules that feed into the allocation procedure. See Section
Section \ref{sec:significance} for further elaboration and significance of the point.} 
In particular, our model has an additional direct application for the seat purchasing policies at Chinese high schools, 
presented in Section  \ref{seatpurchasing}.  Another promising application is the allocation of school seats with or without financial aid (cf. \cite{sonmez/switzer:13}, \cite{artemov/che/he:20}, \cite{Hassidim/Romm/Shorrer:21}, \cite{shorrer/sovago:21}). 

\subsection{Organization of the Rest of the Paper}

The rest of this paper is organized as follows.  Section \ref{sec:min} describes the main elements of
minimalist market design and relates the evolution of the US Army's branching system to this design paradigm.
Section \ref{sec:model} introduces the formal model. This section also includes
Section \ref{sec-axioms} which formalizes Army's desiderata as rigorous axioms.
Section  \ref{sec:USMA-2020} presents the USMA-2020 mechanism and its shortcomings 
which convinced the Army to reconsider a version of the cumulative offer mechanism previously
advocated by \cite{sonmez/switzer:13}. 
Section \ref{subsec:MPCOADSO} presents the multi-price cumulative offer mechanism and our 
main theoretical result characterizing it as the unique mechanism that satisfies the Army's desiderata.
Section \ref{sec:evolution} describes how the Army used the new mechanism to refine
the trade-off between retention and talent alignment by changing the main parameters of this mechanism.
Section \ref{sec:otherapplications} provides details on another real-life application of our model.
Section \ref{sec:conclusion} provides some general lessons for broader efforts in institution design.
All proofs, independence of the axioms in our main characterization result,
an in-depth analysis of the USMA-2020 mechanism, additional data analysis, results from cadet surveys, and other potential applications 
are presented in the Online Appendix.

\section{Minimalist Market Design} \label{sec:min}

\subsection{Overview}
This section summarizes some essential elements of  the minimalist approach to market design formulated in 
\cite{sonmez:23}. The approach integrates research and policy efforts to influence the design of real-life allocation systems.  
We describe the framework to explain some of our modeling choices for the US Army branch assignment process and to
elaborate on our strategy for convincing stakeholders to adopt a new system.   
We first review the general framework and then describe how it relates to the Army's branching system.
Though minimalist market design has evolved through other experiences in other resource allocation
problems, our integrated research and policy efforts to reform the US Army's branching process 
is its first proof-of-concept where the approach is directly tested and  deployed systematically. That is, from its inception in \cite{sonmez/switzer:13},
the effort to influence the design of the US Army's branching process followed the framework, and 
eventually succeeded.

The minimalist approach to institution design or reform can be particularly useful when there is no consensus for reform, 
and the design economist is initially an outsider.
When a consensus for reform exists and the design economist is commissioned to guide a reform, 
the cautious approach underlying the minimalist framework may not be necessary.  In this case, 
stakeholders often commission the market designer to devise a system and delegate the critical design decisions to her expertise.  
 To propose a design, the market designer can deploy the usual economic design tools, including game-theoretic models, constrained optimization approaches, simulations, lab experiments, or computational heuristics.

When stakeholders need convincing, however, a redesign is aspirational, and the minimalist approach helps.
When a redesign effort starts with a criticism of an existing institution, stakeholders are often defensive, so gaining trust is imperative.
Instead of challenging the main features of the existing institution,
the premise of the minimalist approach is that the existing institution reflects the objectives and constraints of stakeholders.
With multiple stakeholders,  a range of views about objectives may exist. 
Some objectives may not even be consequentialist, and a single objective function in which everyone agrees may not exist, 
especially when the problem has incommensurable dimensions. 
Indeed, the involvement of the market designer itself may interfere with delicate balances across constituents, including
those purposely left implicit.

The first step of minimalist market design is to identify the objectives that are key to stakeholders in designing their current institution. 
An in-depth understanding of the historical evolution of a system aids in determining these considerations. 
It is essential to recognize the multi-faceted nature of these principles. 
Situations with delicate social and distributional elements may necessitate respecting a range of viewpoints  \citep{hitzig:20}. 
Incomplete or superficial knowledge about the origin or rationale for certain institutional practices risks undermining the credibility of reform efforts. 
During the initial design of an institution, formal tools may not have been available to policymakers and system operators. 
Such designs may be improved with a rigorous formulation of the underlying objectives and constraints of the institution. 
In our experience, even when stakeholders are able to verbalize their objectives, they still need help with operationalizing them with a procedure. 
If the policy proposals of a design economist are in line with the stakeholder objectives, stakeholders may be 
open to reform.  
It is important to emphasize that a standard optimization framework may not be possible when principles are complex or conflicting. 
Therefore, mainstream approaches from traditional mechanism design may not receive a favorable reception.

The next step is to examine whether the existing institution satisfies stakeholders' fundamental objectives 
or if there is a discord between the intention and the practice. 
Here, the aim is to provide the stakeholders with a critique of the existing institution on the terms laid out by themselves, 
rather than with a critique that is based on primary considerations in mainstream economics 
(such as preference utilitarianism).\footnote{Our approach is consistent with 
\cite{chassang/ortner:22}, who make a similar point in the context of regulating collusion: ``In 
addition, we try to do justice to the peculiarities of the legal system: modeling the courts
as they are, rather than as economists think they should be, is essential for economic analysis to 
improve the way collusion is regulated.''}   The inconsistency between the aims of the institution and its practice in the field 
can take many forms and be more than just incentive and participation issues central to mechanism design. 
Identification of such an inconsistency creates an opening for the market designer. If some stakeholder objectives are not satisfied by 
the existing institution,  then a strong case can be made for reform, provided that 
the market designer advocates for an alternative institution that satisfies all objectives. 

When a  disconnect is identified between the intention and the practice of an institution, 
stakeholders are more likely to be receptive to reform if it involves minimal interference with the existing system. 
To design such an alternative, it is imperative to find the root causes of the failures. 
Once the culprits are identified in the discord between the objectives and the practice of an institution, 
the aim is to correct those issues only and otherwise leave the rest of the system untouched.  
That is, any interference with the existing institution should target the root causes of the issues and aspire to a surgical fix.
This allows stakeholders to position the reform as a relatively small tweak. It also reduces the risk of changing aspects
of the institution in which the market designer might be unaware and upsetting implicit compromises between stakeholders.
The aim is for stakeholders to see the proposal as representing what they wanted to do in the first place, 
but did not have the technical know-how to formulate.

The model's realism is important if criticizing an existing system because stakeholders are often wedded to the status quo. 
Models with abstractions, even those intentionally made to isolate specific intuitions, can be easily dismissed as being unrealistic. 
Furthermore, if particular design choices have normative implications, they should be made transparent. \cite{li:17} calls this 
maintaining \textit{informed neutrality} between reasonable normative principles.
For establishing trust and mitigating concerns about ulterior motives or hidden agendas, it is in the best interest of a market designer to elucidate implications for an aspired reform than take positions on tradeoffs.\footnote{Pro-bono assistance also helps to build trust.} 
Beyond the pragmatic considerations, it is also a good practice to be completely transparent about various normative implications
of any design. A minimalist market designer should aim to provide stakeholders with tools to examine the implications of 
particular design choices and help to facilitate an open and informed debate about their system. 

Since the starting point of the aspired reform is to find a way to accommodate stakeholders' key principles, the axiomatic approach is a natural methodology \citep{moulin:88,moulin:04,thomson:01,thomson:11}.  
In the axiomatic methodology, the researcher formalizes principles as mathematical properties and examines their implications.  
In some cases, only a unique system or a family of systems satisfies all requirements, 
a result known as an axiomatic characterization. If such a characterization exists, it provides a natural candidate for practical implementation. 
To the extent it is technically possible, 
finding all systems that meet the objectives is the best practice because it describes the landscape of possibilities, including the identification of 
systems that may have unintended effects. 

Since the minimalist framework starts with the existing institution, it is best suited for pursuing incremental changes within the system. Once a market designer has shown apparent deficits with the existing system, it may be possible to consider more substantial changes. At this stage, questions regarding the implications of taking some aspects of the problem as given or whether they can be modified are most fruitful. In the ideal scenario, the market designer partners with stakeholders and can jointly design the institution. Stakeholders can rely on the market designer for technical expertise and delegate any formal analysis of specific design changes based on their expertise. Through this iterative process, it is possible to move from local changes to more substantial changes.\footnote{This process of continuous improvement has been emphasized in other policy contexts, including the \cite{duflo:17} metaphor of plumbing in development economics.}
 
\subsection{Minimalist Market Design to Reform the US Army's Branching Process}\label{history}

This section provides background on cadet branch assignment in the US Army
and the relationship to the minimalist approach. 
For decades, the Army has offered cadets choice over their branch assignment and has used a
cadet's performance ranking, known as the \textit{order of merit list (OML)}, as a factor in
determining assignments.\footnote{The OML was first formalized in 1818 when the Army's Secretary of War
approved USMA's criteria.   Army documents from that period describe the importance
of respect priority, stating that ``the distribution of cadets, into the branches of the army,
be made in accordance with their qualifications, talents, and without violating the principle of 
order of merit'' \citep{topping}. }
Through the late 1970s, cadet assignment was an in-person process, where
cadets convened in an auditorium.  Cadet names were called in OML order,
and each cadet selected their most preferred branch with available capacity.\footnote{
\cite{atkinson:09} provides a vivid account of the process for the West Point Class
of 1966.} Starting in the 1980s, cadets submitted preferences over the set of branches, and a branching board convened to match cadets to branches \citep{usma:82}.  
In the mechanism, the highest-OML cadet was assigned her most-preferred branch, 
the second highest-OML cadet was assigned her most-preferred branch among branches with remaining positions, and so on.   
This mechanism, the \textit{simple serial dictatorship induced by OML\/} (SSD-OML), 
established several foundational components of the assignment system, and formed the basis for further reforms.

A new objective of encouraging retention arose due to declining junior officer retention rates during the late 1990s and early 2000s. The Army
offered a menu of retention incentives to cadets at USMA and ROTC through the Officer Career Satisfaction Program, 
first implemented in 2006 \citep{colarusso/lyle/wardynski:10}.
The most popular incentive, which involved a reform of the branching mechanism, was the  \textit{BRADSO\/} program. 
The BRADSO program gives higher priority for a fraction of positions in each branch to cadets willing to extend their Active Duty Service Obligation (ADSO) by three years if assigned to  that branch.  We call these \textit{flexible-price positions} and say a cadet who ranks such a position is willing to pay the increased price. 
By creating these new types of positions, the BRADSO program altered the role of the OML for these slots.
To infer which cadets were willing to pay the increased price,
USMA required cadets to report the set of branches they were willing to serve the additional years
through a new message space under a new mechanism, the USMA-2006 mechanism.

\subsubsection{USMA-2006}

\cite{sonmez/switzer:13} formulate and analyze the USMA-2006 mechanism.   The USMA-2006 mechanism extends the SSD-OML
to accommodate the treatment of the flexible-price positions. 
When a cadet expresses willingness to pay the increased price for the flexible-price positions at any branch, that cadet is given priority over any other cadet unwilling to pay the increased price at these positions. When two cadets are willing to pay the increased price, they are ordered according to their  OML. If a cadet is assigned a base-price position, she is charged the base price. If a cadet is assigned one of the flexible-price positions, she is charged the increased price if she is willing to pay the increased price at the branch. Otherwise, she is charged the base price. Appendix \ref{sec:usma2006}  provides a formal mechanism definition.

Whether the Army should override the OML to increase retention was subject to intense debate.  \cite{colarusso/lyle/wardynski:10}
write: ``Devoted supporters of the ROTC and West Point Order of Merit (OML)
system for allocating branches and posts objected that low OML cadets could ``buy'' their branch or post of choice ahead of higher
OML cadets. Since branch and post assignments represent a zero-sum game, the ability of 
cadets with lower OML ranking to displace those above them was viewed by some as unfair or as undermining the OML
system.''  This discussion illustrates that stakeholders had different views
on the appropriate balance between retention incentives and merit, an issue subject to ongoing debate.

In USMA-2006, cadets only submit their preferences over branches alone and ``signal'' their willingness to pay the increased price at any branch rather than over branch-price pairs.  A direct mechanism would solicit cadet preferences over branch-price pairs. \cite{sonmez/switzer:13} describe two main failures of USMA-2006. First, cadet $i$ can prefer cadet $j$'s assignment to her own, and cadet $i$ can have a higher OML score than cadet $j$.  We refer to this situation as a \textit{priority reversal}.     Computation of \textit{all} priority reversals depends on knowledge of cadet preferences over all branch-price pairs.  Detection of \textit{some} priority reversals only requires information on cadet preference collected under the USMA-2006 mechanism.   We denote these as \textit{detectable priority reversals.} Second,  under USMA-2006, a cadet who is assigned an increased-price position at branch $b$ can potentially receive that position at a base price by declaring that she is unwilling to pay the increased price at the branch.  We refer to this as a \textit{failure of BRADSO-incentive compatibility}.  After introducing the model in Section \ref{sec:model}, we define these concepts formally.

The issue with the message space and a disconnect between branch and price assignments are the 
two root causes of the problem with the USMA-2006 mechanism.  When a cadet volunteers for BRADSO
at her top choice, the mechanism cannot tell whether she prefers her first choice branch at increased price over her second choice branch at cheaper base price.    \cite{sonmez/switzer:13} proposed fixing the first issue by simply changing the message space of the mechanism. 
The disconnect between branch and price assignments were then addressed via the \textit{cumulative offer process\/} \citep{hatfield/milgrom:05}. 

Despite the shortcomings of the USMA-2006, for many years the Army did not embrace the \cite{sonmez/switzer:13} proposal.
The Army did not change its mechanism for three main reasons:
\begin{enumerate}
\item The Army can manually correct a failure of BRADSO-incentive compatibility or a detectable priority reversal ex-post. Both issues involve a cadet needlessly paying the increased price at her assigned branch. The Army can resolve either issue by manually reducing the charged price to the base price.

\item Even though the USMA-2006 mechanism allows for additional priority reversals, which cannot be manually corrected ex-post,  
verifying any such theoretical failure relies on cadet preferences over branch-price pairs.  Since USMA-2006 is not a direct mechanism, information on cadet preferences over branch-price pairs is unavailable.

\item Failures of BRADSO-incentive compatibility and detectable priority reversals
have been relatively rare in practice. 
\end{enumerate}

The Army initially thought that the issues identified by \cite{sonmez/switzer:13} are not significant enough to justify 
adopting a mechanism that has a more complex message space. 
Any possible failure of the desiderata above could either be manually corrected ex-post or could not
be verified with data solicited under the message space for the USMA-2006 mechanism. Therefore the Army concluded, 
the failures identified by \cite{sonmez/switzer:13} were not visible or significant enough to lead to a change. The introduction of a new program aimed at improved talent alignment altered these tradeoffs and triggered an adjustment in the mechanism, which we describe next.

\subsubsection{Talent-Based Branching and the USMA-2020 Mechanism}\label{sec:usma2020}

In 2012, the US Army introduced Talent-Based Branching to develop a ``talent market'' where additional information about each cadet influences the priority a cadet receives at a branch \citep{colarusso2016starting}. Instead of relying only on the OML, 
Talent-Based Branching allowed branches and cadets to align their interests and fit better with one another. Under Talent-Based Branching, 
branches prioritize cadets into three tiers: high, medium, and low. Before the Class of 2020, these rating categories did not influence baseline branch priorities at USMA. The Army used these ratings as part of talent assessments to help cadets learn which branches would be a good fit for them. The Army also made rare, ex-post adjustments to a cadet's branch assignment based on ratings. After several years and much debate, the Army decided to use ratings to adjust the underlying OML-based prioritization for the Class of 2020. The slow pace of reform was due partly to ongoing debates between a faction in favor of granting branches more power to directly influence branch assignments and another faction concerned about diluting the power of the OML (see, \cite{garcia:20}).

Just as the introduction of the BRADSO program triggered a reform in the branching mechanism, the full integration of the TBB program with the
branching process resulted in another adjustment. The Army operated under an abbreviated timeline, and their perspective focused on coming up with an algorithm rather than issues brought about by the new structure of claims for branches created by the TBB program. The \textit{US Army News} suggests that National Residency Matching Program inspired the design of USMA-2020 mechanism.\footnote{\cite{oconnor:19}
states: ``The cadets’ branch rankings and the branches’ cadet preferences will then determine a cadets’ branch using a modified version of the National Resident Matching Program’s algorithm, which won a Nobel Prize for Economics in 2012 and pairs medical school graduates with residency programs.''} The Army replaced the USMA-2006 mechanism with another quasi-direct mechanism based on the 
individual-proposing deferred acceptance algorithm,\footnote{Section  \ref{sec:da} in the Online Appendix defines the DA algorithm.} 
where branches have heterogeneous baseline priorities over cadets according to a tiered price responsiveness policy described in Section \ref{subsec:tiered}. This procedure separated the assignment of branches from the charging cadet base or increased prices. After branch assignments were determined, a cadet's willingness to pay the increased price determined price charges. The Army charged the increased price to willing cadets in reverse priority order, stopping when 25 percent of cadets assigned to the branch were charged the increased price. For example, if 100 cadets are assigned to a branch and 50 of the cadets volunteer for the increased price, the Army would charge the increased price to the 25 lowest priority cadets of the 50 willing to pay the increased price. We formally define this mechanism after introducing the model in the next section.

The Army created two less-than-ideal theoretical possibilities in the USMA-2020 mechanism in their design. 
First, a cadet can be charged BRADSO under the USMA-2020 mechanism even if she does not need heightened priority to receive a position at that branch.
While this was also possible under USMA-2006, it was nearly four times as frequent under USMA-2020. 
Second, under USMA-2020, a cadet's willingness to BRADSO for a branch can improve priorities even for base-price positions.
Surveys of cadets showed that these aspects potentially undermined trust in the branching system, 
and led the Army to reconsider the cumulative offer mechanism, despite its more complex message space. 

We next introduce our model before formally describing  USMA-2020, and elaborating on its exacerbated failures.

\section{Model}\label{sec:model}

A set of individuals  $I$ seek placement at one of a set of institutions $B$.  
Since our primary application is US Army branching, we refer to an individual as a \textbf{cadet} 
and  an institution as a \textbf{branch}.\footnote{Section \ref{sec:otherapplications} presents other direct applications of our model outside
of the US Army context.} 
At any given branch $b \in B$, there are $q_b$ identical positions.
Each cadet wants at most one position
and can be assigned a branch under multiple contractual terms. 
Let $T = \{t^0,t^1, \ldots, t^h\}$ denote a finite set of contractual terms or ``prices''  where 
\begin{enumerate}
\item $t \in \mathbb{R}^+$ for each $t\in T$, and 
\item $t^0 < t^1< \cdots t^h$. 
\end{enumerate}
Here,  $t^0$  denotes the \textbf{base price}, and it represents the default arrangement. 
In our Army application, it corresponds to $t^0$ years of mandatory service upon completion 
of the USMA Military program or the ROTC program. 
 Let $T^+ = T\setminus\{t^0\}$ denote the \textbf{set of increased prices}.\footnote{We assume
 that the set of contractual terms $T$ is a finite subset of real numbers due to the price interpretation of the contractual terms
 in our main application. Our entire analysis directly extends to any finite and strictly ordered set of contractual terms $T$. }
 For the Army application, a single increased price corresponds to $t^h = t^1$ years of mandatory service through
the  BRADSO program.\footnote{In our US Army application,
the base price corresponds to three to five years of mandatory service
and the increased price corresponds to three additional years of mandatory service. USMA graduates incur a five-year service obligation upon graduation. ROTC graduates incur a three or four-year service obligation upon graduation. Incurring the increased price through the BRADSO program extends the initial service obligation for USMA and ROTC cadets by three years (Army Regulation 350-100).\nocite{ar350_100}}

For any branch $b\in B$, at most $q^f_b \in [0,q_b]$ positions can be assigned to cadets at any increased price in $T^+$.  
We refer to these positions as \textbf{flexible-price} positions. For any branch $b\in B$,  
let $q^0_b = (q_b-q^f_b)$ denote the number  of \textbf{base-price} positions which can be assigned only at the default price of $t^0$. 

\subsection{Cadet Preferences and Baseline Branch Priorities}

Each cadet has a  strict preference relation on branch-price pairs and remaining unmatched, represented by a 
linear order on $(B \times T) \cup \{\emptyset\}$.
We assume that, for any cadet $i\in I$ and  branch $b\in B$,  cadet $i$ strictly
prefers a cheaper position at branch $b$ to a more expensive position at branch  $b$.
Let $\calq$ denote the set of linear orders  on $(B \times T) \cup \{\emptyset\}$  identified by this assumption.   
Therefore, for any cadet $i \in I$,\, preference relation $\succ_i\;\in\calq$, branch $b \in B$,  and pair of prices $t, t' \in T$, 
\[ t < t' \quad \implies \quad (b,t) \; \succ_i  (b,t').
\]
For any strict preference relation $ \succ_i \, \in \calq$, let $\succeq_i$ denote the induced weak preference relation.

Let $\Pi$ denote the set of all linear orders on the set of cadets $I$.
Each branch $b \in B$ has a strict priority order $\pi_b \in \Pi$ on the set of cadets $I$. 
We refer $\pi_b$ as the \textbf{baseline priority order} at branch $b$. 
The baseline priority order represents the ``baseline claims'' of cadets for positions
at the branch. 

\subsection{Price Responsiveness Policies}

Given any branch $b\in B$, the overall claims of cadets for positions at the branch depend on 
both the baseline priority order $\pi_b$ and how much cadets are willing to pay for a position at the branch.  
The Army policy fully specifies the scenarios under which the baseline priority order at a branch can be overturned 
due to cadets who are willing to incur higher prices.
This tradeoff is captured by a \textit{price responsiveness policy\/},
which specifies the priority advantage any given cadet gains against other cadets if she is willing to bear 
a higher price.\footnote{A price responsiveness policy in our model is similar to the marginal rates of substitution from price theory.} 

Formally, for a given branch $b\in B$ and a baseline priority order $\pi_b \in\Pi$, a \textbf{price responsiveness policy}  
is a linear order $\omega_b$  on $I\times T$  with the following two properties:
\begin{enumerate}
\item for any pair of cadets $i,j \in I$ and price $t \in T$, 
\[ (i, t) \; \omega_b \; (j,t) \quad \iff \quad i \; \pi_b \; j \quad \mbox{ and }\]
\item for any cadet $i \in I$ and price pair $t,t'\in T$, 
\[  t < t' \quad \implies \quad  (i, t') \; \omega_b \; (i,t). \] 
\end{enumerate}
For the positions in which the price responsiveness policy is invoked at branch $b$, (i) the relative priority order of cadets 
who are willing to pay the same price
is the same as in their baseline priority order $\pi_b$, and (ii) any given cadet 
has a higher claim with a higher price compared to her claims at a lower price. 
A price responsiveness policy can be invoked at a branch for some or all of its positions.

Let $\Omega_b$ be the set of all linear orders on $I\times T$  which satisfy these two conditions. 
The set $\Omega_b$ denotes the set of all price responsiveness policies at branch $b$.

The advantage a price responsiveness policy gives to cadets in securing a position at branch $b$
due to their willingness to pay higher prices differs between distinct price responsiveness policies. 
Given two distinct price responsiveness policies $\omega_b, \nu_b \in \Omega_b$, policy $\nu_b$ 
\textbf{is more responsive to a price increase} than policy $\omega_b$, if
\[ \mbox{ for any } i,j \in I \mbox{ and } t,t'\in T \mbox{ with } t'>t, \qquad (i, t') \; \omega_b \; (j,t) \; \implies \;  (i, t') \; \nu_b \; (j,t).
\]

We next present three price responsiveness policies used in practice.

\subsubsection{Ultimate Price Responsiveness Policy} \label{subsec:ultimate}

Given a branch $b\in B$ and a baseline priority order $\pi_b \in\Pi$, define the \textbf{ultimate price responsiveness policy} 
$\overline{\omega}_b \in \Omega_b$ 
as one where willingness to pay  any higher price overrides any differences in cadet ranking under the
baseline priority order $\pi_b$ at branch $b$. That is, for  any pair of cadets $i,j \in I$ and pair of prices $t,t'\in T$, 
\[ t'>t \quad \implies \quad (i,t') \; \overline{\omega}_b \; (j,t).
\]

As we have indicated earlier, the Army application has only one increased price. 
For the Classes of 2006-2019, the USMA used the ultimate price responsiveness policy. During these years, the USMA capped the positions that could be 
assigned at the increased price at 25 percent of total positions within each branch.  
At each branch $b\in B$, any
cadet who is willing to pay the increased price for branch $b$ had higher priority for the 
$q_b^f$ flexible-price positions than any cadet who is not willing to pay the increased price for branch $b$.

\subsubsection{Tiered Price Responsiveness Policy} \label{subsec:tiered}

Fix a branch $b\in B$ and a baseline priority order $\pi_b \in\Pi$. 
To define our second price responsiveness policy, partition cadets into $n$ tiers $I_b^1, I_b^2, \ldots, I_b^n$
so that for any  two tiers $\ell, m \in \{1,\ldots,n\}$ and pair of cadets $i,j \in I$,  
    \[ \left. \begin{array}{l}
  \ell < m,\\   
  i \in I_b^{\ell} \;, \mbox{ and}\\ 
  j \in I_b^m  \end{array}  \right\}
\implies \; i \; \pi_b \; j.
\]
In this partition, any cadet in  tier $I_{b}^{\ell}$ has higher baseline priority at branch $b$ than a cadet
in tier $I_{b}^{m}$ for $\ell < m$.  

Under a \textbf{tiered price responsiveness policy\/} $\omega^T_b \in \Omega$, for any  tier $\ell \in \{1,\ldots,n\}$, triple of cadets $i, j, k\in I$, 
and pair of prices $t,t'\in T$ with $t'>t$,  
    \[ \left. \begin{array}{l}
 i \; \pi_b \; k ,\\   
j \; \pi_b \; k, \; \mbox{ and}\\ 
  i,j \in I_b^{\ell}  \end{array}  \right\}
\quad \implies \qquad  \Bigg((k, t') \; \omega^T_b \; (i,t) \quad \iff \quad  (k, t') \; \omega^T_b \; (j,t)\Bigg). 
\]
That is, given two cadets $i,j\in I$ in the same tier and  
a third cadet $k\in I$ with lower  $\pi_b$-priority than both $i$ and $j$,  cadet $k$ can gain priority over cadet $i$ 
through willingness to pay a higher price if and only if  cadet $k$ can gain priority over cadet $j$  through willingness to pay  a higher price.

For the Classes of 2020 and 2021, the USMA used two different tiered price responsiveness policies.  
In both years, cadets were prioritized by each branch into one of three tiers, which we 
denote high, middle, and low.\footnote{Branch rating categories are known to cadets and finalized before cadets submit their preferences for branches.}  
In 2020, when a cadet expressed a willingness to pay the increased price, she had higher priority among
cadets in the same tier. For example, a middle-tier cadet who
was willing to pay the increased price would not obtain higher
priority than a high-tier cadet who was unwilling to pay the increased price. 
Therefore,  under the 2020 policy, the willingness to  pay the higher price overrides any difference in 
cadet ranking under $\pi_b$ only among cadets in the same tier.
The price responsiveness policy for the Class of 2021 granted cadets more advantage 
in securing a position. In 2021, when a cadet expressed a willingness 
to pay the increased price,  she had higher priority over all other cadets if she was in the medium or high-tier categories. 
Low-tier cadets who expressed a willingness to pay, in contrast,   only received higher priority among other low-tier cadets.  
The ultimate policy is more responsive to a price increase than the 2021 policy, which is in turn
more responsive to a price increase than the 2020 policy.

\subsubsection{Scoring-Based Price Responsiveness Policy} \label{subsec:scoringbased}

Our third price responsiveness policy is when the baseline priority ranking is based on an underlying score 
(such as from a standardized test). 
Under the \textbf{scoring-based price responsiveness policy}, each level of increased price 
increases the total score by a given amount.   

Given a branch $b\in B$ and individual $i \in I$, let $m^b_i \in \mathbb{R}^+$ denote the merit score of individual  
$i$ at branch $b$.\footnote{Suppose that any ties between two distinct individuals is broken with a tie-breaking rule, so that  
no two distinct individuals have the same merit score at any given branch.} The baseline priority order $\pi_b \in \Pi$
is such that, for any pair of individuals $i,j \in I$, 
\[   i \, \pi_b \, j \quad \iff \quad m^b_i > m^b_j.
\]
Given a branch $b\in B$, let $S^b :  T \rightarrow \mathbb{R}^+$ be a \textbf{scoring rule} such that, 
\[ 0 = S^b(t^0) < S^b(t^1) < \cdots < S^b(t^{h-1}) < S^b(t^h). 
\]

Under a \textbf{scoring-based price responsiveness policy\/} $\omega^S_b \in \Omega$, for any two
individual-price pairs $(i,t), (j,t') \in I\times T$, 
\[   (i, t) \; \omega^S_b \; (j,t') \quad \iff \quad m^b_i + S^b(t) > m^b_j + S^b(t').\footnote{Suppose that any ties 
are broken with a given tie-breaking rule.} 
\]

Drawing upon an analysis  in \cite{wang/zhou:21}, in Section \ref{seatpurchasing} we present 
a real-world application of the scoring-based price responsiveness policy 
for public high school admissions in China. 
Under this policy, student merit scores receive a boost for 
a fraction of seats, if they are willing to pay higher tuition.

\subsection{Formulation through the Matching with Contracts Model}
To introduce the outcome of an economy and some of the mechanisms analyzed in the paper, 
we use the following formulation through the \textit{matching with contracts} model by \cite{hatfield/milgrom:05}. 

For any $i \in I$,  $b \in B$, and $t \in T$, 
the triple $x=(i,b,t)$ is called a \textbf{contract}. It represents
a bilateral match between cadet $i$ and branch $b$ at the price of $t$. 
Let 
\[ \calx = I \times B \times T \] 
denote the set of  all contracts.  
Given a contract $x \in  \calx$, let $\i(x)$ denote the cadet, $\b(x)$ denote the branch, and $\t(x)$ denote the price of the contract $x$.  
That is, $x = \big(\i(x), \b(x), \t(x)\big).$

For any  cadet $i\in I$,  let 
\[ \calx_i = \{x \in \calx : \i(x) = i\} \] 
denote the set of contracts that involve cadet $i$. 
Similarly, for any branch $b \in B$, let  
\[ \calx_b = \{x \in \calx : \b(x) = b\} \] 
denote the set of contracts that involve branch $b$. 
For any cadet $i\in I$, preferences $\succ_i \, \in \calq$ defined over $B \times T \cup \{\emptyset\}$ 
can be redefined over $\calx_i \cup \{\emptyset\}$ (i.e. her contracts and remaining unmatched) by simply 
interpreting a branch-price pair $(b,t) \in B\times T$ in the original domain as a contract between cadet $i$ and branch $b$ at price $t$ in the new domain.

An \textbf{allocation} is a (possibly empty) set of contracts $X \subset  \calx$, such that
\[
\begin{array}{ll}
\mbox{(1)\; for any } i \in I, \qquad   & |\{x \in X : \i(x) = i\}| \leq 1,\\ 

\mbox{(2)\; for any } b \in B, \qquad   & |\{x \in X : \b(x) = b\}| \leq q_b, \quad   \mbox{ and }\\ 

\mbox{(3)\; for any } b \in B, \qquad  & |\{x \in X : \b(x) = b  \mbox{ and } \t(x) \in T^+\}| \leq q^f_b.  
\end{array}
\]
That is, under an allocation  $X$, no individual can appear in more than one contract,  no branch $b$ can appear in more contracts 
than the number of its positions $q_b$, and 
no branch $b$ can appear in more contracts than  $q_b^f$  along with an increased price in $T^+$. 
Let $\cala$ denote the set of all allocations. 

For a given allocation $X \in \cala$ and cadet $i\in I$, the \textbf{assignment} $X_i$ of cadet $i$ under allocation $X$ is defined as
\[ X_i =  \left\{ \begin{array}{cl}
         (b,t)  & \mbox{ if } (i,b,t) \in X,\\
        \emptyset & \mbox{ if } X \cap \calx_i = \emptyset. \end{array} \right.
\]
For the latter case, i.e. if $X_i = \emptyset$, we say that cadet $i$ is \textbf{unmatched} under $X$. 

Similarly, for a given allocation $X \in \cala$ and branch $b$,  define
\[ X_b = \left\{(i,t) \in I\times T \; : \; (i,b,t) \in X\right\}.  
\]

Given an allocation $X \in \cala$ and a cadet $i\in I$, with a slight abuse of the notation,\footnote{The abuse of notation is due to the fact that
while the argument of the functions $\b(.)$, $\t(.)$ are previously introduced as a contract, here it is an assignment. Since a cadet and an assignment 
uniquely defines a (possibly empty) contract, the notational abuse is innocuous.} 
define the \textbf{branch assignment} $\b(X_i)$ of cadet $i$ as
\[ \b(X_i) =  \left\{ \begin{array}{cl}
         b  & \mbox{ if } (i,b,t) \in X,\\
        \emptyset & \mbox{ if } X \cap \calx_i = \emptyset. \end{array} \right.
\]
	
Given an allocation $X \in \cala$ and a cadet $i\in I$, with a slight abuse of the notation,
define the \textbf{price assignment} $\t(X_i)$ of cadet $i$  as
\[ \t(X_i) =  \left\{ \begin{array}{cl}
         t  & \mbox{ if } (i,b,t) \in X,\\
        \emptyset & \mbox{ if } X \cap \calx_i = \emptyset. \end{array} \right.
\]	
	
A \textbf{mechanism} is a message space $\cals_i$ for each cadet $i \in I$	along with an outcome function 
\[ \varphi : \prod_{i\in I} \cals_i \rightarrow \cala\]
that selects an allocation for each strategy profile.  Let $\cals = \prod_{i\in I} \cals_i$. 

A \textbf{direct mechanism} is a mechanism where  $\cals_i = \calq$ for each cadet $i \in I$. 
We denote a direct mechanism with its outcome function only, suppressing its message space
which is always $\calq^{|I|}$.

Given a mechanism $\big(\cals, \varphi\big)$, the resulting \textbf{assignment function} $\varphi_i: \cals \rightarrow B\times T \cup \{\emptyset\}$
for cadet $i\in I$ is defined as follows: For any $s \in \cals$ and $X = \varphi(s)$,
\[   \varphi_i(s) = X_i. \]

\subsection{Primary Desiderata for Allocations and Mechanisms} \label{sec-axioms}	

The history of cadet branch assignment in Section \ref{history} describes some of the system's goals and origins.  
Using the notation introduced in the last section,  we next formulate these goals as formal axioms. 

Our first axiom is \textit{individual rationality\/}.   
The Army cannot compel an assignment because a cadet always has the option to leave the Army. When cadets fail to complete their initial service obligation, they must reimburse the government's education cost according to Army Regulation 150-1 \citep{ar150_1}. For West Point Graduates in the Class of 2018, this cost was \$236,052 \citep{usma:19}. When a cadet voluntarily leaves the Army and pays the fine, we denote this outcome as unmatched. In the last two decades, between 5-10\% of West Point graduates have not fulfilled their commitment.\footnote{In some cases, like a medical or health issue, a cadet does not need to reimburse Army for early separation.}  
At the ROTC, unmatched cadets are placed in reserve duty. For that application, the unmatched outcome corresponds to reserve duty. For the Classes of 2022 and 2023, about 10\% of ROTC cadets are unassigned and placed in reserve duty.

\medskip

\begin{definition} \label{IR}
An allocation $X\in \cala$  satisfies \textbf{individual rationality} if, for any $i\in I$, 
\[ X_i  \succeq_i  \emptyset.
\]
Likewise, a mechanism  $\big(\cals, \varphi\big)$ satisfies \textbf{individual rationality} if 
the allocation $\varphi(s)$  satisfies individual rationality for any strategy profile $s \in \cals$. 
\end{definition}

Each year, the Army carefully regulates the number of positions in each
branch to ensure adequate staffing and effective deployment of the Army's human resources \citep{ar614_100}. Given this, if a branch has a vacant slot, there shouldn't be an unassigned cadet who would like the branch at the base price.  The Army is keen not to waste valuable slots when a cadet is otherwise willing to take that assignment.  This consideration leads to the next axiom.

\begin{definition} \label{non-wastefulness}
An allocation $X\in \cala$  satisfies satisfies \textbf{non-wastefulness} if, for any  $b \in B$ and $i\in I$, 
\[ \left.  \begin{array}{c}
 \big|\{x \in X : \b(x) = b\}\big| <  q_b \, , \; \mbox{ and}\\ 
X_i = \emptyset \end{array}  \right\}
\implies \;  \emptyset \; \succ_i \; (b,t^0).
\]
Likewise, a mechanism $\big(\cals, \varphi\big)$ satisfies \textbf{non-wastefulness} if	
the allocation $\varphi(s)$  satisfies non-wastefulness for any strategy profile $s \in \cals$. 
\end{definition}
\noindent Non-wastefulness is a mild efficiency axiom that requires that no position remains unfilled while
an unassigned cadet who would rather receive the position at the branch at the base price $t^0$ exists.

As we have described, a cadet's OML ranking forms the basis 
of a cadet's claims at a branch.  If cadet $i$ prefers the assignment of cadet $j$ to her own, then cadet $i$
should not have higher priority at cadet $j$'s assigned branch.  When the OML is the only
source of prioritization, a cadet with a higher OML score should not prefer the assignment
of a cadet with a lower OML.  In that situation, the cadet's priority for a branch is reversed.  
Our next  axiom formalizes this consideration.\footnote{This axiom is called \textit{fairness\/} in \cite{sonmez/switzer:13}.  
Here we use the Army's terminology. See Section \ref{sec:quasi-direct} for an additional reason that justifies the use of 
a terminology different from \textit{fairness}.} 

\begin{definition} \label{def:priorityreversal}
An allocation $X\in \cala$   satisfies
\textbf{no priority reversal} if, for any $i,j \in I$, and $b\in B$,
    \[ \left. \begin{array}{c}
  \b(X_j) = b, \; \mbox{ and}\\ 
  X_j  \succ_i  X_i \end{array}  \right\}
\implies \; j \; \pi_b \; i.
\]
Likewise, a mechanism $\big(\cals, \varphi\big)$ satisfies  \textbf{no priority reversal} if
the allocation $\varphi(s)$  satisfies no  priority reversals for any strategy profile $s \in \cals$. 
\end{definition}

This axiom captures the idea that once the price is fixed at $t \in T$,  
cadets with higher baseline priorities at any given branch $b\in B$ have higher claims for a position at branch $b$. 
Therefore, whenever cadet $i$ strictly prefers another cadet $j$'s assignment to her own, 
cadet $j$ must have higher baseline priority at her assigned branch than cadet $i$.  
Otherwise,  if cadet $i$ strictly prefers cadet $j$'s assignment even though
cadet $j$ has lower baseline priority than cadet $i$, then we say that there is a \textbf{priority reversal}.\medskip

The axiom \textit{no priority reversal\/} reduces to the  axiom \textit{no justified envy\/}
in the simpler settings of \cite{balinski/sonmez:99}  and \cite{abdulkadiroglu/sonmez:03}.\footnote{The axiom \textit{no justified envy\/}
is called \textit{fairness\/} in \cite{balinski/sonmez:99}, and \textit{elimination of justified envy\/} in  \cite{abdulkadiroglu/sonmez:03}.} 
The essence of the axiom no justified envy has to do with the following two questions: 
\begin{enumerate}
\item Is there envy? 
\item If there is envy, is it justified? 
\end{enumerate}
A cadet can envy the assignment of another cadet if she
prefers it over her own assignment. The envy is justified relative to the property rights structure, 
which describes which cadet's claims on any given assignment are most deserving. In the simplest case, when there
is a single  performance metric, like the OML, envy is justified if the envious cadet has a higher claim
to the assignment based on this single performance metric.\footnote{It is worth pointing out that the simplest form  of axiom no justified envy by 
\cite{balinski/sonmez:99} and \cite{abdulkadiroglu/sonmez:03}
is technically related to the lack of pairwise blocking in definitions of core-stability in two-sided matching models. 
However, the conceptual justification for the axiom no justified envy is different because it is a completely normative axiom
based on enforcing property rights rather that the traditional positive considerations related to core-stability.}
The introduction of the BRADSO program in 2006 changed this basic structure of claims. 

Under the BRADSO program, the structure of the property rights over positions at a branch does not merely depend on the OML, but also on 
the price cadets are willing to pay for a position in the branch.  Stakeholders have opposing
views on whether and how often a lower OML-ranked cadet should be able to displace a higher OML-ranked cadet at a branch, according to \cite{colarusso/lyle/wardynski:10}.

To formulate these trade-offs rigorously, we consider whether
a cadet may have a legitimate claim on a position awarded to other cadets, but rather than for a given price as in our previous axiom, 
this time with a different price due to the price responsiveness policy.  In what follows, we break this down into two separate cases.
 
\begin{definition} \label{price-reduced}
Let allocation  $X\in \cala$  and cadet $i\in I$ be such that, $X_i = (b,t) \in B\times T^+$. 
Then, cadet $j\in I\setminus\{i\}$ has a \textbf{legitimate claim for
a price-reduced version} of cadet $i$'s assignment $X_i$, if there exists a lower price $t'<t$, such that 
\[  \begin{array}{c}
\big(b,t'\big) \succ_j X_j  \quad  and \\  
(j,t') \; \omega_{b} \;  (i,t). \end{array} 
\]
\end{definition} 
We say that cadet $j$'s claim for a position at branch $b$ at  a lower price $t'$ is \textit{legitimate}\/ because
the price responsiveness policy $\omega_{b}$ does not overturn her claim in favor of cadet $i$ even when cadet $i$ pays  a higher price $t$.

\begin{definition} \label{price-increased}

Let allocation  $X\in \cala$  and cadet $i\in I$ be such that, $X_i = (b,t) \in B\times T\setminus\{t^h\}$. 
Then, cadet $j\in I\setminus\{i\}$ has a \textbf{legitimate claim for
a price-elevated  version} of cadet $i$'s assignment $X_i$, if there exists a  higher price $t'>t$, such that 
\[  \begin{array}{c} 
\big(b,t' \big) \succ_j X_j \, ,\\    
(j,t') \; \omega_{b} \;  (i,t) \, , \quad  and \\
\Big|\big\{(k,t^+)\in I\times T^+ \; : \; (k,b,t^+)\in X_b \big\}\Big|  < q^f_b.
\end{array}  \]
\end{definition} 
Even if cadet $i$ has a higher baseline priority at branch $b$ than cadet $j$, 
cadet $j$'s claim for a position at branch $b$  is legitimate with a  higher price $t'$ because 
\begin{enumerate}
\item the price responsiveness policy $\omega_{b}$ overturns the baseline priority in favor of cadet $j$ as long as 
cadet $j$ pays  the higher price  $t'$, and
\item awarding the position originally given to cadet $i$ instead to cadet $j$ at the higher price $t'$ is feasible
and does not result in exceeding the cap $q^f_{b}$ for flexible-price  positions at branch $b$.\footnote{Observe that
awarding a position at a higher price can be potentially infeasible only when the original price is equal to the base price.} 
\end{enumerate}

Legitimate claims for price-reduced and price-elevated versions of another cadet's
assignment are conceptually similar, but they have one technical difference due to the feasibility of 
changing a price of a position.\footnote{This is why the last condition in Definition \ref{price-increased} is absent in Definition \ref{price-reduced}.} 
Given a pair of prices $t, t'$ with $t>t'$, it is always feasible to replace the higher price contract $(i,b,t)$  of a cadet $i$ with a  lower price contract 
 $(j,b,t')$  for another cadet $j$. 
 In contrast,  it is not always feasible  to replace a lower price contract $(i,b,t)$  of a cadet $i$ with a  higher price contract 
 $(j,b,t')$  for another cadet $j$. In particular, the replacement is not possible if $t=t^0$ and there are already $q^f_{b}$ positions
 at branch $b$ which are awarded at an increased price in $T^+$ under allocation $X$.

The absence of either type of legitimate claim defines the role of the price responsiveness policy in our model,
as we formulate next. 

\begin{definition} \label{BRADSO-policy}
An allocation  $X \in \cala$ satisfies \textbf{enforcement of the price responsiveness policy} if 
no cadet $j\in I$ has a legitimate claim for either a price-reduced version or a price-elevated version of 
the assignment $X_i$ of another cadet $i\in I\setminus\{j\}$. \medskip

\noindent Likewise, a mechanism $\big(\cals, \varphi\big)$ satisfies \textbf{enforcement of the price responsiveness policy} if	
the allocation $\varphi(s)$  satisfies enforcement of the price responsiveness policy for any strategy profile $s \in \cals$. 
\end{definition}

A mechanism vulnerable to ``gaming'' could erode cadets' trust in the Army's branching process.
A mechanism that erodes trust is unlikely to persist in the US Army, where trust is seen as the foundation of 
their talent management strategy.\footnote{For example, in \textit{The Army Profession}, the US Army's Training and Doctrine Command identifies trust as an essential characteristic that defines the Army as a profession \citep{army:19}.   The Army's People Strategy describes one of the Army's strategic outcomes as building a professional Army that retains the trust and confidence of the American people and its members \citep{army:19b}.} Maintaining trust is especially important since cadets may find themselves relying on other cadets for their own security in life-and-death combat situations.
Perhaps unsurprisingly, when considering potential reforms to the USMA-2020 mechanism, the manager of the Talent-Based Branching program stated the Army prefers a mechanism that incentivizes honest preference submissions.\footnote{Lieutenant Colonel Riley Post, the Talent-Based Branching Program Manager, said ``cadets should be honest when submitting preferences for branches, instead of gaming the system'' in a statement in West Point's official newspaper \citep{garcia:20}.} 

Our next axiom is the gold standard for incentive compatibility in direct mechanisms.
\begin{definition} \label{strategy-proofness}
A direct mechanism $\varphi$ is \textbf{strategy-proof} if, for any $\succ \, \in \calq^{|I|}$, any $i \in I$, and any $\succ'_i \, \in \calq$, 
\[ \varphi_i(\succ) \, \succeq_i \, \varphi_i(\succ_{-i}, \succ'_i).
\]
\end{definition}

In a strategy-proof mechanism, truthful preference revelation is always in the best interests of the cadets.

\section{USMA-2020 Mechanism and Its Shortcomings} \label{sec:USMA-2020}

During the first 15 years of the BRADSO program,  the US Army did not use a direct mechanism for the branching process. 
Cadets do not submit their full preferences over branch-price pairs under USMA-2006 or USMA-2020.  
To describe and analyze these two mechanisms, we next introduce the class of \textit{quasi-direct\/} mechanisms. 
A quasi-direct mechanism is defined for a version of the problem with a single increased price, and thus, two prices in total. 
As such, throughout this section we assume that $T= \{t^0,t^h\}$,  or equivalently $T^+ = \{t^h\}$. 

\subsection{Quasi-Direct Mechanisms and their Desiderata} \label{sec:quasi-direct}

A \textbf{quasi-direct mechanism} is a mechanism where the message space is $\cals_i = \calp \times 2^B$ for each cadet $i \in I$.  
For any strategy $s_i = (P_i, B_i) \in \cals_i$ of cadet $i\in I$,  
the first component $P_i\in\calp$ of the strategy is the cadet's preference ranking over branches (when they are awarded at the base price $t^0$)
and remaining unmatched.  The second component,  $B_i\in2^B$, is the set of branches the cadet is willing to pay the increased price $t^h$.

With the exception of \textit{strategy-proofness\/}, which is only defined for direct mechanisms, all other 
axioms are well-defined for quasi-direct mechanisms. 
However,  a subtle issue arises in the verification of two of these axioms under quasi-direct mechanisms. 
Both \textit{no priority reversals\/} and \textit{enforcement of the price responsiveness policy\/} 
rely on knowing if there is a cadet who prefers another cadet's assignment to her own assignment.
Unlike a direct mechanism, this information is not fully solicited under a quasi-direct mechanism. 
Hence, for any given cadet, determining all priority reversals that adversely affect her or her
legitimate claims for price-reduced versions of other cadets' assignments is not possible in a quasi-direct mechanism.
However, some priority reversals can still be detected even under the restricted message space of 
quasi-direct mechanisms. This is the motivation for our next definition.

\begin{definition} \label{detectable-priorityreversal}
A quasi-direct mechanism $\varphi$  has \textbf{no detectable priority reversal}  if, 
for any $s = \big(P_j, B_j\big)_{j\in I} \in (\calp \times 2^B)^{|I|}$, $b \in B$, and
$i,j \in I$,   
    \[ \left. \begin{array}{c}
  \varphi_j(s) = (b,t^0), \; \mbox{ and }\\ 
    \varphi_i(s) =  (b,t^h) \quad \mbox{ or } \quad b \; P_i \; \b\big(\varphi_i(s)\big)
  \end{array}  \right\}
\implies \; j \; \pi_b \; i.
\]
\end{definition}
\noindent Under this axiom, if cadet $j$ is assigned a base-price position at branch $b$
and another cadet $i$ receives a less desired assignment by 
\begin{itemize}
\item[(i)] either receiving an increased-price position at the same branch  or 
\item[(ii)] by receiving  a position at a strictly less preferred (and possibly empty) branch based on cadet $i$'s submitted preferences $P_i$ on $B\cup\{\emptyset\}$, 
\end{itemize}
then cadet $j$ must have higher baseline priority under branch $b$ than cadet $i$.  \medskip

When a quasi-direct mechanism has detectable priority reversals, 
there is a cadet $i$ who strictly prefers the assignment of another cadet $j$ no matter what cadet $i$'s preferences $\succ_i\in\calq$ over 
branch-price pairs are (provided that they are consistent with her submitted preferences $P_i\in\calp$ over branches alone). 
For this reason, detectable priority reversals can be verified under a quasi-direct mechanism.  Verification
of the absence of all priority reversals, in contrast, requires knowledge of cadet $i$'s preferences over branch-price pairs.

To study incentive compatibility of quasi-direct mechanisms, 
we can no longer consider \textit{strategy-proofness} because that concept is only defined
for direct mechanisms.  We instead tailor variants of this axiom that accord with the quasi-direct message space.
We next  formulate two incentive compatibility axioms that do not rely on preferences over branch-price pairs.

The Army created flexible price positions to allow some cadets to obtain priority over other cadets
who may have a higher OML, but are unwilling to extend their service commitment.  Our next
axiom captures the idea that a cadet should not be charged an increased price
for a position when the price responsiveness policy has not been pivotal in securing a branch.

\begin{definition} \label{BRADSO-IC}
A quasi-direct mechanism $\varphi$  satisfies \textbf{BRADSO-incentive compatibility} (or 
\textbf{BRADSO-IC}) if, for any $s = \big(P_j, B_j\big)_{j\in I} \in (\calp \times 2^B)^{|I|}$, 
$i \in I$, and  $b \in B$, 
\[  \varphi_i(s) = (b, t^h) \; \implies \;  \varphi_i\big((P_i, B_i\setminus\{b\}),\; s_{-i}\big) \not= (b, t^0). 
\]
\end{definition}
\noindent A cadet $i$ who receives an increased-price position at branch $b$ under $\varphi$ 
should not be able to profit by receiving a position at the same branch at the base price
by dropping branch $b$ from the set of branches $B_i$ for which she's willing to pay the increased price.

A cadet also should not benefit by declaring a willingness to pay the increased price
to obtain an assignment at the branch at the base price. Failure of this desideratum 
undermines the idea behind the BRADSO system, which is to use information on the willingness to serve extended 
service commitments in exchange for priority.  
Our last axiom formulates this desideratum. 

\begin{definition} \label{strategic-BRADSO}
A quasi-direct mechanism $\varphi$  is \textbf{immune to strategic BRADSO}  if, for any $s = \big(P_j, B_j\big)_{j\in I} \in (\calp \times 2^B)^{|I|}$, 
$i \in I$, and  $b \in B$, 
\[  \varphi_i(s) = (b, t^0) \; \implies \;  \varphi_i\big((P_i, B_i\setminus\{b\}),\; s_{-i}\big) = (b, t^0). 
\]
\end{definition}
\noindent A cadet $i$ who receives a base-price position at branch $b$ under $\varphi$ 
should still do so  upon dropping branch $b$ from the set of branches $B_i$ for which she has indicated willingness to pay the increased price 
(in case $b\in B_i$).\footnote{This statement holds vacuously if $b\not\in B_i$.}
If this axiom fails,
cadet $i$ could strategically indicate a willingness to pay the increased price at branch $b$
and receive an otherwise unattainable base-price position at this branch.

\subsection{USMA-2020 mechanism}
The USMA-2020 mechanism is a quasi-direct mechanism
with message space $\cals^{2020}_i = \calp \times 2^{B}$ for each cadet $i\in I$. 
Given a strategy profile $s = (P_i, B_i)_{i\in I}$, for any branch $b\in B$, construct the following adjusted
priority order $\pi^+_b \in \Pi$ on the set of cadets $I$.
For any $i, j \in I$,
\begin{enumerate}
\item $b\in B_i$ and $b\in B_j \quad \implies \quad i \; \pi^+_b \; j \; \iff i \; \pi_b \; j$,
\item $b\not\in B_i$ and $b\not\in B_j \quad \implies \quad i \; \pi^+_b \; j \; \iff i \; \pi_b \; j$, \mbox{ and} 
\item $b\in B_i$ and $b\not\in B_j \quad \implies \quad i \; \pi^+_b \; j \; \iff (i,t^h) \; \omega_b \; (j,t^0)$.
\end{enumerate}	
Under the priority order $\pi^+_b$, 
any two cadets are rank ordered using the baseline priority order $\pi_b$
if they have indicated the same willingness to pay the increased price for branch $b$, and using the price responsiveness policy $\omega_b$ otherwise.\footnote{When (i) the baseline priority order $\pi_b$ is fixed as OML at each branch $b\in B$, 
and (ii) the price responsiveness policy $\omega_b$ is fixed as the ultimate price responsiveness policy
${\overline{\omega}}_b$ at each branch $b\in B$, this construction gives the same adjusted priority order constructed for the USMA-2006 mechanism.}

For any strategy profile $s = (P_i, B_i)_{i\in I}$, 	let $\mu$ be the outcome
of the  \textit{individual-proposing deferred acceptance algorithm\/} for submitted cadet preferences $(P_i)_{i\in I}$ and constructed
branch priorities $\big(\pi^+_b\big)_{b\in B}$.

For any strategy profile $s = (P_i, B_i)_{i\in I}$, the outcome  $\varphi^{2020}(s)$ of the \textbf{USMA-2020 mechanism} is given as follows:  	
For any cadet $i \in I$, 
\[ \varphi_i^{2020}(s) =   \left\{ \begin{array}{cl}  
\emptyset & \mbox{if } \; \mu(i)=\emptyset, \\
 \big(\mu(i), t^0\big)   & \mbox{if } \; \mu(i)\not\in B_i  \mbox{ or }   \big|\big\{j\in I : \mu(j)=\mu(i), \; \mu(j)\in B_j, \mbox{ and } i \; \pi_{\mu(i)} \;j\big\}\big| \geq q^f_{\mu(i)},\\
 \big(\mu(i), t^h\big)   & \mbox{if } \; \mu(i)\in B_i  \mbox{ and }  \big|\big\{j\in I  : \mu(j)=\mu(i), \; \mu(j)\in B_j, \mbox{ and } i \; \pi_{\mu(i)} \;j\big\}\big| < q^f_{\mu(i)}.
\end{array} \right.
\]	

Under the USMA-2020 mechanism, each cadet $i\in I$ is asked to submit a preference relation $P_i \in \calp$ along with a 
(possibly empty) set of branches 
$B_i\in 2^B$ for which she indicates her willingness to pay the increased price $t^h$ to receive preferential admission. 
A priority order $\pi_b^+$ of cadets is constructed for each branch $b$ by adjusting  the baseline priority order $\pi_b$  using the
price responsiveness policy $\omega_b$ whenever a pair of cadets submitted different willingness to pay the increased price $t^h$  at branch $b$. 
Cadets' branch assignments are determined by the individual-proposing deferred acceptance algorithm  using the submitted 
profile of cadet preferences $(P_i)_{i\in I}$
and the profile of adjusted priority rankings  $(\pi^+_b)_{b\in B}$. 
A cadet pays the base price for her branch assignment  if either she has not declared willingness to pay the increased price for her assigned branch or 
the capacity for the flexible-price positions of the branch is already filled with cadets who have lower baseline priorities. 
With the exception of those who remain unmatched,  all other cadets pay the increased price for their branch assignments.

\subsection{Shortcomings of the USMA-2020 Mechanism} \label{sec:shortcomings2020}

USMA-2020 has perverse incentives in large part because it determines who is charged
the increased price for their assignments only after the completion of branch assignments.  We present an in-depth
equilibrium analysis of the USMA-2020 mechanism in Appendix \ref{sec:singlebranch}.  
Among our results in this analysis,   
Example \ref{knifeedge} in  Appendix  \ref{subsec:NE}   shows that the USMA-2020 mechanism fails BRADSO-IC 
and admits strategic BRADSO even at equilibrium.   That example also illustrates
the ``knife-edge'' aspects of equilibrium strategies in this mechanism.
When there is a minor change in the underlying economy involving
 the lowest baseline priority cadet changing her preferences and this
 only affects her assignment, it nonetheless affects the equilibrium strategies of several higher
priority cadets. Example \ref{Bayesian} in  Appendix  \ref{subsec:NE}  further shows that the USMA-2020 mechanism can admit
detectable priority reversals even under its Bayesian Nash equilibrium outcomes. 

The fragility of the equilibrium strategies helps to understand some of the  failures observed under the USMA-2020 mechanism in the field, which we describe next.
These observations provide support for the relevance of the formal axioms we discussed in Section \ref{sec:quasi-direct}.  

After announcing the mechanism to cadets, USMA leadership recognized
the possibility of detectable priority reversals under the USMA-2020 mechanism due to either failure of BRADSO-IC
or presence of strategic BRADSO. 
In a typical year, the number of cadets willing to pay the increased price for traditionally oversubscribed branches like Military Intelligence 
greatly exceeds 25 percent of the branch's positions. 
Therefore, by volunteering  to pay the increased price for an oversubscribed branch, some cadets could receive a priority upgrade even though they may not be charged for it, making detectable priority reversals possible.
Moreover, unlike the detectable priority reversals under the USMA-2006 mechanism, some of these detectable priority reversals can affect cadet branch assignments, thereby making manual ex-post adjustments infeasible.

Failures of BRADSO-IC, the possibility of strategic BRADSO,  or the presence of detectable priority reversals, 
especially when not manually corrected ex-post,  could erode cadets' trust in the Army's branching process. 
Consider, for example, a comment from a cadet survey administered to the 
Class of 2020:\footnote{The survey was administered to the Class of 2020 immediately 
before they submitted their preferences for branches under the USMA-2020 mechanism.
The response rate to this survey was 98\%. Appendix \ref{survey} contains specific questions and results.} 
 
\begin{quote}
    ``\textit{I believe this system fundamentally does not trust cadets to make the best choice for themselves. It makes it so that we cannot choose what we want and have to play games to avoid force branching}.''
\end{quote}

To mitigate these concerns, USMA leadership executed a simulation using preliminary cadet preferences to 
inform cadets of the potential cutoffs for each branch.\footnote{Cadets in the Class of 2020 submitted preliminary preferences one month before submitting final preferences. USMA ran the USMA-2020 mechanism on these preliminary preferences to derive results for the simulation, which USMA provided to cadets 6 days prior to the deadline for submitting final preferences.} 
The goal of this simulation was to help cadets to optimize their submitted strategies.  Army leadership was quoted as follows \citep{oconnor:19}: 
\begin{quote}
 ``\textit{We're going to tell all the cadets, we're going to show all of them, here's when the branch would have went out, here's the bucket you're in, here's the branch you would have received if this were for real. You have six days to go ahead and redo your preferences and look at if you want to BRADSO or not.'' Sunsdahl said. ``I think it's good to be transparent. I just don't know what 21-year-olds will do with that information}.''
\end{quote}

Several open-ended survey comments from USMA cadets in the Class of 2020 mirrored USMA leadership's concern about the USMA-2020 mechanism.
We present three additional comments articulating concerns related to the lack of BRADSO-IC, the presence of  strategic BRADSO, and the
difficulty of navigating a system with both shortcomings:

\begin{itemize}
    \item[1)] ``\textit{Volunteering for BRADSO should only move you ahead of others if you are actually charged for BRADSO. 
    By doing this, each branch will receive the most qualified people. Otherwise people who are lower in class rank 
    will receive a branch over people that have a higher class rank which does not benefit the branch. Although those who 
    BRADSO may be willing to serve longer, if they aren't charged then they can still leave after their 5 year commitment 
    so it makes more sense to take the cadets with a higher OML.}"
    \item[2)]    ``\textit{I think it is still a little hard to comprehend how the branching process works. For example, I do not know if I put a BRADSO for my preferred branch that happens to be very competitive, am I at a significantly lower chance of getting my second preferred if it happens to be something like engineers? Do I have to BRADSO now if I want engineers??? Am I screwing myself over by going for this competitive branch now that every one is going to try to beat the system????}''
    \item[3)] ``\textit{Releasing the simulation just created chaos and panicked cadets into adding a BRADSO who otherwise wouldn't have.''}
\end{itemize}

Empirical evidence on the extent of failures of these desiderata under USMA-2020, and how they
compare with the failures under USMA-2006 is presented in Appendix \ref{failureprevelance}. 

\section{Multi-Price Cumulative Offer Mechanism and Its Characterization} \label{subsec:MPCOADSO}

The integrated research and policy strategy under minimalist market design \citep{sonmez:23}  revolves around the following 
three endeavors:
\begin{enumerate}
\item Identify key objectives for the stakeholders. 
\item Establish whether the institution in place satisfies all these objectives. 
\item If the current institution fails to meet some of these objectives, 
provide an alternative mechanism that satisfies them if possible. 
\end{enumerate}
In addition, the third endeavor should identify the root causes of the failures
of the existing mechanism, and address these root causes in designing an alternative institution.

In our application, the main stakeholders are Army officers in charge of the branching process and the cadets. 
In Sections \ref{sec-axioms} and \ref{sec:quasi-direct}, we formulated the key objectives of these stakeholders 
as rigorous axioms. In Appendix \ref{sec:singlebranch} and Section \ref{sec:shortcomings2020}, we have shown that USMA-2020 fails three 
key desiderata and identified the following two culprits for the failures:
\begin{enumerate}
\item The message space is not sufficiently rich to capture cadet preferences over branch-price pairs. 
\item The two elements of an assignment, the branch assignment and the price assignment, are determined 
sequentially rather than jointly.   
\end{enumerate}
All three failures under USMA-2020, the presence of priority reversals,  the lack of BRADSO-IC and the presence of  strategic BRADSO,
are directly tied to these flawed aspects of the mechanism. 
In this section, we formulate an alternative mechanism for the Army by directly addressing these two
root causes of the failures of USMA-2020, and leaving the rest of the system untouched.
This makes our intervention one that is minimalist.

\subsection{Changing the Message Space} \label{sec:messagespace}

To resolve the problems with the USMA-2020 mechanism, most notably its failure of BRADSO-IC, the possibility of strategic BRADSO, and the 
resulting detectable priority reversals, the Army established a partnership with the two civilian co-authors of this paper to redesign their branching mechanism.
Critical to achieving these objectives was the Army's decision to permit cadets in the Class of 2021 to submit preferences over branch-price pairs,
thus paving the way to adopt a direct mechanism. 

This decision was aided by evidence from a  cadet survey that mitigated concerns that rating branch-price pairs would be overly complex or unnecessary. 
Indeed, some of the cadets indicated the need for a system that would allow them to rank order branch-price pairs.\footnote{One cadet wrote 
``I wish there was an option to  pick your second choice over your first if your first choice mandated a branch detail.''
Resonating this sentiment, another cadet wrote:
\begin{quote}
``I am indifferent to the alternative or current bradso system.  However, 
[$\ldots$] \textit{I believe that DMI (Department of Military Instruction) could elicit a new type of ranking list. 
Within my proposed system, people could add to the list of 17 branches BRADSO slots and rank them within that list. 
For example: `AV (Aviation) $>$ IN (Infantry) $>$ AV:B (Aviation with BRADSO).}'... 
BRADSO slots are considered almost different things.''
\end{quote}}
More generally, the survey revealed that more than twice as many cadets prefer a mechanism that allows them to submit preferences over branch-price pairs relative to a mechanism that requires them to submit preferences over branches and then separately indicate their willingness to pay an increased price for each branch as in the USMA-2006 and USMA-2020 mechanisms.\footnote{A question on the survey asked cadets whether they prefer a mechanism that allows them to submit preferences over branch-price pairs or a mechanism that requires them to submit preferences over branches alone while separately indicating willingness to pay an increased price, or BRADSO, for each branch.  \fig{Appendix \ref{survey}}  shows that
50 percent of respondents preferred the mechanism that permitted ranking branch-price pairs, 21 percent preferred the mechanism without the option to rank branch-price pairs, 24 percent were indifferent, and 5 percent did not understand.}

\subsection{Coordinating the Branch Assignment and the Price Assignment} 

The central mechanism in the matching with contracts literature is the \textit{cumulative offer mechanism\/} \citep{hatfield/milgrom:05}, 
a direct mechanism which is based on a procedure that involves a sequence of cadet proposals and branch responses.
Cadet proposals simulated under this procedure are based on their submitted preferences.
The \textbf{multi-price cumulative offer (MPCO)} mechanism is a refinement of the cumulative offer mechanism, 
where, for each branch $b\in B$, the branch response takes a specific form determined by the following choice rule $\calc^{MP}_b$.\footnote{The MPCO mechanism 
is a generalization of the COSM mechanism proposed by \cite{sonmez/switzer:13}
for the case of  the ultimate price responsiveness policy $\overline{\omega}_b$ for each branch $b\in B$, and a
refinement of the cumulative offer mechanism for the matching with slot-specific priorities
model by \cite{kominers/sonmez:16}.}

\begin{quote}
 
\noindent \textbf{Multi-Price Choice Rule}  {\boldmath $\calc^{MP}_b$}

\noindent Given $b \in B$ and $X \in \calx_b$,  select (up to) $q_b$ contracts with distinct cadets as 
follows:\smallskip 

\noindent \textbf{Step 1. (Selection for the base-price   positions)}   Let $X^1$ be the set of all base-price contracts in $X$ if there are
no more than $q^0_b$  base-price contracts in $X$, and the set of base-price contracts in $X$
with $q^0_b$ highest $\pi_b$-priority cadets otherwise. 
Pick contracts in $X_1$  for the base-price   positions, and proceed to Step 2. \smallskip

\noindent \textbf{Step 2. (Selection for the flexible-price  positions)} Construct the set of contracts $Y$ from $X$ by
first removing the lower   $\omega_b$-priority contract 
of any cadet who has two contracts in $X$, and next removing  all contracts of any cadet who has a contract already selected in $X_1$.  
Let $X^2$ be the set of all  contracts in $Y$ if there are
no more than $q^f_b$ contracts in $Y$, and the 
set of $q^f_b$ highest $\omega_b$-priority contracts in $Y$ otherwise. 
Pick contracts in $X_2$  for the flexible-price  positions, and terminate the procedure.  \smallskip

The outcome of the multi-price choice rule is  $\calc^{MP}_b(X) = X_1 \cup  X_2$.  \medskip
\end{quote}

\noindent Intuitively, the multi-price choice rule  $\calc^{MP}_b$  first allocates the base-price   positions 
following the  baseline priority order $\pi_b$ , and next allocates the  flexible-price  positions 
following the  price responsiveness policy  $\omega_b$.

We  are ready to formally define the multi-price cumulative offer mechanism.
Given a profile of baseline  priority orders $(\pi_b)_{b\in B}$ and a profile of 
price responsiveness policies $(\omega_b)_{b\in B}$, 
let $\calc^{MP} = \big(\calc^{MP}_b\big)_{b\in B}$ denote the profile of multi-price choice rules defined above. 
Since the MPCO mechanism is a direct mechanism,
the message space for each cadet $i \in I$ is
$\cals^{MPCO}_i = \calq$, where $\calq$ is the set of linear orders on $(B\times T) \cup \{\emptyset\}$.
The second element of the MPCO mechanism, its outcome function $\phi^{MPCO}$ is given by the following \textbf{multi-price cumulative offer (MPCO)} procedure, 
which is simply the celebrated cumulative offer procedure \citep{hatfield/milgrom:05} implemented with the MPCO choice rule for each branch. 

\begin{quote}
\noindent \textbf{Multi-Price Cumulative Offer Procedure}\smallskip

\noindent Fix a linear order of cadets $\pi \in \Pi$.\footnote{By Kominers and S\"{o}nmez (2016), 
the outcome of this procedure is independent of this linear order.} 
For a given profile of cadet preferences $\succ = (\succ_i)_{i\in I} \in \calq^{|I|}$, 
cadets propose their acceptable contracts to branches in a sequence of steps $\ell = 1, 2, \ldots $:\smallskip 

\noindent \textbf{Step 1.}  Let $i_1\in I$ be the highest $\pi$-ranked cadet who has an acceptable contract. 
Cadet $i_1 \in I$ proposes her most preferred contract $x_1 \in \calx_{i_1}$ to branch $\b(x_1)$. 
Branch $\b(x_1)$ holds $x_1$ if $x_1 \in  \calc^{MP}_{\b(x_1)}\big(\{x_1\}\big)$ and rejects $x_1$ otherwise. Set 
$A^2_{\b(x_1)} = \{x_1\}$ and set $A^2_{b'}=\emptyset$ for each $b'\in B\setminus\{\b(x_1)\}$; 
these are the sets of contracts available to branches at the beginning of step 2.\smallskip

\noindent \textbf{Step}  {\boldmath $\ell$.}  {\boldmath ($\ell  \geq 2$)} 
Let $i_{\ell} \in I$ be the  highest $\pi$-ranked cadet  for whom no contract is currently held by any branch,
and let $x_{\ell} \in \calx_{i_{\ell}}$ be her most preferred acceptable contract that has not yet been rejected.  
Cadet $i_{\ell}$ proposes contract   $x_{\ell}$ to branch $\b(x_{\ell})$. 
Branch $\b(x_{\ell})$ holds the contracts in $\calc^{MP}_{\b(x_{\ell})}\big(A^{\ell}_{\b(x_{\ell})}\cup \{x_{\ell}\}\big)$ 
and rejects all other contracts in $A^{\ell}_{\b(x_{\ell})}\cup \{x_{\ell}\}$. 
Set  $A^{\ell +1}_{\b(x_{\ell})} = A^{\ell}_{\b(x_{\ell})}\cup \{x_{\ell}\}$ and 
set $A^{\ell +1}_{b'} = A^{\ell}_{b'}$ for each $b'\in B\setminus\{\b(x_{\ell})\}$; 
these are the sets of contracts available to branches at the beginning of step $\ell +1$.\smallskip

The procedure terminates at a step when either no cadet remains with an acceptable contract that has not been rejected, or
when no contract is rejected. 
\smallskip
\end{quote}

Given a profile of cadet preferences $\succ = (\succ_i)_{i\in I} \in \calq^{|I|}$, 
all the contracts on hold in the final step of the multi-price cumulative offer 
procedure are finalized as the outcome $\phi^{MPCO}(\succ)$ of the  
\textbf{multi-price cumulative offer (MPCO)} mechanism.\footnote{As it is customary in the literature, we denote a direct mechanism with its outcome function and use $\phi^{MPCO}$ to denote both the outcome function and the resulting direct mechanism.}   

Our  main theoretical result shows that MPCO  is the only direct mechanism that satisfies all five desiderata of the Army
formulated in Section \ref{sec-axioms}. 

\begin{theorem} \label{cosm}
Fix a profile of baseline priority orders $(\pi_b)_{b\in B} \in \Pi^{|B|}$ and a profile of price responsiveness policies $\big(\omega_b\big)_{b\in B} \in \prod_{b\in B}\omega_b$. 
A direct mechanism $\varphi$ satisfies
\begin{enumerate} 
\item \textit{individual rationality\/},  
\item \textit{non-wastefulness\/},  
\item \textit{no priority reversal\/}, 
\item \textit{enforcement of the price responsiveness policy\/}, and
\item \textit{strategy-proofness\/}
\end{enumerate}
if and only if it is the MPCO mechanism $\phi^{MPCO}$. 
\end{theorem}

Apart from singling out the MPCO mechanism as the unique mechanism that satisfies the Army's desiderata, 
to the best of our knowledge Theorem \ref{cosm} is the first characterization of an allocation mechanism
(i.e. the cumulative offer mechanism)  which pins down a specific choice rule (i.e. the multi-price choice rule) endogenous to the
policy objectives of the central planner.
We next elaborate on the significance of this result, and how it relates to earlier literature.

\subsection{Related Literature and the Significance of the Main Characterization Result} \label{sec:significance} 
Without the BRADSO program, our model reduces to the standard  model on priority-based and unit demand indivisible goods allocation
(or simply \textit{priority-based allocation\/})
problem. In this version, our axiom \textit{enforcement of the price responsiveness policy\/} becomes vacuous, our
axiom \textit{no priority reversals\/} reduces to the axiom \textit{no justified envy\/} in its most basic form, and the MPCO mechanism
reduces to the celebrated \textit{individual-proposing deferred acceptance mechanism} by \cite{gale/shapley:62}. 
As such,  Theorem \ref{cosm} extends the following well-known result: 

\begin{corollary}[\cite{alcalde/barbera:94, balinski/sonmez:99}] \label{cor:DA} 
Fix a profile of baseline priority orders $(\pi_b)_{b\in B} \in \Pi^{|B|}$. 
A direct mechanism $\varphi$ satisfies
\begin{enumerate} 
\item \textit{individual rationality\/},  
\item \textit{non-wastefulness\/},  
\item \textit{no justified envy\/}, and
\item \textit{strategy-proofness\/}
\end{enumerate}
if and only if it is the individual-proposing deferred acceptance mechanism. 
\end{corollary}

Since claims of individuals over  positions at any given institution are represented with a baseline priority order
in \cite{balinski/sonmez:99} and \cite{abdulkadiroglu/sonmez:03}, the only axiom in these papers that ``enforces''  the underlying basic
structure of property rights is  \textit{no justified envy\/}.
In our more complex setting,  in contrast, a cadet may increase her claims on a position at any given branch  by paying a higher price than its base price. 
The role of our axiom  \textit{enforcement of the price responsiveness policy\/} is to regulate how that happens in our setting. 
Taken together our two axioms  \textit{no priority reversal\/} and \textit{enforcement of the price responsiveness policy\/} can be therefore
interpreted as a generalization of the  \textit{no justified envy\/}  axiom for our richer setting.\footnote{We are grateful
to a referee for suggesting we elaborate on this connection.}
Thus, not only does Theorem  \ref{cosm} extend Corollary \ref{cor:DA} in a technical sense, but also it is in the same spirit.

As discussed in \cite{sonmez:23}, the roots of minimalist market design goes back to  \cite{balinski/sonmez:99} and \cite{abdulkadiroglu/sonmez:03}
in the context of school choice. 
Motivated by Corollary \ref{cor:DA},  \cite{abdulkadiroglu/sonmez:03} proposed the individual-proposing deferred acceptance mechanism
as an alternative to the Boston mechanism. In line with minimalist market design, the subsequent reform efforts at Boston Public Schools (BPS)
by our team of design economists did not interfere with the structure of student priorities at public schools. 
The initial focus was simply on the reform of the allocation mechanism for a given list of priority orders at each school. 
This strategy paid off because it enabled our team to remain impartial in a politically
charged environment with interest groups with strong opinions on the structure of school priorities. 
Without touching this sensitive issue, BPS leadership, and subsequently the school committee supported 
reforming Boston's school choice mechanism in 2005 \citep{aprs:05}. 
In a reform that guided the formulation of minimalist market design, BPS adopted  the individual-proposing deferred acceptance mechanism
at the expense of the Boston mechanism. 

A choice rule is a single-institution solution concept that regulates who deserves the positions at an institution. 
Under some technical conditions, this solution concept easily integrates with the individual-proposing deferred acceptance mechanism and its generalization 
the cumulative offer mechanism, thus extending its scope for multi-institution settings.\footnote{Two technical conditions on choice rules that enable this integration are the \textit{substitutes condition\/} \citep{kelso/crawford:82, hatfield/milgrom:05} and \textit{independence of rejected individuals} \citep{aygun/sonmez:13}.} 
In other words, a natural interface between two major components of a resource allocation system becomes available in these settings. 
Consequently,  the acceptance of the individual-proposing deferred acceptance mechanism as a plausible mechanism for priority-based 
allocation in mid 2000s resulted in a rich literature on analysis of choice rules that implement various social policies. 
Papers in this literature include \cite{pycia:12}, \cite{hafalir/yenmez/yildirim:13}, 
\cite{echenique/yenmez:15}, \cite{kominers/sonmez:16}, \cite{dogan:17},  
\cite{dur_boston}, \cite{kojima/tamura/yokoo:18}, \cite{erdil/kumano:19}, \cite{dur/pathak/sonmez:20}, \cite{imamura:20}, \cite{pathak/rees-jones/sonmez:20}, 
\cite{aygun/bo:21}, \cite{dogan/yildiz:22}, \cite{sonyen22, sonmezyenmez2022b} and \cite{sonmez-unver2022}. 
Some of these papers assume a single institution. In others, 
the foundations for various choice rules have been developed assuming
that the underlying allocation mechanism is either the individual-proposing deferred 
 acceptance mechanism or the cumulative offer mechanism.\footnote{An exception is \cite{sonmezyenmez2022b}, which follows 
our research strategy,  and  establishes the foundations for the individual-proposing deferred acceptance mechanism along 
with a choice rule formulated in  \cite{sonyen22}.} 
Our paper, in contrast, establishes the foundations for both parts of the mechanism together 
from the first primitives of the problem.\footnote{ At this point, it is important to emphasize that all our axioms formulate the Army's policy objectives, thus 
reflecting a signature feature of minimalist market design. None of our axioms are imposed 
as technical conditions for the sake of obtaining an axiomatic characterization.
This aspect of our main result is important to avoid introducing a normative gap between intended and implemented policies \citep{hitzig:20, sonmez:23} 
or various distributional biases in the recommended procedure \citep{li:17, vanBasshuysen:2022, sonmez:23}.}

As in the case of the individual-proposing deferred acceptance mechanism, prior axiomatic characterizations for the cumulative 
offer mechanism also exist in the literature. Most related to Theorem  \ref{cosm} are
\cite{hirata/kasuya:17} and \cite{hatfield/kominers/westkamp:21}, who present  
axiomatic characterizations of the cumulative offer mechanism based on conceptually relevant axioms. 
Our characterization, however, differs from theirs in one fundamental aspect.  
Consistent with the approach in priority-based allocation problems,  starting from a given choice rule for each institution is also
a near universal assumption in the matching with contracts literature.  
Following this tradition, 
in  both \cite{hirata/kasuya:17} and \cite{hatfield/kominers/westkamp:21},  each institution is endowed with an exogenously given choice rule that satisfies 
various technical conditions.  
In our characterization, in contrast,  the multi-price choice rule--one of the two pillars of the MPCO mechanism-- emerges
endogenously from the Army's policy objectives formulated by our desiderata.
Indeed, the very concept of a choice rule is only used in our model to describe the MPCO mechanism. 
Our axioms do not place any structure or assume any functional form of potential branch choice rules.  In fact, we do not even assume the existence of a well-defined choice rule that dictates behavior for any given branch.  Instead, the multi-price choice rule emerges jointly 
with the cumulative offer mechanism as a \textit{collective\/} implication of our five axioms.

\section{Iterative Design: Trading-Off Talent Alignment and Retention}\label{sec:evolution}

In this section, drawing on our experience with the US Army's branching reform, 
we present an example of the iteration in the design after  a partnership is formed with the system operators. 
At this stage in the reform process, the market designer is no longer an outsider, and therefore she has much more flexibility 
to tinker on various aspects of the design. 

After adopting the USMA-2020 mechanism, Army and USMA leadership had several discussions about the potential price responsiveness policy for the Class of 2021 and possibly increasing the share of flexible-price positions. As described in the excerpt below from a news article describing an interview with the Talent-Based Branching Program Manager, selecting these parameters presented the Army with a trade-off between retention and talent alignment \citep{garcia:20}: 
\begin{quote}
   ``\textit{A key question the Army considered when designing this year's mechanism was how much influence to give cadets who are willing to BRADSO. 
    If every cadet who volunteers to BRADSO can gain priority, or ``jump'' above, every cadet who did not volunteer to BRADSO, 
    then that could improve Army retention through more cadets serving an additional three years, but it could also result in 
    more cadets being assigned to branches that do not prefer them.}''
    \end{quote} 

It is possible to formally analyze this tradeoff by focusing on the choice rule $\calc^{MP}_b$
in the new mechanism.  For a given number of total positions, 
if the number of flexible-price positions increases,
then the baseline priority order $\pi_b$ is used for fewer positions
and the price responsiveness policy  $\omega_b$ is used for more positions. 
Likewise, when a price responsiveness policy
becomes more responsive to a price increase,
increased-price contracts receive weakly higher priorities.
Under both of these scenarios, the number of increased-price contracts selected by the choice rule $\calc^{MP}_b$ weakly increases.
We collect these two straightforward observations in the following result.

 \begin{proposition} \label{prop:BRADSOcomparativestatics}
Fix a branch $b\in B$, the total number of branch-$b$ positions at $q_b$, and 
a set of branch-$b$ contracts $X\subset \calx_b$.  Then, 
\begin{enumerate}
\item the number of price-elevated contracts selected under $\calc_b^{MP}(X)$ weakly increases as the number of flexible-price positions $q_b^f$ increases, and 
\item the number of price-elevated contracts selected under $\calc_b^{MP}(X)$ weakly increases as the price responsiveness policy $\omega_b$ gets
more responsive to a price increase. 
\end{enumerate}
\end{proposition}

\begin{figure}[htp]
    \begin{center}
    \caption{\textbf{Number of BRADSOs Charged Across price responsiveness policies and Cap Sizes}\\}
    \label{fig:bradsocap}
    \includegraphics[scale = 0.8]{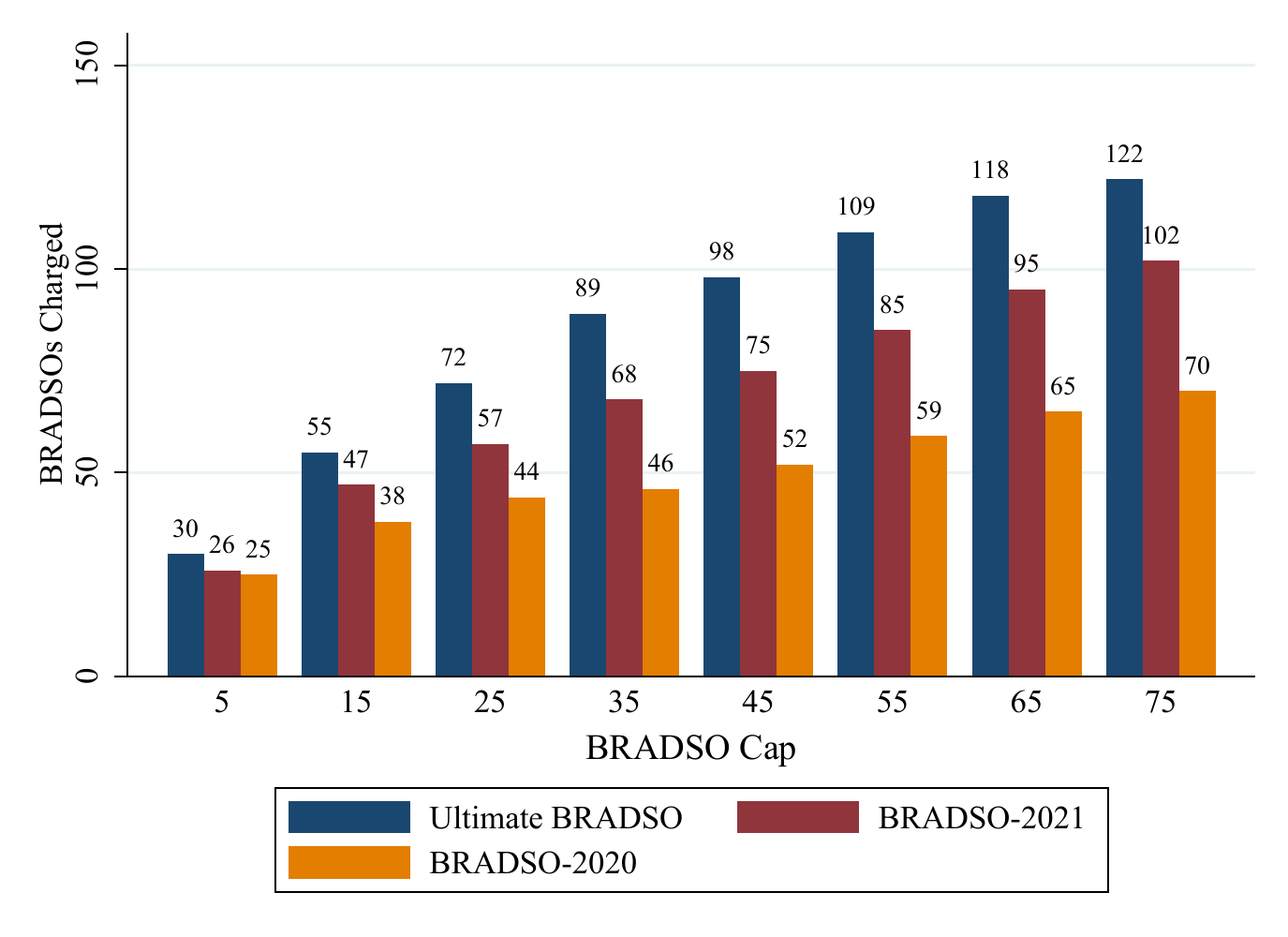}
    \end{center}
\scriptsize{\textbf{Notes.} This figure reports on the number of BRADSOs charged for three price responsiveness policies: Ultimate BRADSO, BRADSO-2020, and BRADSO-2021 using data from the Class of 2021.  The BRADSO cap ranges from 5\% to 75\% of slots at each branch.  Each outcome is computed by running MPCO mechanism  given stated cadet preferences under different price responsiveness policies and cap sizes.}
\end{figure}

While the results on the BRADSO collected (i.e. the flexible-price  positions awarded at elevated prices)
given in  Proposition \ref{prop:BRADSOcomparativestatics} hold for a given
branch under the multi-price choice rule, in theory this result may not hold in aggregate across all branches under the MPCO mechanism.\footnote{The fact that
a global comparative static result does not hold in matching models with slot-specific priorities has been explored in other work,
including \cite{dur_boston} and \cite{dur/pathak/sonmez:20}. Both papers contains examples showing that how a comparative static across all branches need not hold.  However, the two papers also show empirically that these theoretical cases do not apply in their applications.}  
However, as we present next and illustrate in  \fig{Figure \ref{fig:bradsocap}},  
the comparative static properties do hold in our simulations with the Class of 2021 data for several price responsiveness policies.

The Army considered three price responsiveness policies: the ultimate price responsiveness policy and two tiered price responsiveness policies.  Under the BRADSO-2020 price responsiveness policy, a cadet who expressed a willingness to
sign a BRADSO contract only obtained priority over other cadets who had the same categorical branch rating.  Under the BRADSO-2021 price responsiveness policy, a cadet who expressed a willingness to 
sign a BRADSO contract obtained higher priority over all other cadets if she was in the medium or high category.  To illustrate the trade-off between talent alignment and retention, \fig{Figure \ref{fig:bradsocap}} uses preferences from the Class of 2021 and re-runs the MPCO mechanism under these three price responsiveness policies for different levels of flexible-price  positions $q_b^f$, where  $q_b^f$ is expressed as a percentage of the total number of positions for branch $b$. 

To measure the effects of price responsiveness policies on BRADSOs collected, \fig{Figure \ref{fig:bradsocap}} shows how the number of BRADSOs charged increases with $q_b^f$ and with the ``closeness'' of the price responsiveness policy to the ultimate price responsiveness policy. That is, for a given $q_b^f$, the BRADSO-2021 policy results in more BRADSOs charged than the BRADSO-2020 policy, but fewer BRADSOs charged than the ultimate price responsiveness policy.  When the fraction of the flexible-price positions is small, there is relatively
little difference between price responsiveness policies.  For example, when the fraction of the
flexible-price positions is 15\% of all positions, 55 BRADSOs are charged under  the ultimate BRADSO policy, 47 BRADSOs are charged under
BRADSO-2021, and 38 BRADSOs are charged under BRADSO-2020.  When the fraction of the flexible-price  positions is larger, 
the price responsiveness policy has a larger effect on BRADSOs collected.
When the fraction of the
flexible-price  positions  is 65\%, 118 BRADSOs are charged under the ultimate BRADSO policy, 95 BRADSOs are charged under BRADSO-2021,
and 65 BRADSOs are charged under BRADSO-2020.

The ability to run this analysis on the effects of price responsiveness policies is an important benefit of a strategy-proof mechanism and
illustrates the iterative step in the minimalist approach.  
At the request of the Army, we conducted a similar analysis using data from the Class of 2020, but this analysis required stronger assumptions on cadet preferences.\footnote{Because cadets in the Class of 2020 did not submit preferences over branch-price pairs, 
we assumed that all BRADSOs are consecutive, and also considered
different assumptions on the prevalence of non-consecutive BRADSOs.  
These assumptions are not needed when cadets can rank branch-price pairs in a strategy-proof mechanism.}
As a result of this analysis, the Army decided to adopt the BRADSO-2021 policy and  increase  the fraction of the
flexible-price  positions from 25 to 35 percent.  These are both policies
that increase the power of BRADSO.  However, USMA decided against adopting the ultimate price responsiveness policy because branches remained opposed to giving more BRADSO power to low-tier cadets.

\section{Applications Beyond the US Army's Branching System} \label{sec:otherapplications}

The individual-proposing deferred acceptance (DA) algorithm of \cite{gale/shapley:62}
plays a prominent role in several market design applications, in particular
for priority-based resource allocation \citep{balinski/sonmez:99, abdulkadiroglu/sonmez:03}.
Our model is perhaps one of the most natural extensions of this approach for settings where the priorities of individuals can
be increased with a costly action for a subset of positions at each institution.
Based on Theorem 1, we believe that the MPCO mechanism is a natural
counterpart of DA for such settings.  Therefore, 
while our paper is mainly motivated by the Army's 2020 branching reform, our model in Section \ref{sec:model} and main characterization result in Theorem \ref{cosm} have other direct 
applications.\footnote{Section \ref{sec:potentialapplication} in the Online Appendix describes other possible applications
of price-responsiveness policies for priority-based assignment markets.}

In this section, we present a direct application  of our analysis in the context of a
school choice policy widely deployed in the recent history of China. 

\subsection{High School Seat Purchasing Policies in China}  \label{seatpurchasing}

In many cities in China, the priority ranking of students at public high schools mainly depends on their exam scores.  
Motivated by a departure from this policy in several  Chinese cities between 1990s and 2015, 
 \cite{wang/zhou:21} present an extension of the school choice model by \cite{abdulkadiroglu/sonmez:03}.  
 In their application, students gain increased priority for a subset of seats at each school by paying higher tuition levels.  \cite{wang/zhou:21} refer to this policy
as the  \textit{Ze Xiao (ZX) policy\/}.

Cities that deployed the ZX policy used a scoring-based price responsiveness policy we formulated in Section \ref{subsec:scoringbased}.
   
Parallel to our main application on the US Army's branching process, the cities of 
Shanghai and Tianjin used a single level of increased tuition for the \textit{ZX positions\/}.\footnote{\cite{wang/zhou:21}
present the following details for Shanghai and Tianjin:
``Shanghai is one of the cities that discontinued the ZX policy immediately after the
announcement from the Ministry of Education in 2012. The total percentage of ZX students
was restricted within 15\% for each school in 2011, which is the percentage for ZX policy
in the previous year. The ZX tuition in Shanghai was charged according to the type of
school. In district-level key high schools, the basic tuition for students was 2,400 Yuan/year,
whereas the ZX tuition was 6,000 Yuan/year before 2011 and 4,266 Yuan/year in 2011. For
the city-level key high schools, the basic tuition was 3,000 Yuan/year, whereas the ZX tuition
was 10,000 Yuan/year before 2011 and 7,000 Yuan/year in 2011. For the boarding schools,
the basic tuition was 4,000 Yuan/year, whereas the ZX tuition was 13,333 Yuan/year before
2011 and 9,333 Yuan/year in 2011.  
[$\cdots$] Tianjin canceled its ZX policy in 2015. Before 2015, the ZX tuition was standardized
across all general high schools at 8,000 Yuan/year, which was a fourfold increase in the basic
tuition (2,000 Yuan/year).''
} 
The empirical analysis in  \cite{wang/zhou:21} is for a city where the ZX policy is more involved
with four prices: A base  price of 1,600 yuan, and three layers of increased price (3,333 yuan, 5000 yuan, and 6000 yuan).
\cite{wang/zhou:21} describes the scoring-based price responsiveness policy used in this city until 2014 as follows:
\begin{quote}
``Three levels of the higher
tuition paid by ZX [increased price] students are based on their exam scores. A ZX student pays a total of
3,333.3 yuan annually if her score is within 10 points of the school’s cut-off, 5,000 yuan if it
is within 11–20 points, and 6,000 yuan per year if it is within 21–30 points.''
\end{quote}
This practice is equivalent to boosting the merit score of a student by 10 points if she is willing to pay a 
tuition of 3,333.3 yuans, by 20 points if she is willing to pay a tuition of 5,000 yuans, and by 30 points if she is willing to pay a 
tuition of 6,000 yuans. That is, for any school $b\in B$, the scoring rule  $S^b : \{t^0,t^1,t^2,t^3\} \rightarrow \mathbb{Z^+} $  
and the price responsiveness policy $\omega^S_b$ for the empirical application in  \cite{wang/zhou:21} are given as follows: 

For any school $b\in B$ and $t \in \{t^0,t^1,t^2,t^3\}$,  
\[ S^b(t) = 10 (t-t^0). 
\]
Given a list of merit scores $(m_i)_{i\in I}$ and a high school $b\in B$, for any two student-tuition pairs $(i,t), (j,t') \in I \times \{t^0,t^1,t^2,t^3\}$,  
\[   (i, t) \; \omega^S_b \; (j,t') \quad \iff \quad m_i + S^b(t) > m_j + S^b(t').
 \]

By  Theorem \ref{cosm}, MPCO is the only  direct mechanism for this application that satisfies  \textit{individual rationality, non-wastefulness, 
no priority reversals, enforcement of price responsiveness policy\/} and \textit{strategy-proofness\/}. 
The interpretation and desirability of all axioms except  \textit{enforcement of price responsiveness policy\/} is 
from standard arguments in the literature. So let us explore to what extent the axiom  \textit{enforcement of price responsiveness policy\/} is 
desirable in this setting.

Consider an allocation $X\in\cala$ and a student $i\in I$ such that $X_i = (b,t) \in B\times \{t^1,t^2, t^3\}$. 
Suppose there exists a student $j\in I\setminus \{i\}$ who has a legitimate claim for a price-reduced version of student $i's$ assignment $(b,t)$. 
Then,  there exists a tuition level $t' < t$ such that
\[ (b,t') \succ_j X_j \quad \mbox{ and }  \quad (j,t')  \; \omega^S_b \; (i,t), 
\]
or equivalently, 
\[ (b,t') \succ_j X_j \quad \mbox{ and }  \quad (m_j - m_i) >  10 (t-t'). 
\]
The last pair of relations directly contradict the city's ZX policy, because the difference between merit scores is too large to justify 
to award the seat to student $i$ at a higher tuition than $t'$ while the higher merit-score student $j$ is eager to receive  the seat
at this tuition level. 

Next,  consider an allocation $X\in\cala$ and a student $i\in I$ such that $X_i = (b,t) \in B\times \{t^0,t^1, t^2\}$. 
Suppose there exists a student $j\in I\setminus \{i\}$ who has a legitimate claim for a price-elevated version of student $i's$ assignment $(b,t)$. 
Then,  there exists a tuition level $t' > t$ such that
\[ (b,t') \succ_j X_j, \qquad  (j,t') \;  \omega^S_b \; (i,t), \quad \mbox{ and } \qquad  \Big|\big\{(k,t^+)\in I\times \{t^1, t^2, t^3\} \; : \; (k,b,t^+)\in X_b \big\}\Big|  < q^f_b, 
\]
or equivalently 
\[ (b,t') \succ_j X_j, \qquad   (m_i - m_j) <  10 (t'-t),  \qquad \mbox{ and } \qquad  \Big|\big\{(k,t^+)\in I\times \{t^1, t^2, t^3\} \; : \; (k,b,t^+)\in X_b \big\}\Big|  < q^f_b. 
\]
The last triple of relations directly contradicts the city's ZX policy, because the difference between merit scores is not large enough to justify 
awarding the seat to student $i$ at a lower tuition level than $t'$  while student $j$ is eager to receive the seat at a tuition level of $t'$ and 
doing so is feasible. 

Therefore, the axiom  \textit{enforcement of price responsiveness policy\/} is also highly plausible in this setting, 
thus making MPCO a highly desirable mechanism in this setting.

\section{Broader Lessons for Institution Redesign and Conclusions}\label{sec:conclusion}

As described in Section \ref{sec:min}, we view our 
experience with the Army's branching process as the first proof-of-concept for  minimalist market design.  
We believe our experience redesigning the cadet branch assignment process has lessons for other institution redesign efforts. 

First, our redesign effort started by understanding what principles the Army's leadership had in mind when they set up their system. Rather than seeing their existing procedure as ad hoc, we saw it as a valuable precedent and used it to discover their (and not our) essential principles. For many decades, the branch assignment process has been based on a merit-based hierarchy where cadets have some choice over their branch assignments. Since retention has been a persistent problem, some stakeholders were open to experimenting with the rigidity of that hierarchy. It is unlikely the military would be sympathetic, at least initially, to a perspective that challenged these broad goals, or more specifically, to one that tinkered with the use of the Order-of-Merit List (OML), eliminating cadet choice, or took a position on the size of the BRADSO cap. With the adoption of the Talent-Based Branching (TBB) program, the US Army made the cadet property right structure more complex because a cadet could obtain priority for a branch in several ways.  

USMA-2020 was an attempt to devise a procedure that accommodates new considerations brought by the TBB program. The solution that the Army devised made several issues apparent to the system operator, participants, and market designers. The Army did not struggle to articulate its new objectives as intuitive 
principles with the introduction of TBB,  but it did struggle to combine these principles with existing ones and then find a procedure 
which satisfies them all. We saw our role as discovering and formalizing these principles and then proposing a mechanism. Our perspective is in line with \cite{li:17}, who sees market designers as uniquely positioned to formalize normative criteria for practitioners.

Second, to turn a criticism into a collaboration, we focused on demonstrating value to the system operators and helping them build the case for reform.  
Real-life evidence related to the issues we raised showed that our concerns are not simply imagined.  Discussion on online Army forums and responses to cadet surveys cemented the concerns.\footnote{In other instances, court cases or stories in the media can create pressure for reform. 
See, for example, \cite{sonyen22}) and \cite{cook:03}.}   
Next, we volunteered to help authorities verify or detect issues related to the criticism.  With the TBB program, detectable priority reversals could no longer be manually corrected ex-post without affecting the assignment of multiple cadets.  Earlier criticisms of the USMA-2006 mechanism by \cite{sonmez/switzer:13} did not justify a costly reform because these issues could be more easily fixed.\footnote{As also emphasized in \cite{sonmez:23}, 
the importance of detectability for driving reform is apparent in the contrast between unsuccessful efforts to reform Turkish college admissions in \cite{balinski/sonmez:99} and the successful effort to reform the Boston mechanism in \cite{abdulkadiroglu/sonmez:03}.  In Turkey, authorities did not see the system as broken enough to warrant a change. In Boston, authorities were more receptive to the initial criticism because of feedback from participants, such as parent groups, who were affected by the incentive compatibility issues.} 
The Army specifically asked us to quantify the extent of priority reversals using data from the USMA-2020 mechanism.  We made several presentations using historical data attempting to measure the extent of the issues we highlighted, including a presentation at West Point in January 2020.\footnote{Descriptive
empirical work identifying issues with existing mechanisms was also important in the Boston school choice redesign effort, see, \cite{apsr:06}.}  

When making a case for reform, it was valuable to emphasize our role as technical experts and avoid taking positions on issues that would upset the balance between competing groups. 
For example, the Army continues to debate the appropriate balance between retention incentives, merit, and talent alignment. 
Our involvement stayed clear of taking positions on these trade-offs.
For some dimensions with broad consensus among all stakeholders, it can be appropriate to take a stronger position. 
For example, all Army groups agreed that cadets should not engage in a high-stakes strategic calculation to obtain their branch assignment, a position with which we agreed. If a system operator intentionally encourages strategic behavior, they should be transparent about it.\footnote{In our experience, systems that purposefully put strategic pressure on participants are often not advertised as such. For instance, as discussed in \cite{pathak/sonmez:13}, 
officials in Seattle Public Schools appear to have adopted the Boston (immediate acceptance) mechanism to embrace a policy objective not discussed openly. 
See also \cite{dur/pathak/sonmez/song:22} for a similar phenomenon in Taiwan.}
The Army was receptive to changing its mechanism because our proposal minimized normative interference. Questions about trade-offs between competing objectives were deferred to later stages of the Army's design effort until after they fixed the apparently broken parts of the system.  

Fourth, the Army experience illustrates the value of iteration in design. Our initial critical appraisal of the Army's system led to a partnership where we could provide technical assistance and eventually assist with other reforms. After the Army's redesign, more
than 3,000 ROTC cadets were assigned to branches on an accelerated timeline using the same assignment mechanism.  
After one year with the system, the Army switched its BRADSO cap and price responsiveness policy. These subsequent policy developments are examples of going from local improvements, the aim of the minimalist approach, towards more substantial changes.\footnote{Here, again, there is a parallel to work on student assignment. Districts often tinker with admissions priorities after adopting a strategy-proof assignment mechanism. Economists are now starting to explore issues related to the design of these priorities and, in some cases, they are influencing policy discussions about their design (see, e.g., \cite{dur_boston}, \cite{celebi/flynn:22}, and \cite{idoux:22}.)}

Finally, the Army's redesign sheds new light on the role of the deferred acceptance algorithm in market design. While rightly celebrated as a significant achievement, it is crucial to understand the primitive settings of any design problem before simply relying on this familiar tool. The focus on the stability axiom in practical design dates to \cite{roth:84}, who showed that the medical match clearinghouse produced a stable outcome, but this aim was undermined by the presence of married couples. For many market design settings, the stability axiom may not be an important or even adequate axiom. 
Its relevance depends on the particular setting and normative goals of the system operator and stakeholders. The USMA-2020 system illustrates the risk of a design
methodology based on a famous algorithm rather than the principles of the stakeholders.  The design of the USMA-2020 mechanism attempted to force the problem into a two-sided matching framework based on stability.  This perspective inadvertently missed essential aspects of the problem, the most important of which was how to design a mechanism to accommodate several  types of property rights. For this reason, it is essential to begin by understanding the first principles involved in the system before turning to a procedure. In the case of the Army, the OML, the BRADSO program, and Talent-Based Branching created a complex structure of property rights that required new conceptual formulations. Some of the desiderata bear mathematical similarities with simpler models of two-sided markets, but their interpretation is entirely different for the Army setting. Because of these different interpretations, the Army required a new mechanism, which required developing and extending theory for this application.

\bibliographystyle{aer}
\bibliography{Army-MMD-March2023-ArXiv}

\newpage

\newpage
\appendix

\setcounter{figure}{0}
\section{Online Appendix: Main Characterization Result}
\label{proofs}

\subsection{Proof of Theorem \ref{cosm}}

\noindent \textbf{Proof of Theorem \ref{cosm}}: Fix $(\pi_b)_{b\in B} \in \Pi^{|B|}$ and $(\omega_b)_{b\in B} \in \prod_{b\in B}\omega_b$.

We first show that the mechanism $\phi^{MPCO}$ satisfies the five  axioms. 
For the proofs of \textit{individual rationality, non-wastefulness, no priority reversal,\/} and
\textit{enforcement of the price responsiveness policy\/}, also  fix $\succ \, \in \calq^{|I|}$. 
\smallskip

\textbf{\textit{Individual rationality\/}}: 
No cadet $i\in I$ ever makes a proposal to a branch $b$ at a price $t\in T$ under the MPCO procedure,  unless
her preferences are such that $(b,t)  \succ_i \emptyset$. Hence, the MPCO mechanism satisfies \textit{individual rationality\/}. \smallskip

\textbf{\textit{Non-wastefulness\/}}: For any branch $b\in B$, 
unless there are  already $q$ contracts with distinct cadets on hold, 
it is not possible for  the base-price contract of any given cadet to be rejected at any step of the MPCO procedure.  
Hence, the  MPCO mechanism satisfies \textit{non-wastefulness\/}. \smallskip

\textbf{\textit{No priority reversal\/}}:  Suppose that $\phi_j^{MPCO}(\succ)\succ_i\phi_i^{MPCO}(\succ)$ for a pair of cadets $i,j\in I.$  Since 
the MPCO mechanism 
is \textit{individually rational\/}, $\phi_j^{MPCO}(\succ)  \not= \emptyset$. Let branch $b\in B$ and price $t\in T$ be such that $\phi_j^{MPCO}(\succ)  = (b,t)$. 
Let $k$ be the final step of the MPCO  procedure. 
Since  $\phi_j^{MPCO}(\succ) \succ_i  \phi_i^{MPCO}(\succ)$, cadet $i$ has proposed
the contract $(i,b,t)$ to branch $b$ at some step of the  MPCO  procedure, which is rejected by branch $b$ (strictly speaking for the first time) 
either immediately or at a later step. 
Since the proposed contracts remain available until the termination  of the MPCO procedure, 
the contract  $(i,b,t)$ is also rejected by branch $b$ at the final Step $k$ of the MPCO procedure.  
In contrast, since $\phi_j^{MPCO}(\succ)  = (b,t)$, contract $(j,b,t)$ is chosen by branch $b$ at the final step $k$ of the MPCO procedure.  
If the contract $(j,b,t)$ is accepted as one of the first $q^0_b$ positions under the choice rule $\calc^{MP}_b$, then $j \; \pi_b \; i$. 
Otherwise, if the contract $(j,b,t)$ is accepted as one of the last $q^f_b$ positions under the choice rule $\calc^{MP}_b$, then $(j,b,t) \; \omega_b \; (i,b,t)$.
In either case, we have $j \; \pi_b \; i$,  proving that the MPCO mechanism satisfies \textit{no priority reversal\/}. \smallskip

\textbf{\textit{Enforcement of the price responsiveness policy\/}}:  Let cadet $i\in I$ be such that $\phi^{MPCO}_i(\succ) = X_i = (b,t)\in B \times T^+$. 
Let price $t' \in T$ be such that $t' < t$. Let cadet $j \in I\setminus\{i\}$ be such that  $(b,t') \succ_j \phi^{MPCO}_j(\succ)$.  
Then cadet $j$ has proposed
the contract $(j,b,t')$ to branch $b$ at some step of the MPCO procedure, which is rejected by branch $b$ either immediately or at a later step. 
Let $k$ be the final step of the  MPCO procedure. 
Since the proposed contracts remain available until the termination  of the procedure, the contract  $(j,b,t')$ is also
rejected by branch $b$ at the final Step $k$ of the  MPCO procedure. 
More specifically, it is rejected by the choice rule $\calc^{MP}_b$ at the final Step $k$  of the procedure both for the first $q^0_b$ positions using the 
baseline priority order $\pi_b$ and for the last $q^f_b$ positions using the price responsiveness policy $\omega_b$. 
In contrast, since $t \in T^+$ by assumption, 
contract $(i,b,t)$ is chosen by branch $b$ at the final Step $k$ of the  MPCO procedure  using the price responsiveness policy $\omega_b$.
Therefore, we have $(i,t) \; \omega_b \; (j,t')$, which in turn implies that cadet $j$ does not have a legitimate claim for a price reduced version of $X_i = (b,t)$. 

\newpage

Let cadet $i\in I$ be such that $\phi^{MPCO}_i(\succ) = X_i = (b,t)\in B \times T$. 
Let price $t' \in T^+$ be such that $t' >t$. Let cadet $j \in I\setminus\{i\}$ be such that  $(b,t') \succ_j \phi^{MPCO}_j(\succ)$ and $(j,t') \; \omega_b \; (i,t)$. 
Further assume that cadet $i$ is the lowest $\pi_b$-priority cadet  with  an assignment of $(b,t)$. 

The relation  $(b,t') \succ_j \phi^{MPCO}_j(\succ)$ implies that cadet $j$ has proposed
the contract $(j,b,t')$ to branch $b$ at some step of the  MPCO procedure, which is rejected by branch $b$ either immediately or at a later step. 
Let $k$ be the final step of the MPCO  procedure. 
Since the proposed contracts remain available until the termination  of the procedure, the contract  $(j,b,t')$ is also
rejected by branch $b$ at the final Step $k$ of the MPCO  procedure. 
More specifically, it is rejected by the choice rule $\calc^{MP}_b$ at the final Step $k$  
even for the last $q^f_b$ positions using the price responsiveness policy  $\omega_b$. 
Therefore, since by assumption we have $(j,t') \; \omega_b \; (i,t)$, cadet $i$  must have received one of the first $q^0$ positions using the 
baseline priority ranking $\pi_b$. But since cadet $i$ is the lowest $\pi_b$-priority cadet  with an assignment of $(b,t) = (b,t^0)$, 
that means no cadet has received any of the last $q^f_b$ positions at the base price of $t^0$. Therefore, since MPCO mechanism satisfies
\textit{non-wastefulness\/}, 
\[  \Big|\big\{(k,t^+)\in I\times T^+ \, : \; (k,b,t^+)\in X_b \big\}\Big|  =  q^f_b. 
\]
That is, the cap for flexible-price positions is already reached at branch $b$.
As such, cadet $j$ does not have a legitimate claim for a price increased version of $X_i  = (b,t)$.

Since no cadet  has a legitimate claim for either  a priced reduced or a priced increased version of 
another cadet's assignment, the MPCO mechanism  satisfies \textit{enforcement of the price responsiveness policy\/}. \smallskip

\indent \textbf{\textit{Strategy-proofness\/}}:  The MPCO mechanism is a special case of 
the cumulative offer mechanism  for  \textit{matching problems with slot-specific priorities\/} formulated in 
\cite{kominers/sonmez:16}. Hence \textit{strategy-poofness\/} of the MPCO mechanism is a direct corollary of
their Theorem 3, which proves \textit{strategy-proofness\/} of the cumulative offer mechanism more broadly for matching problems with slot-specific priorities. \medskip

\textbf{\textit{Uniqueness\/}}: We prove uniqueness via two lemmata.

\begin{lemma} \label{lemma3} 
Let  $X, Y \in \cala$ be two distinct allocations that satisfy  \textit{individual rationality, non-wastefulness, no priority reversal,\/} and 
\textit{enforcement of the price responsiveness policy\/}. 
Then there exists a cadet $i \in I$ who receives non-empty and distinct assignments under $X$ and $Y$. 
\end{lemma}

\noindent \textit{Proof of Lemma \ref{lemma3}\/}: The proof is by contradiction.  Fix $\succ \; \in \calq^{|I|}$. 
Let  $X, Y\in \cala$ be two distinct allocations that 
satisfy  \textit{individual rationality, non-wastefulness, no priority reversal,\/} and 
\textit{enforcement of the price responsiveness policy\/}. 
 To derive the desired contradiction, suppose that,  for any cadet $i \in I$,
 \begin{equation} \label{lem3-eqn0}
 X_i \not= Y_i \quad  \implies \quad X_i = \emptyset \; \mbox{ or  } \; Y_i =\emptyset. 
\end{equation}

Pick any branch $b\in B$ such that $X_b \not= Y_b$. Let $j\in I$ be the highest $\pi_b$-priority cadet who
is assigned to branch $b$  either under  $X$ or under $Y$, but not both. 
W.l.o.g.,  let cadet  $j$ be assigned to branch $b$ under allocation $X$ but  not under allocation $Y$. By relation \eqref{lem3-eqn0}, 
\begin{equation} \label{Yj}
Y_j = \emptyset.
\end{equation}
Since allocation $Y$ satisfies \textit{non-wastefulness\/}, there exists a cadet $k\in I$ who is assigned to branch $b$ under allocation $Y$
but not under allocation $X$. By relation \eqref{lem3-eqn0},  
 \begin{equation} \label{Xk}
X_k = \emptyset,
\end{equation}
and therefore, by the choice of cadet $j$, we have
\begin{equation} \label{j-k-priority}
j \; \pi_b \; k. 
\end{equation}

Let $X_j = (b,t)$ and $Y_k = (b,t')$. 
Since allocations $X$ and $Y$ satisfy \textit{individual rationality\/}, we must have $t' \not= t$, for otherwise
one of the allocations $X, Y$ would fail \textit{no priority reversal\/}. 
Moreover, since we either have  $(j,t) \; \omega_b \; (k,t')$ or $(k,t') \; \omega_b \; (j,t)$, 
one of the two prices $t, t'$ has to be equal to $t^0$, for otherwise one of the allocations $X, Y$ would fail \textit{enforcement of the price responsiveness policy\/}. 
Therefore, by relation \eqref{j-k-priority},
\begin{equation} \label{Yk-Tplus}
Y_k = (b,t') \quad \mbox{for some } t'\in T^+,
\end{equation}
for otherwise (i.e., if $Y_k = (b,t^0)$) allocation $Y$ would fail \textit{no priority reversal\/}. Hence, 
\begin{equation} \label{Xj-t0}
X_j = (b,t^0).
\end{equation}

Since allocation $Y$ satisfies \textit{enforcement of the price responsiveness policy\/}, relations \eqref{Yj}  and \eqref{Yk-Tplus} imply
 \begin{equation} \label{lem3-eqn7}
(k, t') \; \omega_b \; \underbrace{(j,t)}_{= (j,t^0)}.
\end{equation} 
 
Define 
\begin{equation} \label{I-star-def}
I^*  = \big\{i\in I:  X_i = (b,t^+_i) \; \mbox{ for some } t^+_i \in T^+ \big\}. 
\end{equation} 
Since allocation $Y$ satisfies  \textit{enforcement of the price responsiveness policy\/}, 
\begin{equation} \label{I-star}
|I^*| = q^f_b, 
 \end{equation}
 for otherwise cadet $k$ would have a legitimate claim for a price increased version of cadet $j's$ assignment $X_j = (b,t^0)$
 by relations \eqref{Xk} and  \eqref{lem3-eqn7}. 
 
 Since allocation $X$ satisfies \textit{enforcement of the price responsiveness policy\/} and 
 \[ \underbrace{(b,t')}_{= Y_k} \; \succ_k \;  \underbrace{X_k}_{= \emptyset}, 
 \]
 for any $i\in I^*$, we have
  \begin{equation} \label{i-k}
(i, t^+_i) \; \omega_b \;  (k,t').
\end{equation}

But since $Y_k = (b,t')$ for some $t'\in T^+$ by relation \eqref{Yk-Tplus} and  $|I^*| = q^f_b$ by relation \eqref{I-star},
there exists a cadet $\ell\in I^*$ with $Y_{\ell} \not\in \{(b,t_{\ell}^+)  : t_{\ell}^+ \in T^+\}$. 
Therefore, by relations  \eqref{lem3-eqn0} and \eqref{I-star-def}, we have
\begin{equation} \label{Y-ell-emptyset}
Y_{\ell} = \emptyset. 
\end{equation}
Since $X$ satisfies \textit{individual rationality\/} and $\ell \in I^*$, we have
\begin{equation} \label{ell-ir}
(b,t_{\ell}^+) \; \succ_{\ell} \; \emptyset.
\end{equation}
Therefore, by relations \eqref{Yk-Tplus}, \eqref{i-k}, \eqref{Y-ell-emptyset} and \eqref{ell-ir} allocation $Y$ fails 
either \textit{no priority reversal\/} or 
\textit{enforcement of the price responsiveness policy\/} (depending on whether $t_{\ell}^+ = t'$ or $t_{\ell}^+ \not= t'$), 
thus giving us the desired contradiction and completing the proof of Lemma \eqref{lemma3}. 
\mbox{}\hfill$\diamondsuit$ \medskip

\begin{lemma} \label{lemma4} There can be at most one direct mechanism that satisfies 
 \textit{individual rationality, non-wastefulness, no priority reversal, enforcement of the price responsiveness policy\/} and
 \textit{strategy-proofness\/}. 
\end{lemma}

\noindent \textit{Proof of Lemma \ref{lemma4}\/}: The proof of this lemma is inspired by a technique introduced by \cite{hirata/kasuya:17}.  
Towards a contradiction, suppose there exists two distinct direct mechanisms $\varphi$ and $\psi$ that satisfy  
 \textit{individual rationality, non-wastefulness, no priority reversal, enforcement of the price responsiveness policy\/} and
 \textit{strategy-proofness\/}. 
Let the preference profile $\succ^* \in \calq^{|I|}$ be such that
\begin{enumerate}
\item $\varphi(\succ^*) \not= \psi(\succ^*)$, and
\item the aggregate number of acceptable contracts between all cadets is minimized 
among all preference profiles    $\widetilde{\succ} \in \calq^{|I|}$ such that  $\varphi(\widetilde{\succ}) \not= \psi(\widetilde{\succ})$. 
\end{enumerate}
Let  $X= \varphi(\succ^*)$ and $Y= \psi(\succ^*)$. By Lemma \ref{lemma3}, there exists a cadet $i\in I$ such that
\begin{enumerate}
\item $X_i \not= \emptyset$, 
\item $Y_i \not= \emptyset$, and
\item $X_i \not= Y_i$. 
\end{enumerate}
Since both allocations $X$ and $Y$ satisfy \textit{individual rationality\/}, 
\[ X_i \; \succ^*_i \; \emptyset \quad \mbox{ and } \;  Y_i \; \succ^*_i \; \emptyset. 
\]
W.l.o.g., assume
\[ X_i \; \succ^*_i \; Y_i \; \succ^*_i \; \emptyset.
\]
Construct the preference relation $\succ'_i \in \calq$ as follows:\smallskip

If $X_i = (b,t^0)$ for some $b\in B$, then 
\[ (b,t^0) \; \succ'_i \; \emptyset \; \succ'_i \; (b',t') \qquad \mbox{ for any } (b',t')\in B\times \big(T \setminus \{(b,t^0)\}\big).
\]
Otherwise, if 
$X_i = (b,t^r)$ for some $b\in B$ and $r \in \{1,\ldots,h\}$, then 
\[ (b,t^0) \; \succ'_i \; \cdots \succ'_i  \; (b,t^{r-1}) \; \succ'_i \;  (b,t^r) \; \succ'_i \;  \emptyset \; \succ'_i \; (b',t') 
\; \mbox{ for any } (b',t')\in B\times  \big(T\setminus\{(b,t^0), \ldots,(b,t^r)\}\big).
\]
Since $X_i \succ^*_i  Y_i \, \succ^*_i  \emptyset$ and 
$(b,t^0) \succ^*_i  \;  \cdots \; \succ^*_i  \; (b,t^{r-1}) \;  \succ^*_i (b,t^r)$, the preference relation 
$\succ'_i$ has strictly fewer acceptable contracts for cadet $i$ than the preference relation $\succ^*_i$.

By \textit{strategy-proofness\/} of the mechanism $\psi$, we have
\[ \underbrace{\psi_i(\succ^*_i,\succ^*_{-i})}_{=Y_i} \; \succeq^*_i \;  \psi_i(\succ'_i,\succ^*_{-i}), 
\]
and since no branch-price pair $(b',t') \in B\times T$ with $Y_i \succeq'_i (b',t')$ is acceptable under $\succ'_i$, 
by \textit{individual rationality\/}  of the mechanism $\psi$ we have
\begin{equation} \label{lemma4-eqn1}
 \psi_i(\succ'_i,\succ^*_{-i}) = \emptyset.
\end{equation}
Similarly, by \textit{strategy-proofness\/} of the mechanism $\varphi$, we have
\[ \varphi_i(\succ'_i,\succ^*_{-i}) \; \succeq'_i  \; \underbrace{\varphi_i(\succ^*_i,\succ^*_{-i})}_{=X_i}, 
\] 
which in turn implies 
\begin{equation} \label{lemma4-eqn2}
 \varphi_i(\succ'_i,\succ^*_{-i}) \not= \emptyset.
\end{equation}
But then, by relations  \eqref{lemma4-eqn1} and \eqref{lemma4-eqn2} we have 
\[  \varphi(\succ'_i,\succ^*_{-i}) \not= \psi(\succ'_i,\succ^*_{-i}),
\]
giving us the desired contradiction, 
since between all cadets the preference profile $(\succ'_i,\succ^*_{-i})$ has strictly fewer acceptable contracts than 
the preference profile $\succ^*$. This completes the proof of Lemma \ref{lemma4}. 
\mbox{}\hfill$\diamondsuit$ \medskip

Since we have already shown that the MPCO mechanism satisfies all five axioms, Lemma \ref{lemma4} establishes the desired
uniqueness result, thus concluding the proof of Theorem  \ref{cosm}. 
\qed \medskip 

\bigskip

\subsection{Independence of  Axioms in Theorem \ref{cosm}} \label{independenceofaxioms} 

We establish the independence of the axioms in Theorem \ref{cosm} by
presenting five direct mechanisms.  Each fails one of our five axioms and satisfies the other four. 
Our result shows that none of the axioms are redundant in Theorem \ref{cosm} and each
is important for the characterization of MPCO mechanism. \smallskip

\subsubsection{A mechanism that satisfies all axioms except individual rationality} Given any preference profile
$\succ \, \in \calq^{|I|}$ and individual $i\in I$, let $\succ_i^0\in \calq$ be the preference relation where  
the relative preference ranking of all branch-price pairs in $B\times T$ is the same as in   $\succ_i$, and remaining unmatched (i.e. $\emptyset$) is  the last choice. 
Define the direct mechanism $\phi^0$ as, for any  preference profile $\succ \, \in \calq^{|I|}$,
\[ \phi^0(\succ) = \phi^{MPCO}\big(\succ^0\big). 
\]
Mechanism $\phi^0$ satisfies all axioms except \textit{individual rationality\/}. 
 
\subsubsection{A mechanism that satisfies all axioms except non-wastefulness} 
Define the direct mechanism $\phi^{\emptyset}$ as,  for any preference profile
$\succ \, \in \calq^{|I|}$, 
\[ \phi^{\emptyset}(\succ) = \emptyset. 
\]
Mechanism $\phi^{\emptyset}$ satisfies all axioms except \textit{non-wastefulness\/}. 

\subsubsection{A mechanism that satisfies all axioms except enforcement of the price responsiveness policy} 
The individual-proposing deferred acceptance mechanism given in  Online Appendix \ref{sec:da} satisfies
all axioms except \textit{enforcement of the price responsiveness policy}.\footnote{More broadly  the MPCO mechanism when implemented with a different profile of price responsiveness policies than the
underlying one also satisfies all axioms except \textit{enforcement of the price responsiveness policy}.} 

\subsubsection{A mechanism that satisfies all axioms except no priority reversal} 

We will assume that there are only two prices. In all other cases, assume that
the outcome of mechanism $\psi$  is same as the outcome of the  MPCO mechanism.
When there are two prices, $t^0$ and $t^h$, 
the outcome of the mechanism $\psi$  is derived from the outcome of the MPCO mechanism  as follows.  

Fix a branch $b\in B$. Given any preference profile $\succ \, \in \calq^{|I|}$,
let $i\in I$ be the lowest $\pi_b$-priority individual with $\b\big(\phi_i^{MPCO}(\succ)\big)=b$. 
Let the preference relation $\succ^{-b}_i \in \calq$ be constructed from $\succ_i$ by
making branch-price pairs $(b,t^0)$ and $(b,t^h)$ 
unacceptable, but otherwise keeping the rest of
the preference order same as in $\succ_i$. Let the outcome of the mechanism $\psi$ be given as
\begin{itemize}
\item  $\psi(\succ) = \phi^{MPCO}(\succ_{-i}, \succ^{-b}_i)$ if all $q^f_b$ flexible-price  positions 
at branch $b$ are awarded at  the increased price $t^h$
under both $\phi^{MPCO}(\succ_{-i}, \succ^{-b}_i)$ and $\phi^{MPCO}(\succ)$, and 
\item  $\psi(\succ) = \phi^{MPCO}(\succ)$ otherwise. 
\end{itemize}
For any given branch $b\in B$, mechanism $\psi$ derives its outcome mostly using the MPCO mechanism, except
it ``ignores'' the lowest $\pi_b$-priority  individual 
who receives  a position at branch $b$ under the MPCO mechanism 
provided that all flexible-price  positions at branch $b$ are awarded at the increased price $t^h$ under the MPCO
mechanism  whether  the lowest $\pi_b$-priority  individual is being ignored or not. 
If in either scenario some of the $q^f_b$ flexible-price  positions are awarded at the base price $t^0$ or remain idle, 
then the  outcome of the mechanism $\psi$ is the same as the outcome of the MPCO mechanism. 

Mechanism $\psi$ satisfies all axioms except the \textit{no priority reversal\/}. The detailed construction 
above assures that it does not also lose \textit{enforcement of the price responsiveness policy\/} or \textit{strategy-proofness\/}
due to the modification.  

\subsubsection{A mechanism that satisfies all axioms except strategy-proofness} 

We assume that there are only two prices. In all other cases, assume that
the outcome of mechanism $\psi$  is same as the outcome of the  MPCO mechanism.
The outcome of the mechanism $\varphi$  is derived from the outcome of the MPCO mechanism 
as follows.  

Fix a branch $b\in B$. Given any preference profile $\succ \, \in \calq^{|I|}$,
let $i\in I$ be the lowest $\pi_b$-priority individual with $\b\big(\phi_i^{MPCO}(\succ)\big)=b$. 
If 
\begin{enumerate}
\item $\phi_i^{MPCO}(\succ) = (b,t^0)$, 
\item $(b,t^h) \; \succ_i \; \emptyset$, and 
\item $\Big(\phi^{MPCO}(\succ)\setminus\big\{(i,b,t^0)\big\}\Big) \bigcup \big\{(i,b,t^h)\big\} \in \cala$, 
\end{enumerate}
then let $\varphi(\succ) = \Big(\phi^{MPCO}(\succ)\setminus\big\{(i,b,t^0)\big\}\Big) \bigcup \big\{(i,b,t^h)\big\}$. 
Otherwise, i.e. if any of the three conditions fail, then let  $\varphi(\succ) = \phi^{MPCO}(\succ)$. 

Compared to the outcome of the MPCO mechanism, the
mechanism $\varphi$ simply increases the charged price for the
lowest $\pi_b$-priority individual who receive a position at branch $b$ under the  MPCO mechanism,
if doing so is \textit{feasible\/} and does not violate \textit{individual rationality\/}.

Mechanism $\varphi$ satisfies all axioms except \textit{strategy-proofness\/}. 
The affected individual  can profit by declaring the branch-price pair $(b,t^h)$ as unacceptable under the mechanism $\varphi$. 
The detailed construction 
above assures that the mechanism does not also lose \textit{individual rationality\/}, 
\textit{no priority reversal}, or \textit{enforcement of the price responsiveness policy\/}  due to the modification. 

\newpage

\section{Formal Analysis of  USMA-2020 Mechanism} \label{sec:singlebranch}
Sections \ref{sec:shortcomings2020} and \ref{field-2006-2020} present the shortcomings of the USMA-2020 mechanism. 
In this section of the Online Appendix, we present a more in-depth analysis of the USMA-2020 mechanism to offer additional insight on why it
resulted in a much more complex branching system than its predecessor USMA-2006 mechanism. 

Since USMA-2020 is defined only when there is a single increased price, throughout this section, we assume that $T^+ = \{t^h\}$. 

As with the USMA-2006 mechanism, truthful revelation of branch preferences is not a dominant strategy under the USMA-2020 mechanism,
thereby making its formal analysis challenging. 
Fortunately, focusing on a simpler version of the model with a single branch is sufficient to illustrate and analyze the main challenges
of the USMA-2020 mechanism. 

Suppose we consider a single branch $b \in B$.  When there is a single branch $b \in B$, there are only two preferences  for any cadet $i \in I$.
The base price contract  $(i,b,t^0)$ is by assumption preferred by cadet $i$ to both its 
increased price version $(i,b,t^h)$ and also to remaining unmatched. Therefore, the only variation in cadet $i$'s preferences
depends on whether the increased price contact $(i,b,t^h)$ is preferred to remaining unmatched. For any cadet $i \in I$, $|\calq| =2$
when there is a single branch $b\in B$, since 
\begin{itemize}
\item indicating willingness  to pay the increased price $t^h$ 
under a quasi-direct mechanism can be naturally mapped to the preference relation  where the  increased price contact $(i,b,t^h)$ is acceptable, whereas 
\item not doing so 
can be naturally mapped to the preference relation  where the  increased price contact $(i,b,t^h)$ is unacceptable, 
\end{itemize}
any quasi-direct mechanism  can be interpreted as a direct mechanism. 
Therefore, unlike the general version of the model, the axioms of BRADSO-IC and elimination of strategic BRADSO are  also
well-defined for direct mechanisms when there is a single branch, and moreover, they are both implied by strategy-proofness.\footnote{BRADSO-IC and elimination of strategic BRADSO together are equivalent to strategy-proofness when there is a single branch. Strategy-proofness of a single branch,  called non-manipulability via contractual terms, also plays an important role in the analysis of \cite{hatfield/kominers/westkamp:21}.}  

\subsection{Single-Branch Mechanism {\boldmath $\phi^{MP}$} and Its Characterization}

We next introduce a single-branch direct mechanism that is key for our analysis of the USMA-2020 mechanism.
The main feature of this mechanism is its iterative subroutine (in Step 2), which determines how many flexible-price  positions 
are assigned at the increased price and which cadets receive these positions.   
\smallskip

\begin{quote}
\noindent \textbf{Mechanism} {\boldmath $\phi^{MP}$} 

\noindent For any given profile of cadet preferences $\succ = (\succ_i)_{i\in I} \in \calq^{|I|}$, 
construct the allocation $\phi^{MP}(\succ)$ as follows:\smallskip

\noindent \textbf{Step 0.} Let $I^0 \subset I$ be the set of  $q^0_b$ highest $\pi_b$-priority cadets in $I$. 
For each cadet $i \in I^0$, finalize the assignment of cadet $i$ as $\phi_i^{MP}(\succ) = (b,t^0)$.  \smallskip

\noindent \textbf{Step 1.} Let $I^1 \subset I\setminus I^0$ be the set of  $q^f_b$ highest $\pi_b$-priority cadets in $I\setminus I^0$. 
\textit{Tentatively\/} assign each cadet in $I^1$  a position at the base price $t^0$. 
Relabel the set of cadets in $I^1$ so that cadet $i^1 \in I^1$ has the lowest $\pi_b$-priority in $I^1$,  
cadet $i^2 \in I^1$ has the second-lowest $\pi_b$-priority in $I^1$, $\ldots$, and  
cadet $i^{q^f_b} \in I^1$ has the highest $\pi_b$-priority in $I^1$. 
Also relabel the lowest  $\pi_b$-priority cadet in $I^0$ as $i^{q^f_b + 1}$.
\smallskip

\noindent \textbf{Step 2.} This step determines how many positions are assigned at the increased price $t^h$. \smallskip

\noindent \textbf{Step 2.0.} 
Let $J^0 \subset I \setminus (I^0 \cup I^1)$ be the set of cadets in $I \setminus (I^0 \cup I^1)$ who declared 
the position at the increased price  $t^h$ as acceptable:
\[ J^0 = \{j \in I\setminus (I^0 \cup I^1) :  (b,t^h) \; \succ_j \; \emptyset\}.
\]
If \[\big|\big\{j \in J^0 : (j, t^h) \; \omega_b \; (i^1, t^0) \big\}\big| = 0,\] then finalize Step 2 and proceed to Step 3.
In this case no position will be assigned at the increased price $t^h$. 

Otherwise, if 
\[\big|\big\{j \in J^0 : (j, t^h) \; \omega_b \; (i^1, t^0) \big\}\big|  \geq 1,\] 
then proceed to Step 2.1.  \smallskip

\noindent \textbf{Step 2.}{\boldmath $\ell$.} {\boldmath $(\ell = 1,\ldots, q^f_b)$} 
Let 
\[J^{\ell} = \left\{ \begin{array}{cl}
         J^{\ell -1}  & \mbox{ if } \; \emptyset \, \succ_{i^{\ell}} \, (b,t^h)\\
       J^{\ell -1} \cup\{i^{\ell}\} & \mbox{ if } \; (b,t^h) \succ_{i^{\ell}} \, \emptyset. \end{array} \right.
\]
If \[\big|\big\{j \in J^{\ell} : (j, t^h) \; \omega_b \; (i^{\ell +1}, t^0) \big\}\big| = \ell,\] then finalize Step 2 
and proceed to Step 3.\footnote{Since  $J^{\ell} \supseteq J^{\ell -1}$ by construction, the fact that the procedure has reached Step 2.$\ell$ 
implies that the inequality  $\big|\big\{j \in J^{\ell} : (j, t^h) \; \omega_b \; (i^{\ell +1}, t^0) \big\}\big| \geq \ell$ must hold.}
In this case $\ell$ positions will be assigned at the increased price $t^h$. 

Otherwise, if 
\[\big|\big\{j \in J^{\ell} : (j, t^h) \; \omega_b \; (i^{\ell +1}, t^0) \big\}\big|  \geq \ell +1,\] 
then proceed to Step 2.$(\ell+1)$, unless $\ell = q^f_b$, in which  case finalize Step 2 and proceed to Step 3.\smallskip 

\noindent \textbf{Step 3.} Let Step 2.$n$ be the final sub-step of Step 2  leading to Step 3. $\{i^1,\ldots,i^n\}\subset I^1$ is the set of cadets in $I^1$ who each
lose their tentative assignment $(b,t^0)$. For each cadet $i \in I^1\setminus \{i^1,\ldots,i^n\}$, 
finalize the assignment of cadet $i$ as $\phi_i^{MP}(\succ) = (b,t^0)$.

For each cadet $i \in J^{n}$ with one of the $n$ highest $\pi_b$-priorities in $J^{n}$, 
finalize the assignment of cadet $i$ as $\phi_i^{MP}(\succ) = (b,t^h)$.
Finalize the assignment of any remaining cadet as $\emptyset$.\medskip
\end{quote}

The key step in the procedure is Step 2 where it is determined how many of the $q^f_b$ flexible-price  positions are  to be awarded at the increased price $t^h$. 
To determine this number, the price responsiveness policy $\omega_b$ is used to check 
\begin{itemize}
\item[(1)] whether there is at least one cadet with a lower baseline priority $\pi_b$ than cadet $i^1$, who is willing to pay the increased price $t^h$ and  
whose increased price contract has higher priority under the price effectiveness  policy $\omega_b$ than the base price contract of cadet $i^1$; 
\item[(2)] whether there are at least two cadets each with a lower baseline priority $\pi_b$ than cadet $i^2$, who are each willing to pay the increased price $t^h$ and  
whose increased price contracts have higher priority under the price effectiveness   policy $\omega_b$ than the base price contract of cadet $i^2$; \\
\mbox{} $\vdots$\\
\item[($q^f_b$)] whether there are at least $q^f_b$ cadets each with a lower baseline priority $\pi_b$ than cadet $i^{q^f_b}$, who are each willing to pay the increased price $t^h$ and  whose increased price contracts have higher priority under the price effectiveness   policy $\omega_b$ than the base price contract of cadet $i^{q^f_b}$.
\end{itemize}
Once the number of positions awarded  through increased price $t^h$ contracts is determined in this way, all other positions are 
assigned to the highest baseline priority cadets  as base price contracts.  The increased price contracts are awarded 
to the remaining highest baseline priority cadets who are willing to pay the increased price $t^h$.\smallskip 

\begin{example} \textbf{(Mechanics of Mechanism $\phi^{MP}$)} \label{ex-mechanics}
There is a single branch $b$ with $q^0_b = 3$ and $q^f_b = 3$. There are eight cadets, with their set given as $I=\{i^1, i^2, i^3, i^4, i^5, i^6,  j^1, j^2\}$.  
The baseline priority order $\pi_b$ is given as
\[  i^6 \; \pi_b \;  i^5 \; \pi_b \;  i^4 \; \pi_b \;  i^3 \; \pi_b \;  i^2 \; \pi_b \;  i^1 \; \pi_b \;  j^1 \; \pi_b \;  j^2,
\]
and the price responsiveness policy is the ultimate price responsiveness policy $\overline{\omega}_b$. 
Cadet preferences are given as
\begin{eqnarray*}
(b,t^0) \; \succ_i \; (b,t^h) \; \succ_i \; \emptyset \qquad  && \mbox{for any } i \in \{i^1,i^3,i^5,j^1\}, \; \mbox{ and} \\
(b,t^0) \; \succ_i \; \emptyset  \; \succ_i \;  (b,t^h) \qquad && \mbox{for any }i\in \{i^2,i^4,i^6,j^2\}.
\end{eqnarray*}
We next run the procedure for the mechanism $\phi^{MP}$. \smallskip

\noindent \textbf{\textit{Step 0\/}}: There are three base-price positions. The three highest $\pi_b$-priority cadets in the set $I$ are $i^6$, $i^5$, and $i^4$.
Let $I^0=\{i^4,i^5,i^6\}$, and finalize
the assignments of cadets in $I^0$ as $\phi^{MP}_{i^6}(\succ) = \phi^{MP}_{i^5}(\succ) = \phi^{MP}_{i^4}(\succ) = (b,t^0)$. \smallskip

\noindent \textbf{\textit{Step 1\/}}: There are three flexible-price  positions. Three highest $\pi_b$-priority cadets in the set $I\setminus I^0$ are $i^3$, $i^2$, and $i^1$.
Let $I^1 = \{i^1,i^2,i^3\}$, and the tentative assignment of each cadet in $I^1$ is $(b,t^0)$.  There is no need to relabel the cadets since cadet $i^1$ is already the 
lowest $\pi_b$-priority cadet in $I^1$,  cadet $i^2$ is the 
second lowest $\pi_b$-priority cadet in $I^1$, and cadet $i^3$ is the highest $\pi_b$-priority cadet in $I^1$.

\noindent \textbf{\textit{Step 2.0\/}}: The set of cadets in $I\setminus(I^0 \cup I^1) = \{j^1,j^2\}$ for whom the assignment $(b,t^h)$ is acceptable is
$J^0 = \{j^1\}$. Since
\[ \underbrace{\big|\big\{j\in J^0 : (j,t^h) \; \overline{\omega}_b \; (i^1,t^0) \big\}\big|}_{=|J^0|=|\{j^1\}|=1}  \geq 1,
\]
we proceed to Step 2.1. \smallskip

\noindent \textbf{\textit{Step 2.1\/}}: Since $(b,t^h) \, \succ_{i^1} \, \emptyset$, we have  $J^1 = J^0 \cup \{i^1\} = \{i^1,j^1\}$.  Since
\[ \underbrace{\big|\big\{j\in J^1 : (j,t^h) \; \overline{\omega}_b \; (i^2,t^0) \big\}\big|}_{=|J^1|=|\{i^1,j^1\}|=2}  \geq 2,
\]
we proceed to Step 2.2. \smallskip

\noindent \textbf{\textit{Step 2.2\/}}: Since $\emptyset \, \succ_{i^2} (b,t^h)$, we have  $J^2 = J^1 = \{i^1,j^1\}$.  Since
\[ \underbrace{\big|\big\{j\in J^2 : (j,t^h) \; \overline{\omega}_b \; (i^3,t^0) \big\}\big|}_{=|J^2|=|\{i^1,j^1\}|=2}  = 2,
\]
we finalize Step 2 and proceed to Step 2.3. \smallskip

\noindent \textbf{\textit{Step 3\/}}: Step $2.2$ is the last sub-step of Step 2. Therefore two lowest $\pi_b$-priority cadets
in $I^1$, i.e cadets $i^1$ and $i^2$, lose their tentative assignments of $(b,t^0)$. In contrast, the only remaining cadet in the set 
$I^1 \setminus \{i^1,i^2\}$,  i.e cadet $i^3$ maintains her tentative assignment, which is finalized as  $\phi^{MP}_{i^3}(\succ) = (b,t^0)$.

The two highest priority cadets in $J^2$ are $i^1$ and $j^1$. Their assignments are finalized as  $\phi^{MP}_{i^1}(\succ) = \phi^{MP}_{j^1}(\succ) = (b,t^h)$.
Assignments of the remaining cadets $i^2$ and $j^2$ are finalized as $\emptyset$. The final allocation is:
\[ \phi^{MP}(\succ) = \left( \begin{array}{cccccccc}
i^1 & i^2 & i^3 & i^4 & i^5 & i^6& j^1 & j^2 \\
(b,t^h) & \emptyset & (b,t^0) & (b,t^0) & (b,t^0) & (b,t^0) & (b,t^h) & \emptyset
\end{array} \right).
\]
\mbox{} \hfill $\blacksquare$
\end{example}

Our next result is  the following characterization of the the single-branch direct mechanism $\phi^{MP}$.

\begin{proposition} \label{thm:singlebranchcharacterization}
Suppose there is a single branch $b$. Fix a baseline priority order $\pi_b \in \Pi$ and a price responsiveness policy $\omega_b \in \Omega_b$. 
A direct mechanism $\varphi$ satisfies
\begin{enumerate}
\item individual rationality, 
\item non-wastefulness, 
\item no priority reversal, 
\item enforcement of the price responsiveness policy, and
\item BRADSO-IC
\end{enumerate}
if and only if $\varphi = \phi^{MP}$. 
\end{proposition}

Since (i) a quasi-direct mechanism becomes a direct mechanism when there is a single branch,  and 
(ii) strategy-proofness implies BRADSO-IC in this environment, 
Theorem \ref{cosm} and Proposition \ref{thm:singlebranchcharacterization} immediately imply the following result. 

\begin{corollary} \label{corollary:phi=cosm}
Suppose there is a single branch $b$. Fix a baseline priority order $\pi_b \in \Pi$ and a price responsiveness policy $\omega_b \in \Omega_b$. 
Then, for any preference profile $\succ \, \in \calq^{|I|}$, 
\[  \phi^{MP}(\succ) = \phi^{MPCO}(\succ).  
\]
\end{corollary}
The mechanism $ \phi^{MP}$ is merely an alternative formulation of the MPCO mechanism that does not
rely on the cumulative offer procedure when there is a single branch. This formulation is helpful for the single-branch equilibrium 
analysis of the USMA-2020 mechanism we present next. 

\subsection{Equilibrium Outcomes under the USMA-2020 Mechanism} \label{subsec:NE}

While the USMA-2020 mechanism is not a direct mechanism in general, when there is a single branch it can be
interpreted a direct mechanism. In this case, for any cadet $i \in I$ the first part of the message space $\cals_i = \calp \times 2^B$
becomes redundant, and the second part simply solicits whether branch $b$ is acceptable by cadet $i$ or not (analogous to a direct mechanism). 

Our next result shows that
when there is a single branch the truthful outcome of the direct mechanism $\phi^{MP}$ is 
the same as the unique Nash equilibrium outcome of the mechanism $\varphi^{2020}$. 

\begin{proposition} \label{prop:2021NashEqm}
Suppose there is a single branch $b$.  
Fix a baseline priority order $\pi_b \in \Pi$, a price responsiveness policy $\omega_b \in \Omega_b$, and a preference profile 
$\succ \, \in \calq^{|I|}$. 
Then the strategic-form game induced by the mechanism $(\cals^{2020},\varphi^{2020})$ has a unique Nash equilibrium outcome that is
equal to the allocation $\phi^{MP}(\succ)$.\footnote{Using the terminology of the 
\textit{implementation theory\/}, this result can be alternatively stated as
follows: When there is a single branch, the mechanism $(\cals^{2020},\varphi^{2020})$ implements the allocation rule   $\phi^{MP}$ in 
Nash equilibrium.} 
\end{proposition}

Caution is needed when interpreting Proposition \ref{prop:2021NashEqm}; 
if interpreted literally, this result can be misleading. 
What is  more consequential for Proposition \ref{prop:2021NashEqm}  is not the result itself, but rather
its proof which constructs  the equilibrium strategies of cadets.  The proof provides insight
into why the failure of BRADSO-IC, the presence of strategic BRADSO, and the presence of detectable priority reversals are
all common phenomena under the real-life implementation of the USMA-2020 mechanism
(despite the outcome equivalence suggested by Proposition \ref{prop:2021NashEqm}).

Given the byzantine structure of the Nash equilibrium strategies even with a single branch, it is perhaps not surprising that
reaching such a well-behaved Nash equilibrium is highly unlikely to be observed under the USMA-2020 mechanism. 
The following example illustrates  the knife-edge structure of the Nash equilibrium  strategies under the USMA-2020 mechanism. 

\begin{example} \textbf{(Knife-Edge Nash Equilibrium Strategies)} \label{knifeedge}

To illustrate how challenging it is for the cadets to figure out their best responses under the USMA-2020 mechanism, 
we present two scenarios.  The scenarios differ from each other minimally, but cadet best responses differ dramatically. 
Our first scenario is same as the one we presented in Example \ref{ex-mechanics}. \smallskip

\noindent \textbf{\textit{Scenario 1\/}}: There is a single branch $b$ with $q^0_b = 3$ and $q^f_b = 3$. 
There are eight cadets, $I=\{i^1, i^2, i^3, i^4, i^5, i^6,  j^1, j^2\}$.  
The baseline priority order $\pi_b$ is given as
\[  i^6 \; \pi_b \;  i^5 \; \pi_b \;  i^4 \; \pi_b \;  i^3 \; \pi_b \;  i^2 \; \pi_b \;  i^1 \; \pi_b \;  j^1 \; \pi_b \;  j^2 \quad \mbox{and}
\]
and the price responsiveness policy is the ultimate price responsiveness policy $\overline{\omega}_b$. 
Cadet preferences are 
\begin{eqnarray*}
(b,t^0) \; \succ_i \; (b,t^h) \; \succ_i \; \emptyset \qquad  && \mbox{for any } i \in \{i^1,i^3,i^5,j^1\}, \; \mbox{ and} \\
(b,t^0) \; \succ_i \; \emptyset  \; \succ_i \;  (b,t^h) \qquad && \mbox{for any }i\in \{i^2,i^4,i^6,j^2\}.
\end{eqnarray*}
Let $s^*$ be a Nash equilibrium strategy for Scenario 1 under the USMA-2020 mechanism. 
Recall that when there is a single branch $b$, the message space for each cadet $i\in I$ is simply $\cals_i =\{b,\emptyset\}$. 
We construct the Nash equilibrium strategies in several phases.\smallskip 

\textit{Phase 1\/}: Consider cadets $i^1$ and $j^1$, each of whom prefers the increased price assignment $(b,t^h)$ to remaining unmatched. 
Since there are six positions altogether and there are five higher $\pi_b$-priority cadets than either of these two cadets, 
at most one of them can receive a position (at any cost) unless each of them submit a strategy of $b$. 
And if one of them submits a strategy of $\emptyset$, the other one has a best response strategy of $b$ assuring a position
at the increased price rather than remaining unmatched. Hence, $s^*_{i^1}=s^*_{j^1}=b$ at any Nash equilibrium.    

\textit{Phase 2\/}: Consider cadet $j^2$ who prefers remaining unmatched to the increased price assignment $(b,t^h)$. 
Since she is the lowest $\pi_b$-priority cadet, she cannot receive an assignment of $(b,t^0)$ regardless of her strategy. 
In contrast, she can guarantee remaining unmatched with a strategy of $s_{j^2} = \emptyset$. While this does not at this
point rule out a strategy of  $s_{j^2} = b$ at Nash equilibrium (just yet), it means $\varphi^{2020}_{j^2}(s^*)=\emptyset$. 

\textit{Phase 3\/}: Consider cadet $i^2$ who prefers remaining unmatched to the increased price assignment $(b,t^h)$. 
She is the fifth highest $\pi_b$-priority cadet, so she secures a position if she submits a strategy of $s_{i^2} = b$, 
but the position will have to be at the increased price $t^h$, since the lowest $\pi_b$-priority cadet $j^2$ is remaining unmatched  from Phase 2,
and therefore  there cannot be three cadets with lower $\pi_b$-priority who receive an assignment of $(b,t^h)$.
But since cadet $j^2$  prefers remaining unmatched to the increased price assignment $(b,t^h)$, she cannot receive an assignment
of $(b,t^h)$ at Nash equilibria. 
Hence, cadet $i^2$'s Nash equilibrium strategy is $s^*_{i^2}=\emptyset$, and her Nash equilibrium assignment is $\varphi^{2020}_{i^2}(s^*)=\emptyset$. 

\textit{Phase 4\/}: Consider the remaining cadets $i^3$, $i^4$, $i^5$ and $i^6$. 
Since cadets $i^2$ and $j^2$ have to remain unmatched (from Phases 2 and 3) at Nash equilibria,
they each receive a position at Nash equilibrium. Since only the two cadets $i^1$ and $j^1$ from Phases 1-3 have Nash equilibrium
strategies of $b$, the lowest $\pi_b$-priority cadet of the four cadets $i^3$, $i^4$, $i^5$, $i^6$ who submit a strategy of $b$
receives an assignment of $(b,t^h)$. But this cannot happen at Nash equilibria since that particular cadet can instead submit a strategy of $\emptyset$
receiving a more preferred assignment of $(b,t^0)$. Hence, $s^*_i = \emptyset$ and $\varphi^{2020}_{i}(s^*)= (b,t^0)$ for any $i\in \{i^3,i^4,i^5,i^6\}$. 

The unique Nash equilibrium strategy $s^*$ and its Nash equilibrium outcome $\varphi^{2020}(s^*)$ for Scenario 1 are given as:
\[  \begin{array}{lcccccccc}
\mbox{Cadet} & i^1 & i^2 & i^3 & i^4 & i^5 & i^6& j^1 & j^2 \\
\hline\mbox{Nash equilibrium strategy} & b & \emptyset & \emptyset &\emptyset &\emptyset & \emptyset & b &\emptyset\\
\mbox{Nash equilibrium assignment} & (b,t^h) & \emptyset & (b,t^0) & (b,t^0) & (b,t^0) & (b,t^0) & (b,t^h) & \emptyset
\end{array}  \smallskip 
\]
Scenario 1 involves  \textit{BRADSO-IC\/} failures for cadets $i^3$ and $i^5$ whose Nash equilibrium strategies
force them into hiding their willingness  to pay the increased price $t^h$.  Any deviation from her Nash equilibrium strategy
by truthfully declaring her willingness to pay the increased price $t^h$ will result in an detectable priority reversal for cadet $i^5$.

\smallskip

\textbf{\textit{Scenario 2\/}}: This scenario differs from Scenario 1 in only the preferences of the lowest $\pi_b$-priority cadet $j^2$
and nothing else. Thus,  cadet preferences for this scenario are given as:
\begin{eqnarray*}
(b,t^0) \; \succ'_i \; (b,t^h) \; \succ'_i \; \emptyset \qquad  && \mbox{for any } i \in \{i^1,i^3,i^5,j^1,j^2\}, \; \mbox{ and} \\
(b,t^0) \; \succ'_i \; \emptyset  \; \succ'_i \;  (b,t^h) \qquad && \mbox{for any }i\in \{i^2,i^4,i^6\}.
\end{eqnarray*}
Let $s'$ be a Nash equilibrium strategy for Scenario 2 under the USMA-2020 mechanism.  \smallskip 

\textit{Phase 1\/}: Identical to Phase 1 for Scenario 1, and thus $s'_{i^1}=s'_{j^1}=b$ at any Nash equilibrium.   

\textit{Phase 2\/}: Consider cadet $i^2$ who prefers remaining unmatched to the increased price assignment $(b,t^h)$, 
and cadets $i^3$ and $j^2$, each of whom prefers the increased price assignment $(b,t^h)$ to remaining unmatched. 
Since (i) there are six positions altogether,  (ii) three cadets with higher $\pi_b$-priority  than  each one of $i^2,i^3,$ and $j^2$,
and (iii) $s'_{i^1}=s'_{j^1}=b$ from Phase 1, at most one of  the cadets $i^2,i^3,j^2$ can receive an assignment of $(b,t^0)$ if any. 
Therefore, submitting a strategy of $s_{i^3} = \emptyset$ is a best response for cadet $i^3$ only if both cadets $i^2$ and $j^2$
also submit a strategy of $\emptyset$ each. But this cannot happen in Nash equilibria, since it gives cadet $j^2$ a profitable
deviation by submitting a strategy of $s_{j^2} = b$ and jumping ahead of cadets $i^2$ and $i^3$ securing her a position. 
Hence $s'_{i^3} = b$ and $\varphi^{2020}_{i^3}(s')= (b,t^h)$.  When cadet $i^3$ joins 
the  two cadets from Phase 1 each also submitting a strategy of $b$,  this assures that exactly three positions
will be assigned at the increased price $t^h$.  Therefore a strategy of f $s_{i^2} = b$ assures 
assures cadet $i^2$ an assignment of $(b,t^h)$, which cannot happen at Nash equilibrium. 
Therefore, $s'_{i^2} = \emptyset$ and $\varphi^{2020}_{i^2}(s')= \emptyset$. 
This not only assures that $\varphi^{2020}_{i^3}(s')= \varphi^{2020}_{i^1}(s') =  \varphi^{2020}_{j^1}(s')=(b,t^h)$,  but
it also means that $s'_{j^2} = b$ at Nash equilibrium, for otherwise with two lower $\pi_b$-priority cadets with strategies
of $\emptyset$, cadet $i^3$ would have an incentive to deviate himself and receiving the position at the base price rather than the increased price. 
  
\textit{Phase 3\/}:  Consider the remaining cadets $i^4$, $i^5$ and $i^6$. 
Of all lower $\pi_b$-priority cadets, only the cadet $i^2$ and has  Nash equilibrium
strategies of $\emptyset$ from Phases 1 and 2.  Therefore
the lowest $\pi_b$-priority cadet of the three cadets  $i^4$, $i^5$, $i^6$ who submit a strategy of $\emptyset$
receives an assignment of $\emptyset$. But this cannot happen at Nash equilibria since that particular cadet can instead submit a strategy of $b$
and receive a more preferred assignment of $(b,t^0)$ since three lower $\pi_b$-priority cadets already receive an assignment of $(b,t^h)$ each from
Phase 2. Therefore, regardless of their preferences $s'_{i^4} = s'_{i^5} = s'_{i^6} = b$, and 
$\varphi^{2020}_{i^4}(s')= \varphi^{2020}_{i^5}(s') = \varphi^{2020}_{i^6}(s') (b,t^0)$. 

The unique Nash equilibrium strategy $s'$ and its Nash equilibrium outcome $\varphi^{2020}(s')$ for Scenario 2 are given as:
\[  \begin{array}{lcccccccc}
\mbox{Cadet} & i^1 & i^2 & i^3 & i^4 & i^5 & i^6& j^1 & j^2 \\ \hline
\mbox{Nash equilibrium strategy} & b & \emptyset & b & b & b & b & b & b\\
\mbox{Nash equilibrium assignment} & (b,t^h) & \emptyset & (b,t^h) & (b,t^0) & (b,t^0) & (b,t^0) & (b,t^h) & \emptyset
\end{array}  \smallskip 
\]
Not only does the Nash equilibrium strategies of cadets $i^4$ and $i^6$ involve
 strategic BRADSO in Scenario 2 and they have to  declare willingness  to pay the increased price $t^h$ even though
 under their true preferences they do not, but any deviation from this Nash equilibrium strategy
by declaring their unwillingness to pay the increased price $t^h$  will result in detectable priority reversals for both cadets. 

Another key insight from this example is the dramatic difference between the Nash equilibrium strategies due to one minor
change in the underlying economy, a preference change in the lowest base priority cadet. 
This minor change only affects the assignment of cadet $i^3$ by changing it from $(b,t^0)$ to $(b,t^h)$. 
It also changes the Nash equilibrium strategy of not only  cadet $i^3$, and also all other higher $\pi_b$-priority cadets $i^4, i^5,$ and $i^6$. 
Moreover, in addition to BRADSO-IC failures and the presence of strategic BRADSO
under Nash equilibria,  any deviation from these strategies result in detectable priority reversals.
The fragility of our equilibrium strategies provides us intuition on the prevalence of these phenomena under the USMA-2020 mechanism. 
\mbox{} \hfill $\blacksquare$
\end{example}

Example \ref{knifeedge} shows that while the failure of BRADSO-IC and the presence of strategic BRADSO
can be observed at Nash equilibria  of the USMA-2020 mechanism,  
the presence of detectable priority reversals is out-of-equilibrium behavior under complete information
when there is a single branch.  Our next example shows that if the complete information assumption is relaxed  
there can also be detectable priority reversals in the
Bayesian equilibria of the USMA-2020 mechanism.

\medskip

\begin{example} \textbf{(Detectable Priority Reversals at Bayesian Equilibria)}  \label{Bayesian}

Suppose there is a single branch $b$ with $q^0_b=q^f_b=1$ and three cadets $i_1, i_2,$ and $i_3$. 
The baseline priority order $\pi_b$ is such that
\[  i_1 \; \pi_b \; i_2 \; \pi_b \; i_3,  
\]
and the price responsiveness policy $\omega_b$ is the ultimate price responsiveness policy $\overline{\omega}_b$.

Each cadet has a utility function that is drawn from a distribution with the following two elements, $u$ and $v$, where:
\[ u(b,t^0) = 10, \; u(\emptyset) = 8, \; u(b,t^h)=0, \quad \mbox{ and } \quad v(b,t^0) = 10, \; v(b,t^h) = 8, \; v(\emptyset)=0.
\]
Let us refer to cadets with a utility function $u(.)$ as type 1 and cadets with a utility function $v(.)$ as type 2. 
All cadets have a utility of 10 for their first choice assignment of $(b,t^0)$,  
a utility of 8 for their second choice assignment, and  a utility of 0 for their last choice assignment. 
For type 1 cadets, the second choice is remaining unmatched whereas for type 2 cadets the second choice
is receiving a position at the increased price $t^h$. 
Suppose each cadet can be of the either type with a probability of 50 percent, and they are all expected utility maximizers.  

The unique Bayesian Nash equilibrium $s^*$ under the incomplete information game induced by the
USMA-2020 mechanism is, for any cadet $i \in \{i_1,i_2,i_3\}$,
\[  s^*_i = \left\{ \begin{array}{cl}
\emptyset & \mbox{ if cadet } i \mbox{ is of type 1,\;  and}\\
b & \mbox{ if cadet } i \mbox{ is of type 2.}   \end{array}
\right.
\]
That is, truth-telling is the unique  Bayesian Nash equilibrium strategy for each cadet. 
However, this unique Bayesian Nash equilibrium strategy results in  detectable priority reversals whenever either
\begin{enumerate}
\item  cadet $i_1$ is of type 1 and cadets $i_2, i_3$ are of type 2, or
\item  cadet $i_1$ is of type 2 and cadets $i_2, i_3$ and are of type 1. 
\end{enumerate}
While cadet $i_2$ receives a position at the base price $t^0$ in both cases, 
the  highest baseline priority cadet $i_1$ remains unassigned in the first case
and receives a position at the increased price $t^h$ in the second case. 
\mbox{} \hfill $\blacksquare$
\end{example}

\subsection{Proofs for Results in Online Appendix Section \ref{sec:singlebranch}}

\noindent \textbf{Proof of Proposition \ref{thm:singlebranchcharacterization}}: Suppose there is only one branch $b\in B$, 
and fix  a profile of cadet preferences $\succ \, \in \calq^{|I|}$.
We first show that the direct mechanism $\phi^{MP}$ satisfies the five axioms.

\textbf{\textit{Individual rationality\/}}: This axiom holds immediately under $\phi^{MP}$, since no cadet $i\in I$ is considered for a position at the increased price $t^h$ unless her submitted preferences is such that $(b,t^h)  \succ_i \emptyset$.

\textbf{\textit{Non-wastefulness\/}}: Since there is only one branch and we already established \textit{individual rationality\/}, we
can focus on cadets who consider a position at the base price to be acceptable. With this observation, 
\textit{non-wastefulness\/} also holds immediately under $\phi^{MP}$, since all positions are allocated at Steps 0 and 1 at the base price $t^0$ either 
as a final assignment or a tentative one. Tentative assignments from Step 1 may be altered later on by increasing their price  to $t^h$ and
possibly changing their recipients, but not by leaving the position unassigned, hence assuring non-wastefulness.

\textbf{\textit{No priority reversal}}:  Under the mechanism $\phi^{MP}$, each of the $q^0_b$ highest $\pi_b$-priority cadets are assigned a position
at the base price $t^0$ at Step 0, and each of the next $q^f_b$ highest $\pi_b$-priority cadets are tentatively assigned a position at the base price $t^0$ at Step 1. 
Tentative positions are lost in Step 2 only if there is excess demand from qualified cadets who are willing to pay the increased price $t^h$, and starting
with the lowest $\pi_b$ priority cadets with tentative assignments. That assures that, for any $i,j \in I$,
\begin{equation} \label{eqn-euj1}
\phi^{MP}_j(\succ) = (b,t^0)  \succ_i \phi^{MP}_i(\succ) \quad \implies \quad j \; \pi_b \; i.
\end{equation} 
Moreover positions at the increased price $t^h$ are offered to cadets with highest $\pi_b$ priorities among those 
(i) who fail to receive a position at the base price $t^0$ and (ii) who declare the expensive assignment $(b,t^h)$ as acceptable. 
Therefore, for any $i,j \in I$,
\begin{equation} \label{eqn-euj2}
\phi^{MP}_j(\succ) = (b,t^h)  \succ_i \phi^{MP}_i(\succ) = \emptyset \quad \implies \quad j \; \pi_b \; i.
\end{equation} 
Relations \eqref{eqn-euj1} and \eqref{eqn-euj2} imply that mechanism $\phi^{MP}$  satisfies  \textit{no priority reversal\/}.

\textbf{\textit{BRADSO-IC\/}}:  Fix a cadet $i \in I$. 
For a given profile of preferences for all cadets except cadet $i$,  
whether cadet $i\in I$ receives an assignment of $(b,t^0)$ under the mechanism $\phi^{MP}$ is
independent of cadet $i$'s preferences under the mechanism $\phi^{MP}$: 
Cadets who are among the $q^0_b$ highest $\pi_b$-priority cadets in $I$ always receive an assignment at the base price $t^0$; 
cadets who are not among the $q$ highest $\pi_b$-priority cadets in $I$ never receive an assignment at the base price $t^0$; 
and for any cadet $i$ who has one of the highest $q$ but not one of the highest $q^0_b$ priorities, whether she receives  an assignment at the base price $t^0$
depends on how many lower $\pi_b$-priority cadets are both willing to pay the increased price $t^h$ and also able to ``jump ahead of''  cadet $i$ through the
price responsiveness policy. Hence if a cadet receives a position under $\phi^{MP}$ at the increased price $t^h$, changing her reported preferences can only 
result in losing the position altogether. Therefore mechanism $\phi^{MP}$ satisfies \textit{BRADSO-IC\/}. 
 
\textbf{\textit{Enforcement of the price responsiveness policy\/}}: The procedure for the mechanism $\phi^{MP}$ initially assigns all positions to the $q_b$ highest 
$\pi_b$-priority cadets at the base price $t^0$, although the assignments of the $q^f_b$-lowest $\pi_b$-priority  cadets among these awardees 
are only tentative.  Step 2 of the procedure for mechanism $\phi^{MP}$ ensures that, if any cadet $j\in I$ loses her tentative assignment $(b,t^0)$ from Step 1, 
then any cadet $i\in I$ who receives an assignment of $(b,t^h)$ is such that $(i,t^h) \; \omega_b \; (j,t^0)$. Therefore, 
\begin{equation} \label{eqn-BRADSO1}
\left. \begin{array}{l}
  \phi^{MP}_i(\succ) = (b,t^h), \; \mbox{ and}\\ 
    (b,t^0) \succ_j \phi^{MP}_j(\succ)  \end{array}  \right\}     \quad  \implies \quad (i,t^h) \; \omega_b \; (j,t^0). 
\end{equation}
Moreover, Step 2 of the same procedure also ensures that, 
for any $\ell \in \{1,\dots, q^f_b\}$, the $\ell^{\footnotesize th}$ lowest $\pi_b$-priority cadet $i^{\ell}$ with a tentative assignment of $(b,t^0)$
cannot maintain this tentative assignment, for as long as  there are  at least $\ell$ lower $\pi_b$-priority cadets  who 
are both willing to pay the increased price $t^h$ and also able to ``jump ahead of'' the cadet $i^{\ell}$ through the price responsiveness policy. 
Therefore,      
\begin{eqnarray} \label{eqn-BRADSO2}
\left. \begin{array}{l}
        \phi^{DP}_i(\succ) = (b,t^0),\\ 
      (b,t^+) \succ_j \phi^{DP}_j(\succ),  \; \mbox{ and}\\
       (j,t^+) \; \omega_b^+ \; (i,t^0) \end{array}  \right\}     \quad  & \implies & \quad 
       \big|\big\{i'\in I : \phi^{DP}_{i'}(\succ)=(b,t^+)\big\}\big|  = q^f_b.
\end{eqnarray}

       Relations \eqref{eqn-BRADSO1} and \eqref{eqn-BRADSO2} imply that mechanism $\phi^{MP}$ satisfies \textit{enforcement of the price responsiveness policy\/}. \\

\textbf{\textit{Uniqueness\/}}: We next show that mechanism $\phi^{MP}$ is the only mechanism that satisfies all five axioms. 

Let the direct mechanism $\varphi$ satisfy \textit{individual rationality,  non-wastefulness,
no priority reversal, enforcement of the price responsiveness policy\/} and \textit{BRADSO-IC}. 
We want to show that 
\[\varphi(\succ) = \phi^{MP}(\succ).
\] 

If there are less than or equal to $q$ cadets for whom the assignment $(b,t^0)$ is acceptable under the preference profile $\succ$, 
all such cadets  must receive an assignment of $(b,t^0)$ by 
\textit{individual rationality, non-wastefulness\/}, and \textit{BRADSO-IC\/}. 
Since this is also the case under the allocation $\phi^{MP}(\succ)$, 
the result holds immediately for this case.
 
Therefore, w.l.o.g assume that there are strictly more than $q$ cadets for whom the assignment $(b,t^0)$ is acceptable under the preference profile $\succ$. 
Let $I^0$ be the set of $q^0_b$ highest $\pi_b$-priority cadets in $I$. 
By \textit{non-wastefulness\/}, all positions are assigned under $\varphi(\succ)$.  Since at most $q^f_b$ positions  can be awarded at the 
increased price $t^h$, at least $q^0_b$ positions has to be allocated at the base price  $t^0$. 
Therefore,  
\begin{equation} \label{eqn1}
\mbox{for any } i\in I^0, \qquad 
\varphi_i(\succ) = (b,t^0) = \phi^{MP}_i(\succ)
\end{equation}
by \textit{no priority reversal\/}. 

Let $I^1$ be the set of $q^f_b$ highest $\pi_b$-priority cadets in $I\setminus I^0$. 
Relabel the cadets in the set $I^1$ so that for any $\ell \in \{1,\ldots, q^f_b\}$, cadet $i^{\ell}$ is the ${\ell}^{\mbox{\footnotesize th}}$-lowest $\pi_b$-priority cadet in $I^1$. 
Let 
\[J^0 = \big\{j \in I\setminus (I^0\cup I^1) : (b,t^h) \succ_j \emptyset\big\}. 
\]
By \textit{individual rationality\/} and the \textit{no priority reversal\/},  
\begin{equation} \label{eqn2}
\mbox{for any } i\in I\setminus(I^0\cup I^1 \cup J^0), \qquad 
\varphi_i(\succ) = \emptyset = \phi^{MP}_i(\succ).
\end{equation}
By relations \eqref{eqn1} and \eqref{eqn2},
the only set of cadets whose assignments are yet to be determined under $\varphi(\succ)$ are cadets in $I^1 \cup J^0$. 
Moreover, by \textit{no priority reversal\/}, cadets in $J^0$ can only receive a position at the increased price $t^h$. 
That is, 
\begin{equation} \label{eqn3}
\mbox{for any } j\in J^0, \qquad 
\varphi_j(\succ) \not=  (b,t^0).
\end{equation}

For the next phase of our proof, we will rely on the sequence of individuals $i^1,\ldots,i^{q^f_b}$ and the
sequence of sets  $J^0, J^1, \ldots$ 
that are constructed for the  Step 2 of the mechanism $\phi^{MP}$. 
Here individual $i^1$ is the $q^{\footnotesize \mbox{th}}$ highest $\pi_b$-priority cadet in set $I$, cadet 
$i^2$ is the $(q-1)^{\footnotesize \mbox{th}}$ highest $\pi_b$-priority cadet in set $I$, and so on. 
The starting element of the second  sequence is  $J^0 = \{j \in I\setminus (I^0\cup I^1) : (b,t^h) \succ_j \emptyset\}$. 
Assuming Step 2.$n$ is the last sub-step of Step 2,
the remaining elements of the latter  sequence  for $n\geq 1$ is given as follows: For any $\ell \in \{1,\ldots, n\}$, 
\[ J^{\ell} = \left\{ \begin{array}{cl}
         J^{\ell -1}  & \mbox{ if } \; \emptyset \,  \succ_{i^{\ell}}  \, (b,t^h)\\
       J^{\ell -1} \cup\{i^{\ell}\} & \mbox{ if } \; (b,t^h) \succ_{i^{\ell}} \, \emptyset \end{array} \right.
\]
We have three cases to consider. \medskip

\noindent \textit{\textbf{Case 1.}\/} $n=0$ \smallskip

For this case, by  the mechanics of the Step 2 of the mechanism $\phi^{MP}$, we have 
\begin{equation} \label{eqn-case1a}
\big|\big\{j \in J^{0} : (j, t^h) \; \omega_b \; (i^1, t^0) \big\}\big| = 0.
\end{equation}
Therefore,  by relations \ref{eqn2}, \ref{eqn3}, and  condition (1) of the axiom \textit{enforcement of the price responsiveness policy\/}, 
\begin{equation}
\mbox{for any }  i \in I\setminus (I^0 \cup I^1), \qquad \varphi_i(\succ) = \emptyset = \phi^{MP}_i(\succ). 
\end{equation}
Hence by \textit{non-wastefulness\/},
\begin{equation} \label{eqn-case1b}
\mbox{for any }  i \in I^1, \qquad \varphi_i(\succ) \in \big\{(b,t^0),(b,t^h)\big\}. 
\end{equation}
But since $\varphi$ satisfies \textit{individual rationality\/}, relation \eqref{eqn-case1b} implies that $\varphi_i(\succ) = (b,t^0)$
for any $i\in I^1$ with $\emptyset \succ_i (b,t^h)$. 
Furthermore for any $i\in I^1$ with $(b,t^h) \succ_i \emptyset$,  instead reporting 
the fake preference relation $\succ'_i \in \calq$ with $\emptyset \succ'_i (b,t^h)$ would guarantee cadet $i$ an assignment of 
$\varphi_i(\succ_{-i},\succ'_i)=(b,t^0)$ due to the same arguments applied for the economy $(\succ_{-i},\succ'_i)$,
and therefore by \textit{BRADSO-IC\/} these cadets too must receive an assignment of $(b,t^0)$ each. Hence  
\begin{equation} \label{eqn-case1c}
\mbox{ for any } i\in I^1, \qquad \varphi_i(\succ) = (b,t^0) = \phi_i^{MP}(\succ). 
\end{equation}
Relations \eqref{eqn1},  and \eqref{eqn-case1c}  imply $\varphi(\succ) = \phi^{MP}(\succ)$,  completing the proof for Case 1.$\blacksquare$ \medskip

\noindent \textit{\textbf{Case 2.}\/} $n \in \{1,\ldots, q^f_b-1\}$ \smallskip

For this case, by the mechanics of the Step 2 of the mechanism $\phi^{MP}$, we have 
\begin{equation} \label{eqn-case2a}
\mbox{for any } \ell \in \{1,\ldots, n\}, \qquad  \big|\big\{j \in J^{\ell -1} : (j, t^h) \; \omega_b \; (i^{\ell}, t^0) \big\}\big| \geq \ell,
\end{equation}
and
\begin{equation} \label{eqn-case2b}
\big|\big\{j \in J^{n} : (j, t^h) \; \omega_b \; (i^{n+1}, t^0) \big\}\big| = n.
\end{equation}
Since mechanism $\varphi$ satisfies  condition (2) of the axiom \textit{enforcement of the price responsiveness policy\/}, 
the \textit{no priority reversal\/} and relation \ref{eqn-case2a} imply
\begin{equation} \label{eqn6}
\mbox{ for any } i\in \{i^1, \dots, i^{n}\}, \qquad \varphi_i(\succ)\not= (b,t^0). 
\end{equation}
Therefore, by \textit{non-wastefulness\/} and relations \eqref{eqn1}, \eqref{eqn2}, \eqref{eqn3}, and \eqref{eqn6}, 
at least  $n$ positions must be assigned at the increased price $t^h$.

Moreover, since mechanism $\varphi$ satisfies \textit{non-wastefulness, no priority reversal\/}, 
and condition (1) of the axiom \textit{enforcement of the price responsiveness policy\/},
relation \eqref{eqn-case2b} implies
\begin{equation} \label{eqn7}
\mbox{ for any } i\in \{i^{n+1}, \dots, i^{q^f_b}\}, \qquad \varphi_i(\succ) \in \big\{(b,t^0),(b,t^h)\big\}. 
\end{equation}
But since $\varphi$ satisfies \textit{individual rationality\/}, relation \eqref{eqn7} implies that $\varphi_i(\succ) = (b,t^0)$
for any $i\in\{i^{n+1}, \dots, i^{q^f_b}\}$ with $\emptyset \succ_i (b,t^h)$. 
Furthermore for any $i\in\{i^{n+1}, \dots, i^{q^f_b}\}$ with $(b,t^h) \succ_i \emptyset$,  instead reporting 
the fake preference relation $\succ'_i \in \calq$ with $\emptyset \succ'_i (b,t^h)$ would guarantee cadet $i$ an assignment of 
$\varphi_i(\succ_{-i},\succ'_i)=(b,t^0)$ due to the same arguments applied for the economy $(\succ_{-i},\succ'_i)$,
and therefore by \textit{BRADSO-IC\/} these cadets must also receive an assignment of $(b,t^0)$ each. Hence  
\begin{equation} \label{eqn8}
\mbox{ for any } i\in \{i^{n+1}, \dots, i^{q^f_b}\}, \qquad \varphi_i(\succ) = (b,t^0) = \phi_i^{MP}(\succ). 
\end{equation}
Since we have already shown that at least  $n$ positions must be assigned at an increased price of $t^h$, 
relation \eqref{eqn8} implies that exactly $n$ positions must be assigned this cost, 
and therefore for any cadet $j \in J^n$ who is one of the $n$ highest $\pi_b$-priority cadets  in $J^n$, 
\begin{equation} \label{eqn9}
\varphi_j(\succ) = (b,t^h) = \phi_i^{MP}(\succ)
\end{equation}
by \textit{no priority reversal\/}. 

Relations \eqref{eqn1}, \eqref{eqn8}, and \eqref{eqn9}  imply $\varphi(\succ) = \phi^{MP}(\succ)$,  completing the proof for Case 2. $\blacksquare$ \medskip

\noindent \textit{\textbf{Case 3.}\/} $n= q^f_b$ \smallskip

For this case, by the mechanics of the Step 2 of the mechanism $\phi^{MP}$, we have 
\begin{equation} \label{eqn-case3}
\mbox{for any } \ell \in \{1,\ldots, q^f_b\}, \qquad  \big|\big\{j \in J^{\ell -1} : (j, t^h) \; \omega_b \; (i^{\ell}, t^0) \big\}\big| \geq \ell.
\end{equation}
Since mechanism $\varphi$ satisfies  condition (2) of the axiom \textit{enforcement of the price responsiveness policy\/}, 
relation \ref{eqn-case3} implies
\begin{equation} \label{eqn11}
\mbox{ for any } i\in \underbrace{\{i^1, \dots, i^{q^f_b}\}}_{=I^1}, \qquad \varphi_i(\succ)\not= (b,t^0). 
\end{equation}
Therefore, by \textit{non-wastefulness\/} and the \textit{no priority reversal}, 
exactly $q^f_b$ positions must be assigned at the increased price $t^h$.
Hence  for any cadet $j \in J^{q^f_b}$ who is one of the $q^f_b$ highest $\pi_b$-priority cadets  in $J^{q^f_b}$, 
\begin{equation} \label{eqn12}
\varphi_j(\succ) = (b,t^h) = \phi_i^{MP}(\succ)
\end{equation}
by \textit{no priority  reversals\/}. 

Relations \eqref{eqn1} and \eqref{eqn12}  imply $\varphi(\succ) = \phi^{MP}(\succ)$,  completing the proof for Case 3, thus finalizing
the proof of the theorem. $\blacksquare$ \qed \medskip
 
\noindent \textbf{Proof of Proposition \ref{prop:2021NashEqm}}: Suppose that there is only one branch $b\in B$.  
Fixing  the profile of cadet preferences $\succ \, \in \calq$, the baseline priority order $\pi_b$, and the price responsiveness policy
$\omega_b$, consider the strategic-form game induced by the USMA-2020 mechanism $(\cals^{2020}, \varphi^{2020})$. 
When there is only one branch,  the first part of the message space becomes redundant and the second part  contains only the two elements
$b$ and $\emptyset$. Hence, for any cadet $i\in I$, the message space of cadet $i \in I$ under the USMA-2020 mechanism is $\cals^{2020}_i = \{\emptyset, b\}$. 

For a given strategy profile $s \in \cals^{2020}$, construct the  priority order $\pi^+_b(s)$ as follows: 
For any $i, j \in I$,
\begin{enumerate}
\item $\; s_i = s_j \qquad \qquad \qquad \quad \implies \qquad  \qquad i \;\, \pi^+_b(s) \; j \; \iff i \; \pi_b \; j$,
\item $\; s_i = b$ and $s_j = \emptyset \qquad \, \implies \qquad i \;\, \pi^+_b(s) \; j \; \iff (i,t^h) \; \omega_b \; (j,t^0)$.\smallskip
\end{enumerate}

Let $I^+(s)$ be the set of $q_b$ highest $\pi^+_b(s)$-priority cadets in $I$. \smallskip
		
For any cadet $i \in I$, the outcome of the USMA-2020 mechanism is given as,
\[ \varphi_i^{2020}(s) =   \left\{ \begin{array}{cl}  
\emptyset & \mbox{if } \;\;  i\not\in I^+(s),\\
 (b, t^0)   & \mbox{if } \;\;   i\in I^+(s) \mbox{ and }   s_i = \emptyset,\\
 (b, t^0)   & \mbox{if } \;\;   i\in I^+(s) \mbox{ and }   s_i = b \; \mbox{ and }  \;  \big|\{j\in I^+(s) : s_j = b \mbox{ and } i \; \pi_b \;j\}\big| \geq q^f_b,\\
 (b, t^h)   & \mbox{if } \;\;   i\in I^+(s) \mbox{ and }  s_i = b \; \mbox{ and } \;  \big|\{j\in I^+(s) : s_j = b \mbox{ and } i \; \pi_b \;j\}\big|  < q^f_b. 
\end{array} \right.
\]	
We first prove a lemma on the structure of Nash equilibrium strategies of the strategic-form game induced by the 
USMA-2020 mechanism $(\cals^{2020},\varphi^{2020})$. 
\begin{lemma} \label{lemma1} 
Let $s^*$  be a Nash equilibrium of the strategic-form game induced by the mechanism $(\cals^{2020},\varphi^{2020})$. Then,
for any $i,j \in I$, 
\[ \varphi^{2020}_j(s^*) \succ_i \varphi^{2020}_i(s^*) \quad \implies \quad j \; \pi_b \; i. 
\]
\end{lemma}

\noindent \textit{Proof of Lemma \ref{lemma1}\/}:   Let $s^*$  be a Nash equilibrium of the strategic-form game 
induced by the USMA-2020 mechanism $(\cals^{2020},\varphi^{2020})$. Contrary to the claim suppose that, there exists $i,j \in I$ such that  
\[ \varphi^{2020}_j(s^*) \succ_i \varphi^{2020}_i(s^*) \quad \mbox{ and } \quad  i \; \pi_b \; j. 
\]
There are three possible cases, where in each case we reach a contradiction by showing that cadet $i$ has a profitable deviation 
by mimicking the strategy of cadet $j$: \smallskip

\noindent \textbf{\textit{Case 1\/}:} $\varphi^{2020}_j(s^*) = (b,t^0)$ and $\varphi^{2020}_i(s^*) = (b,t^h)$. \smallskip

Since by assumption $\varphi^{2020}_i(s^*) = (b,t^h)$, 
\[ s^*_i = b.
\]
Moreover the assumptions $\varphi^{2020}_j(s^*) = (b,t^0)$,\; $\varphi^{2020}_i(s^*) \not= (b,t^0)$,\; and  $i \; \pi_b \; j$ imply
\begin{equation} \label{case1-eqn2}
j \in I^+(s^*) \quad \mbox{ and } \quad   s^*_j = \emptyset.
\end{equation}
But then, relation \eqref{case1-eqn2} and the assumption $i \; \pi_b \; j$ imply that, 
 for the alternative strategy $\hat{s}_i = \emptyset$ for cadet $i$, 
\[ i \in I^+(s_{-i}^*, \hat{s}_i),
\]
and thus 
\[ \varphi_i^{2020}(s_{-i}^*, \hat{s}_i) = (b,t^0) \succ_i \varphi^{2020}_i(s^*),
\] 
contradicting  $s^*$ is a Nash equilibrium strategy.  This completes the proof  for Case 1. $\blacksquare$  \smallskip

\noindent \textbf{\textit{Case 2\/}:} $\varphi^{2020}_j(s^*) = (b,t^0)$ and $\varphi^{2020}_i(s^*) = \emptyset$. \smallskip

Since by assumption $\varphi^{2020}_j(s^*) = (b,t^0)$,\; $\varphi^{2020}_i(s^*) = \emptyset$,\; and  $i \; \pi_b \; j$, we must have
\begin{equation} \label{lemma1case2-eqn1}
j \in I^+(s^*) \quad \mbox{ and } \quad   s^*_j = b \quad \mbox{ and } \quad   \big|\{k\in I^+(s^*) : s^*_k = b \mbox{ and } j \; \pi_b \;k\}\big| \geq q^f_b, 
\end{equation}
and 
\[ s^*_i = \emptyset. 
\]
But then, relation \eqref{lemma1case2-eqn1} and the assumption  $i \; \pi_b \; j$ imply that, 
for the alternative strategy $\hat{s}_i = b$ for cadet $i$, 
\[
i \in I^+(s_{-i}^*, \hat{s}_i) \quad \mbox{ and } \quad   \hat{s}_i = b \quad \mbox{ and } \quad   \big|\{k\in I^+(s_{-i}^*, \hat{s}_i) : s^*_k = b \mbox{ and } i \; \pi_b \;k\}\big| \geq q^f_b, 
\]
and thus 
\[ \varphi_i^{2020}(s_{-i}^*, \hat{s}_i) = (b,t^0) \succ_i \varphi^{2020}_i(s^*),
\] 
contradicting  $s^*$ is a Nash equilibrium strategy.  This completes the proof for Case 2.  $\blacksquare$  \smallskip

\noindent \textbf{\textit{Case 3\/}:} $\varphi^{2020}_j(s^*) = (b,t^h)$ and  $\varphi^{2020}_i(s^*) = \emptyset$. \smallskip

Since by assumption $\varphi^{2020}_j(s^*) = (b,t^h)$, 
\begin{equation} \label{case3-eqn1}
j \in I^+(s^*) \quad \mbox{ and } \quad s^*_j = b.
\end{equation}
Moreover, since $  \varphi^{2020}_i(s^*) = \emptyset$ by assumption, 
\[  i\not\in  I^+(s^*).
\]
Therefore, since  $i \; \pi_b \; j$ by assumption, 
\[ j \in I^+(s^*) \;  \mbox{ and } \; i\not\in  I^+(s^*) \qquad \implies \qquad s^*_i = \emptyset.  
\]
But then, again thanks to assumption  $i \; \pi_b \; j$, the relation \eqref{case3-eqn1} implies that, 
for the alternative strategy $\hat{s}_i = b$ for cadet $i$, 
\[
i \in I^+(s_{-i}^*, \hat{s}_i), 
\]
and thus 
\[ \underbrace{\varphi_i^{2020}(s_{-i}^*, \hat{s}_i)}_{\in \{(b,t^0),(b,t^h)\}} \succ_i \varphi^{2020}_i(s^*),
\] 
contradicting  $s^*$ is a Nash equilibrium strategy,\footnote{Unlike the first two cases, in this case cadet $i$ may even get a better assignment than
cadet $j$ (i.e. cadet $i$ may receive an assignment of $(b,t^0)$) by mimicking cadet $j$'s strategy. }  
completing the proof for Case 3, and concluding the proof of Lemma \ref{lemma1}. $\blacksquare$ \mbox{}\hfill$\diamondsuit$\smallskip 
 
For the next phase of our proof, we rely on the construction in  the  Step 2 of the mechanism $\phi^{MP}$:
Let $I^0$ be the set of $q^0_b$ highest $\pi_b$-priority cadets in $I$, and  
$I^1$ be the set of $q^f_b$ highest $\pi_b$-priority cadets in $I\setminus I^0$. 
Relabel the set of  cadets in $I^1$, so that $i^1$ is the lowest $\pi_b$-priority cadet in $I^1$, 
$i^2$ is the second lowest $\pi_b$-priority cadet in $I^1$,\ldots, and $i^{q^f_b}$ is the highest  $\pi_b$-priority cadet in $I^1$. 
Note that, cadet $i^1$ is the $q^{\footnotesize \mbox{th}}$ highest $\pi_b$-priority cadet in set $I$, cadet 
$i^2$ is the $(q-1)^{\footnotesize \mbox{th}}$ highest $\pi_b$-priority cadet in set $I$, and so on.
Let $J^0 = \{j \in I\setminus (I^0\cup I^1) : (b,t^h) \succ_j \emptyset\}$. 
Assuming Step 2.$n$ is the last sub-step of Step 2  of the mechanism $\phi^{MP}$, 
for any $\ell \in \{1,\ldots, n\}$, let 
\[ J^{\ell} = \left\{ \begin{array}{cl}
         J^{\ell -1}  & \mbox{ if } \; \emptyset \, \succ_{i^{\ell}} \, (b,t^h)\\
       J^{\ell -1} \cup\{i^{\ell}\} & \mbox{ if } \; (b,t^h) \succ_{i^{\ell}} \, \emptyset \end{array} \right.
\]
Recall that, under the mechanism $\phi^{MP}$, exactly $n$ cadets receive an assignment of $(b,t^h)$. 
We will show that, the same is also the case under the Nash equilibria of the strategic-form game 
induced by the USMA-2020 mechanism $(\cals^{2020},\varphi^{2020})$.

Let $s^*$  be a Nash equilibrium of the strategic-form game induced by the USMA-2020 mechanism $(\cals^{2020},\varphi^{2020})$.
We have three cases to consider:

\noindent \textit{\textbf{Case 1}\/}: $n=0$

Since by assumption  $n=0$ in this case, 
\begin{equation} \label{eqm-n=0a}
\big\{j\in J^0 : (j,t^h) \; \omega_b \; (i^1,t^0)\big\} = \emptyset. 
\end{equation}
Towards a contradiction, suppose there exists a cadet  $i \in I\setminus(I^0\cup I^1)$ such that $i\in I^+(s^*)$. 
Since cadet $i^1$ is the $q^{\mbox{\footnotesize th}}$ highest $\pi_b$-priority cadet in $I$, the assumption  $i\in I^+(s^*)$ and relation \eqref{eqm-n=0a} imply
\begin{equation}  \label{eqm-n=0b}
i \not\in J^0 \implies \emptyset \; \succ_i (b,t^h). 
\end{equation} 
Moreover, since cadet $i$ is not one of the $q$ highest $\pi_b$-priority cadets in $I$, 
\begin{equation}  \label{eqm-n=0c}
i\in I^+(s^*) \implies s^*_i = b. 
\end{equation}
But this means cadet $i$ can instead submit an alternative strategy $\hat{s}_i = \emptyset$, assuring that she remains unmatched, 
contradicting $s^*$ is a Nash equilibrium.
Therefore, 
\begin{equation}  \label{eqm-n=0d}
\mbox{for any } i\in I\setminus (I^0 \cup I^1), \qquad (i,t^h) \; \omega_b \; (i^1,t^0) \implies s^*_i = \emptyset, 
\end{equation}
which in turn implies 
\begin{equation}  \label{eqm-n=0e}
I^+(s^*) = I^0 \cup I^1. 
\end{equation}
Hence all cadets in $I^0 \cup I^1$ receive a position under $\varphi^{2020}(s^*)$.
Next consider the lowest $\pi_b$-priority cadet $i\in I^0 \cup I^1$ such that $\varphi_i^{2020}(s^*) = (b,t^h)$. 
This can only happen if $s^*_i =b$. 
But this means cadet $i$ can instead submit an alternative strategy $\hat{s}_i = \emptyset$, assuring that $\varphi_i^{2020}(s_{-i}^*,\hat{s}_i) = (b,t^0)$
by relation \eqref{eqm-n=0d},  contradicting $s^*$ is a Nash equilibrium.
Hence
\begin{equation}
\mbox{for any } i\in I^0 \cup I^1, \qquad \varphi_i^{2020}(s^*) = (b,t^0) = \phi^{MP}_i(\succ), 
\end{equation}
and therefore $ \varphi^{2020}(s^*) = \phi^{MP}(\succ)$. 

Finally observe that the strategy profile $s'$ where $s'_i = \emptyset$  for any cadet $i\in I$ is a Nash equilibrium, with an outcome 
$\varphi^{2020}(s') = \phi^{MP}(\succ)$, showing that there exists a Nash equilibrium  completing the proof 
for Case 1. $\blacksquare$ \medskip

For any $\ell \in \{1,\ldots, n\}$, let $\overline{J^{\ell}}$ be the set of $\ell$ highest $\pi_b$-priority cadets in the set $J^{\ell}$:
\[ \overline{J^{\ell}} = \Big\{j \in J^{\ell}  : \; \big|\{i \in J^{\ell} : \; i  \; \pi_b \; j\}\big|<\ell\Big\}
\]
Before proceeding with the next two cases, we prove the following lemma that will be helpful for both cases.  

\begin{lemma} \label{lemma2}
Suppose there are $n>0$ positions allocated at the increased price $t^h$ under the allocation $\phi^{MP}(\succ)$. Then, for any  Nash equilibrium
$s^*$ of the strategic-form game induced by the USMA-2020 mechanism $(\cals^{2020},\varphi^{2020})$ and $\ell \in \{1,\ldots,n\}$,
\begin{enumerate}
\item $\varphi^{2020}_{i^{\ell}}(s^*) = (b,t^h) \quad \iff \quad (b,t^h) \succ_{i^{\ell}} \emptyset$, \; and \smallskip
\item  $\varphi^{2020}_i(s^*) = (b,t^h)$ \quad for any $i\in \overline{J^{\ell}}$.
\end{enumerate}
\end{lemma}

\noindent \textit{Proof of Lemma \ref{lemma2}\/}: Let $s^*$ be a Nash equilibrium 
of the strategic-form game induced by the USMA-2020 mechanism $(\cals^{2020},\varphi^{2020})$. 
First recall that,  
\[ \mbox{for any } j \in I \setminus (I^0 \cup I^1), \qquad  \varphi_j^{2020}(s^*) \in \big\{(b,t^h),\emptyset\big\},
\]
and therefore, since any cadet $j \in I \setminus (I^0 \cup I^1 \cup J^0)$ prefers remaining unmatched to receiving a position at the increased price $t^h$ 
and she can assure remaining unmatched by submitting the strategy $s_j = \emptyset$,
\begin{equation} \label{lemma2-eqn1}
\mbox{for any } j \in I \setminus (I^0 \cup I^1 \cup J^0),  \qquad \varphi_j^{2020}(s^*) = \emptyset.
\end{equation}
Also, by the mechanics of the Step 2 of the mechanism $\phi^{MP}$,
\begin{equation} \label{lemma2-eqn2}
\mbox{for any } \ell \in \{1,\ldots, n\}, \qquad  \big|\big\{j \in J^{\ell -1} : (j, t^h) \; \omega_b \; (i^{\ell}, t^0) \big\}\big| \geq \ell. 
\end{equation}

The proof of the lemma is by induction on $\ell$.  We first prove the result for $\ell = 1$. 

Consider the highest $\pi_b$-priority cadet $j$ in the set  $\big\{j \in J^0 : (j, t^h) \; \omega_b \; (i^1, t^0) \big\}$. 
By relation \ref{lemma2-eqn2}, such a cadet exists. 

First assume that $(b,t^h) \succ_{i^1} \emptyset$. 
In this case, $J^1 = J^0 \cup \{i^1\}$ and
cadet $i^1$ is the highest $\pi_b$-priority cadet in $J^1$. Hence $\overline{J^1} = \{i^1\}$ in this case. 
Consider the Nash equilibrium strategies of cadet $i^1$ and cadet $j$. 
If $s^*_{i^1} = \emptyset$, then  by relation \eqref{lemma2-eqn1} her competitor
cadet $j$ can secure himself an assignment of $(b,t^h)$ by reporting a strategy of $s_j =b$, 
which would mean cadet $i^1$ has to remain unassigned, since by Lemma \ref{lemma1} no cadet in $I^0 \cup I^1$  can envy 
the assignment of cadet $i^1$  at Nash equilibria. 
In contrast, reporting a strategy of $s_{i^1}=b$ assures that cadet $i^1$ receives a position, which is preferred at any price to 
remaining unmatched by assumption $(b,t^h) \succ_{i^1} \emptyset$. 
Therefore, $s^*_{i^1} = b$, and  hence
\begin{equation} \label{lemma2-eqn3}
(b,t^h) \succ_{i^1} \emptyset \quad \implies  \quad
\left\{ \begin{array}{ll}
& \varphi^{2020}_{i^1}(s^*) = (b,t^h),  \; \mbox{ and }\\  
& \varphi^{2020}_{i}(s^*) = (b,t^h)  \quad \mbox{ for any } i\in \overline{J^{1}}=\{i^1\}. \end{array} \right.
\end{equation}

Next assume that $\emptyset \succ_{i^1} (b,t^h)$. 
In this case $J^1 = J^0$ and cadet $j$ is the highest $\pi_b$-priority cadet in $J^1$.  Hence $\overline{J^1} = \{j\}$ in this case.
By Lemma \ref{lemma1},  no cadet in $(I^0\cup I^1)\setminus\{i^1\}$  can envy the assignment of cadet $i^1$ at Nash equilibria. 
Therefore, a strategy of $s_{i^1}=b$ means that cadet $i$ receives an assignment of $(b,t^h)$, which is inferior to  
remaining unmatched by assumption. 
Therefore $s^*_{i^1} = \emptyset$. 
Moreover reporting a strategy of $s_j=\emptyset$ means that cadet $j$ remains unmatched, whereas
reporting a  strategy of $s_j=b$ assures that she  receives an assignment of $(b,t^h)$, which is preferred to  
remaining unmatched since $j \in J^0$. Therefore, $s^*_{i^1} = \emptyset$, and  hence
\begin{equation} \label{lemma2-eqn4}
\emptyset \succ_{i^1} (b,t^h) \quad \implies  \quad
\left\{ \begin{array}{ll}
& \varphi^{2020}_{i^1}(s^*) = \emptyset,  \; \mbox{ and }\\  
& \varphi^{2020}_{i}(s^*) = (b,t^h)  \quad \mbox{ for any  } i\in \overline{J^{1}}=\{j\}. \end{array} \right.
\end{equation}
Relations  \eqref{lemma2-eqn3} and  \eqref{lemma2-eqn4} complete the proof for $\ell = 1$. \smallskip

Next assume that the inductive hypothesis holds for $\ell = k<n$. We want to show that the result holds for $\ell = (k+1)$ as well. 

By the inductive hypothesis,
\begin{equation} \label{lemma2-eqn5}
 \mbox{for any } i\in \overline{J^k}, \quad  \varphi^{2020}_{i}(s^*) = (b,t^h).
\end{equation}
By relation \ref{lemma2-eqn2}, there are at least $k+1$ cadets in the set $\big\{j \in J^k : (j, t^h) \; \omega_b \; (i^{k+1}, t^0) \big\}$.
Therefore, since there are $k$ cadets in the set $\overline{J^k}$, there is at least one cadet in the set 
\[\big\{j \in J^k : (j, t^h) \; \omega_b \; (i^{k+1}, t^0) \big\}\setminus \overline{J^k}.
\]
Let $j$ be the highest $\pi_b$-priority cadet in this set. 

First assume that $(b,t^h) \succ_{i^{k+1}} \emptyset$. 
In this case $J^{k+1} = J^k \cup \{i^{k+1}\}$ and
cadet $i^{k+1}$ is the highest $\pi_b$-priority cadet in $J^{k+1}$. Hence $\overline{J^{k+1}} = \overline{J^k} \cup \{i^{k+1}\}$ in this case. 
Consider the Nash equilibrium strategies of cadet $i^{k+1}$ and cadet $j$. 
If $s^*_{i^{k+1}} = \emptyset$, then by relation \eqref{lemma2-eqn1}
cadet $j$ can secure herself an assignment of $(b,t^h)$ by reporting a strategy of $s_j =b$, 
which would mean cadet $i^{k+1}$ has to remain unassigned, since by Lemma \ref{lemma1} no cadet in $(I^0 \cup I^1)\setminus \{i^1,\ldots, i^k\}$  
can envy the assignment of cadet $i^{k+1}$ at Nash equilibria 
and by relation \eqref{lemma2-eqn5} all cadets in $\overline{J^k}$ receive an assignment of $(b,t^h)$.\footnote{Since 
$\left|(I^0 \cup I^1)\setminus \{i^1,\ldots, i^k\}\right|=(q-k)$
and $\left|\overline{J^k}\right| = k$, this basically means cadets
$i^{k+1}$ and $j$ are competing for a single position.} 
In contrast, reporting a strategy of $s_{i^{k+1}}=b$ assures that cadet $i^{k+1}$ receives a position, which is preferred at any price to 
remaining unmatched by assumption $(b,t^h) \succ_{i^{k+1}} \emptyset$. 
Therefore,  $s^*_{i^{k+1}} = b$, and  hence
\begin{equation} \label{lemma2-eqn6}
(b,t^h) \succ_{i^{k+1}} \emptyset \; \implies  \;
\left\{ \begin{array}{ll}
& \varphi^{2020}_{i^{k+1}}(s^*) = (b,t^h),  \; \mbox{ and }\\  
& \varphi^{2020}_{i}(s^*) = (b,t^h)  \; \mbox{ for any } i\in \overline{J^{k+1}}= \overline{J^k}\cup\{i^{k+1}\}. \end{array} \right.
\end{equation}
Next assume that $\emptyset \succ_{i^{k+1}} (b,t^h)$. 
In this case $J^{k+1} = J^k$ and  $\overline{J^{k+1}} = \overline{J^k} \cup \{j\}$.
By Lemma \ref{lemma1},  no cadet in $I^0\cup I^1 \setminus \{i^1,\ldots, i^k\}$  can envy the assignment of cadet $i^{k+1}$ at Nash equilibria. 
Therefore, since all cadets in $\overline{J^k}$ receive an assignment of $(b,t^h)$  by relation \eqref{lemma2-eqn5},
a strategy of $s_{i^{k+1}}=b$ means that cadet $i^{k+1}$ receives an assignment of $(b,t^h)$, which is inferior to  
remaining unmatched by assumption. Therefore $s^*_{i^{k+1}} = \emptyset$. 
Moreover reporting a strategy of $s_j=\emptyset$  means that cadet $j$ remains unmatched, whereas
reporting a  strategy of $s_j=b$ assures that she  receives an assignment of $(b,t^h)$, which is preferred to  
remaining unmatched since $j \in J^k$. Therefore, $s^*_{i^{k+1}} = \emptyset$, and  hence
\begin{equation} \label{lemma2-eqn7}
\emptyset \succ_{i^{k+1}} (b,t^h) \quad \implies  \quad
\left\{ \begin{array}{ll}
& \varphi^{2020}_{i^{k+1}}(s^*) = \emptyset,  \; \mbox{ and }\\  
& \varphi^{2020}_{i}(s^*) = (b,t^h)  \quad \mbox{ for any  } i\in \overline{J^{k+1}}=\overline{J^k}\cup\{j\}. \end{array} \right.
\end{equation}
Relations  \eqref{lemma2-eqn6} and  \eqref{lemma2-eqn7} complete the proof for $\ell = k+1$, 
and conclude the proof of Lemma \ref{lemma2}. \mbox{}\hfill$\diamondsuit$ \medskip

We are ready to complete prove the theorem for our last two cases:

\noindent \textit{\textbf{Case 2.}\/} $n \in \{1,\ldots, q^f_b-1\}$ \smallskip

For this case, by the mechanics of the Step 2 of the mechanism $\phi^{MP}$, 
\begin{equation} \label{case2-eqn1}
\big|\big\{j \in J^{n} : (j, t^h) \; \omega_b \; (i^{n+1}, t^0) \big\}\big| = n.
\end{equation}
Consider cadet $i^{n+1}$. There are $q-(n+1)$ cadets with higher $\pi_b$-priority, and
by relation \eqref{case2-eqn1} there are $n$ cadets in $J^n$ whose increased price assignments have higher $\omega_b$ priority under the price responsiveness policy
than the base-price assignment for cadet $i^{n+1}$.
For any other cadet $i \in I\setminus \Big(J^n \cup I^0 \cup \big(I^1 \setminus \{i^1,\ldots,i^{n+1}\}\big)\Big)$ with $(i, t^h) \; \omega_b \; (i^{n+1}, t^0)$, 
we must have $\emptyset \succ_i (b,t^h)$ since  $J^n \supseteq J^0$. 
Therefore none of these individuals can receive an assignment of $(b,t^h)$ under a Nash equilibrium strategy, and hence 
the number of cadets who can have higher $\pi^+_b(s^*)$-priority than cadet is $i^{n+1}$ is at most 
$q-(n+1)+n = q-1$ under any Nash equilibrium strategy. That is, cadet $i^{n+1} \in I^+(s^*)$ regardless of her submitted strategy, 
and therefore,
\begin{equation}  \label{case2-eqn2}
\varphi^{2020}_{i^{n+1}}(s^*) = (b,t^0), 
\end{equation}
since  her best response $s^*_{i^{n+1}}$ to $s^*_{-i^{n+1}}$ results in an assignment of $(b,t^0)$. 
Moreover, Lemma \ref{lemma1} and relation \eqref{case2-eqn2} imply that,  
for any cadet $i \in I^0 \cup \big(I^1 \setminus  \{i^1,\ldots,i^{n+1}\}\big)$,
\begin{equation}  \label{case2-eqn3}
\varphi^{2020}_i(s^*) = (b,t^0).
\end{equation}
Hence Lemma \ref{lemma2} and relations \eqref{case2-eqn2}, \eqref{case2-eqn3} imply
$\varphi^{2020}(s^*) = \phi^{MP}(\succ)$. 

Finally, the strategy profile $s'$ where $s'_i = b$  for any cadet $i\in J^n$ and $s'_j = \emptyset$
for any cadet  $j \in I\setminus J^n$ is a Nash equilibrium, with an outcome 
$\varphi^{2020}(s') = \phi^{MP}(\succ)$, showing that there exists a Nash equilibrium  completing the proof 
for Case 2. $\blacksquare$ \medskip

\noindent \textit{\textbf{Case 3.}\/} $n =  q^f_b$ \smallskip

Since at most $q^f_b$ positions can be assigned at the increased price $t^h$, Lemma \ref{lemma1} and Lemma \ref{lemma2} immediately imply
$\varphi^{2020}(s^*) = \phi^{MP}(\succ)$. 

Finally the strategy profile $s'$ where $s'_i = b$  for any cadet $i\in J^{q^f_b} \cup I^0$ and $s'_j = \emptyset$
for any cadet  $j \in I\setminus \big(J^n \cup I^0\big)$ is a Nash equilibrium, with an outcome 
$\varphi^{2020}(s') = \phi^{MP}(\succ)$, showing that there exists a Nash equilibrium  completing the proof  for Case 3, 
and the proof of the proposition. $\blacksquare$ \qed \medskip

\noindent \textbf{Proof of Corollary \ref{corollary:phi=cosm}}: Since \textit{BRADSO-IC\/} is implied by \textit{strategy-proofness\/}, 
Corollary  \ref{corollary:phi=cosm} is a direct implication of Theorem  \ref{cosm} and Proposition \ref{thm:singlebranchcharacterization}. \qed 

\newpage

\section{Formal Description of USMA-2006 Mechanism and Individual-Proposing Deferred Acceptance Algorithm}
\subsection{USMA-2006 Mechanism}\label{sec:usma2006}

The USMA-2006 mechanism is a quasi-direct mechanism with the following 
message space:
\[ \cals^{2006} = \big(\calp \times 2^{B} \big)^{|I|}. 
\]
The following construction is useful to formulate the outcome function  for the USMA-2006 mechanism:  

Given an OML $\pi$ and a strategy profile $s= (P_i, B_i)_{i\in I} \in \cals^{2006}$, for any branch $b\in B$
construct the following adjusted priority order $\pi^+_b \in \Pi$ on the set of cadets $I$.
For any pair of cadets $i, j \in I$,
\begin{enumerate}
\item $b\in B_i$ and $b\in B_j \quad \implies \quad i \; \pi^+_b \; j \; \iff i \; \pi \; j$,
\item $b\not\in B_i$ and $b\not\in B_j \quad \implies \quad i \; \pi^+_b \; j \; \iff i \; \pi \; j$, \mbox{ and} 
\item $b\in B_i$ and $b\not\in B_j \quad \implies \quad i \; \pi^+_b \; j$.
\end{enumerate}	
Under the adjusted priority order $\pi^+_b$, 
any pair of cadets are rank ordered through the OML $\pi$
if they have indicated the same willingness to pay the increased price for branch $b$, and 
through the ultimate price responsiveness policy $\overline{\omega}_b$ (which 
gives higher priority to the cadet who has indicated to pay the increases price) otherwise.  

Given an OML $\pi\in\Pi$ and  a strategy profile $s= (P_i, B_i)_{i\in I} \in \cals^{2006}$,
the outcome $\varphi^{2006}(s)$ of the 
\textbf{USMA-2006 mechanism\/} is obtained with the following sequential procedure:\smallskip
\begin{quote}
\textbf{\textit{Branch assignment\/}}: At any step $\ell \geq 1$ of the procedure, the highest $\pi$-priority cadet $i$ who is not tentatively on hold for a position at any branch
applies to her highest-ranked acceptable branch $b$ under her submitted branch preferences $P_i$ that 
has not rejected her from earlier steps.\footnote{The USMA-2006 mechanism can also 
be implemented with a variant of the algorithm where each cadet who is not tentatively holding a position simultaneously apply to
her next choice branch among branches that has not rejected her application.}  
 
Branch $b$ considers cadet $i$ together with all cadets it has been tentatively holding both for 
its $q^0_b$ base-price positions and also for its $q^f_b$ flexible-price  positions, and 
\begin{enumerate}
\item it tentatively holds (up to) $q^0_b$ highest  $\pi$-priority applicants for one of its $q_b^0$ base-price positions, 
\item among the remaining applicants it tentatively holds  (up to) $q^f_b$ highest  $\pi^+_b$-priority applicants for one of its $q_b^f$ 
flexible-price  positions, and
\item it rejects any remaining applicant. 
\end{enumerate}
The procedure terminates when no applicant is rejected.   Any cadet who is not tentatively on hold
at any brach remains unmatched, and all tentative branch assignments are finalized. 

\textbf{\textit{Price assignment\/}}: For any branch $b\in B$,
\begin{enumerate}
\item any cadet $i\in I$ who is assigned one of the $q^0_b$ base-price positions at branch $b$ is charged the base price $t^0$, and
\item any cadet  $i\in I$ who is assigned one of the $q^f_b$  flexible-price  positions is charged 
\begin{enumerate}
\item the increased price $t^h$ if $b\in B_i$, and 
\item the base price $t^0$ if $b\not\in B_i$.
\end{enumerate} 
\end{enumerate}
\end{quote}

\subsection{Individual-Proposing Deferred Acceptance Algorithm}\label{sec:da}

The USMA-2020 mechanism was based on the individual-proposing deferred acceptance
algorithm \citep{gale/shapley:62}.  Given a ranking over branches, 
the individual-proposing deferred acceptance algorithm  (DA)  produces a matching as follows.

\medskip
	\begin{quote}
        \noindent{}{\bf Individual-Proposing Deferred Acceptance Algorithm ($\mathbf{DA}$)}

		\noindent{}{\bf Step 1:}
            Each cadet applies to her most preferred branch.
            Each branch $b$ tentatively assigns
            applicants with the highest priority until
            all cadets are chosen or all $q_{b}$ slots
            as assigned and permanently rejects the rest. If there are no rejections, then stop.
	
		\noindent{}{\bf Step k:}
			Each cadet who was rejected in Step k-1 applies to her next
			preferred branch, if such a branch exists. Branch $b$ tentatively assigns cadets
            with the highest priority until all all cadets are chosen or all $q_b$ slots
            are assigned and permanently rejects the rest. If there are no rejections,
            then stop.\smallskip

            The algorithm terminates when there are no rejections, at which point all tentative
            assignments are finalized.
	\end{quote}

\newpage

\section{Additional Results}

\subsection{Empirical Evidence on the Failure of Desiderata under the USMA-2006 and USMA-2020 Mechanisms} \label{failureprevelance}

In this section, we report on the failure of BRADSO-IC, presence of strategic BRADSO, and presence
of detectable priority reversals under the USMA-2006 and USMA-2020 mechanisms.
We show that the USMA-2020 mechanism exacerbated challenges compared to the USMA-2006 mechanism.
We use actual  data submitted under these mechanisms and also simulated data generated 
from the MPCO mechanism for the USMA Class of 2021. 

\subsubsection{USMA-2006 and USMA-2020 Mechanisms in the Field} \label{field-2006-2020}

BRADSO-IC failures were much more common under USMA-2020 than under USMA-2006.
\fig{Figure \ref{fig:failures}} shows that
nearly four times (85 versus 22) as many cadets from the Class of 2020 (which used the USMA-2020 mechanism) were part of BRADSO-ICs than were cadets from the Classes of 2014 to 2019 (which used the USMA-2006 mechanism).  
Strategic BRADSOs must be more common under USMA-2020 because
they are not possible under USMA-2006.
For the Class of 2020, 18 cadets were part of strategic BRADSOs under the USMA-2020 mechanism.  
Importantly, fixing these instances ex-post would have required a change in branch assignments (rather than merely foregoing a BRADSO charge). Finally, nearly four times as many cadets were part of detectable priority reversals under the USMA-2020 mechanism than under the USMA-2006 mechanism (75 versus 20).

\subsubsection{USMA-2006 and USMA-2020 Mechanisms with Simulated Data} \label{simulated-2006-2020}

Our comparison of prior mechanisms has so far been based on
preferences submitted under those mechanisms.  
We can also use cadet preference data on branch-price pairs generated by the strategy-proof 
MPCO mechanism to simulate the outcome of
 USMA-2006 and USMA-2020 mechanisms under truthful strategies.   
This is valuable because for cadet preferences submitted under the USMA-2006 and USMA-2020 mechanisms, we could only measure detectable priority reversals (reported in 
\fig{Figure \ref{fig:failures}}) and not all priority reversals.  

To measure all priority reversals, 
we use preferences over branch-price pairs under the MPCO mechanism to construct a truthful strategy, denoted $s_i = (P_i,B_i)$, under a quasi-direct mechanism by using the branch rank ordering for $P_i$ and assuming that if a cadet ever expresses a willingness to pay the increased price at a branch, 
then the cadet is willing to pay the increased price under $B_i$.   Taking this constructed strategy as input, we then simulate the USMA-2006 and USMA-2020 mechanism
using the branch capacities and priorities from the Class of 2021.   Under the USMA-2006 mechanism simulation,
there are 29 priority reversals and 20 are detectable priority reversals.  Under the USMA-2020 mechanism simulation, there are 204 priority reversals and 197 are detectable priority reversals.  This
suggests that, in practice, detectable priority reversals likely constitute the majority of priority reversals among the Classes of 2014-2019, which used the USMA-2006 mechanism,
and the Class of 2020, which used the USMA-2020 mechanism.

Using truthful strategies to evaluate the USMA-2006 and USMA-2020 mechanism, \fig{Figure \ref{fig:2021sim}} shows that there are nearly seven times as many BRADSO-IC failures under the USMA-2020 mechanism compared to the USMA-2006 mechanism (146 vs. 21) and seven times as many priority reversals under the USMA-2020 mechanism compared to the USMA-2006 mechanism (204 vs. 29).  This pattern of 
behavior suggests that the comparison reported in \fig{Figure \ref{fig:failures}} potentially understates the dramatic increase in BRADSO-IC failures and priority reversals stemming from
the adoption of the USMA-2020 mechanism because the \fig{Figure \ref{fig:failures}} comparison is based on strategies submitted under the message space of quasi-direct mechanisms and not underlying cadet preferences.

One reason the comparison between USMA-2006 and USMA-2020 in \fig{Figure \ref{fig:failures}} 
is not as striking as the comparison in \fig{Figure \ref{fig:2021sim}} is that, 
as we have presented in Section \ref{sec:shortcomings2020}, many cadets 
were well-aware of the necessity to strategically make their increased price willingness choices under the USMA-2020 mechanism.  
Our analysis in Appendix \ref{sec:singlebranch} illustrates the perverse incentives in the USMA-2020
mechanism.  For the Class of 2020, a dry-run of the mechanism where cadets submitted indicative rankings of branches and learned about their assignment  took place.
After observing their dry-run assignment, cadets were allowed to submit a final set of rankings under USMA-2020, 
and therefore had the opportunity to revise their strategies in response
to this feedback.   \fig{Figure \ref{fig:usma2020indicative}} tabulates strategic BRADSOs, BRADSO-IC failures, and detectable priority
reversals under indicative and final preferences.  Final preferences result in fewer strategic BRADSOs, BRADSO-IC failures, and detectable
priority reversals.  This pattern is consistent with some cadets responding to the dry-run by ranking branch choices in response to these issues.

In general, cadets form their preferences over branches over time as they acquire
more information about branches and their own tastes.  Therefore, the change documented in \fig{Figure \ref{fig:usma2020indicative}} may simply
reflect general preference formation from acquiring information about branches, and not revisions to preferences in response to the specific mechanism.  We briefly investigate this possibility by looking at the presence of strategic BRADSOs,
BRADSO-IC failures, and priority reversals using data on the indicative and final preferences from the Class of 2021.  This class participated in the 
strategy-proof MPCO mechanism.  We take indicative and final cadet preferences under MPCO mechanism and construct truthful strategies, following
the approach described above, for the USMA-2020 mechanism.  \fig{Figure \ref{fig:usma2021indicative}} shows that with preferences
constructed from a strategy-proof mechanism, there are only modest differences in strategic BRADSOs, BRADSO-IC failures, and priority reversals between the indicative and final rounds.
This comparison supports our claim that revisions of rank order lists in response to a dry-run of the USMA-2020 mechanism might understate
the issues this mechanism created, and why these issues became so pronounced with the USMA-2020 mechanism relative to the USMA-2006 mechanism.

\subsection{Cadet Utilization of the Richer message space of the MPCO Mechanism}

Preference data from the Class of 2021 confirm that cadets used the flexibility to express preferences over branch-price pairs.  \fig{Figure \ref{fig:nonconsecutive_bradso}} provides details on the extent to which cadets did not rank a branch with increased price immediately after
the branch at base price. For each of 994 cadet first branch choices, 272 cadets rank that branch with increased price as their second choice and 36 cadets rank that 
branch with increased price as their third choice or lower.  
These 36 cadets would not have been able to express this preference under the message space of a quasi-direct mechanism 
like the USMA-2006 mechanism
or the USMA-2020 mechanism.  When we consider the next branch on a cadet's rank order list, cadets also value the flexibility of the new mechanism.  
For the branch that appears next on the rank order list, 78 cadets rank that branch with  increased price as their immediate next highest choice and 24 cadets rank that branch
with increased price two or more places below on their rank order list.  
These 24 cadets also would not have been able to express this preference under a quasi-direct mechanism.

\section{Cadet Data and Survey Appendix}
\label{subsec:dataappendix}

\subsection{Data Appendix}
\label{subsec:datasubappendix}

\setcounter{table}{0}
\renewcommand{\thetable}{E\arabic{table}}
\renewcommand{\thefigure}{E\arabic{figure}}

Our data cover the West Point Classes of 2014 through 2021.  
We present two tables about data processing.  The first table
reports summary statistics on branches for the Class of 2020 and Class of 
2021.  The second table presents summary information about
mechanism replication for the Classes of 2014-2021.

\begin{table}[htp]
	\begin{center}
    \caption{\textbf{Branches and Applications for Classes of 2020 and 2021}}
    \label{fig:tableCapacity}
    \includegraphics[scale = 0.6]{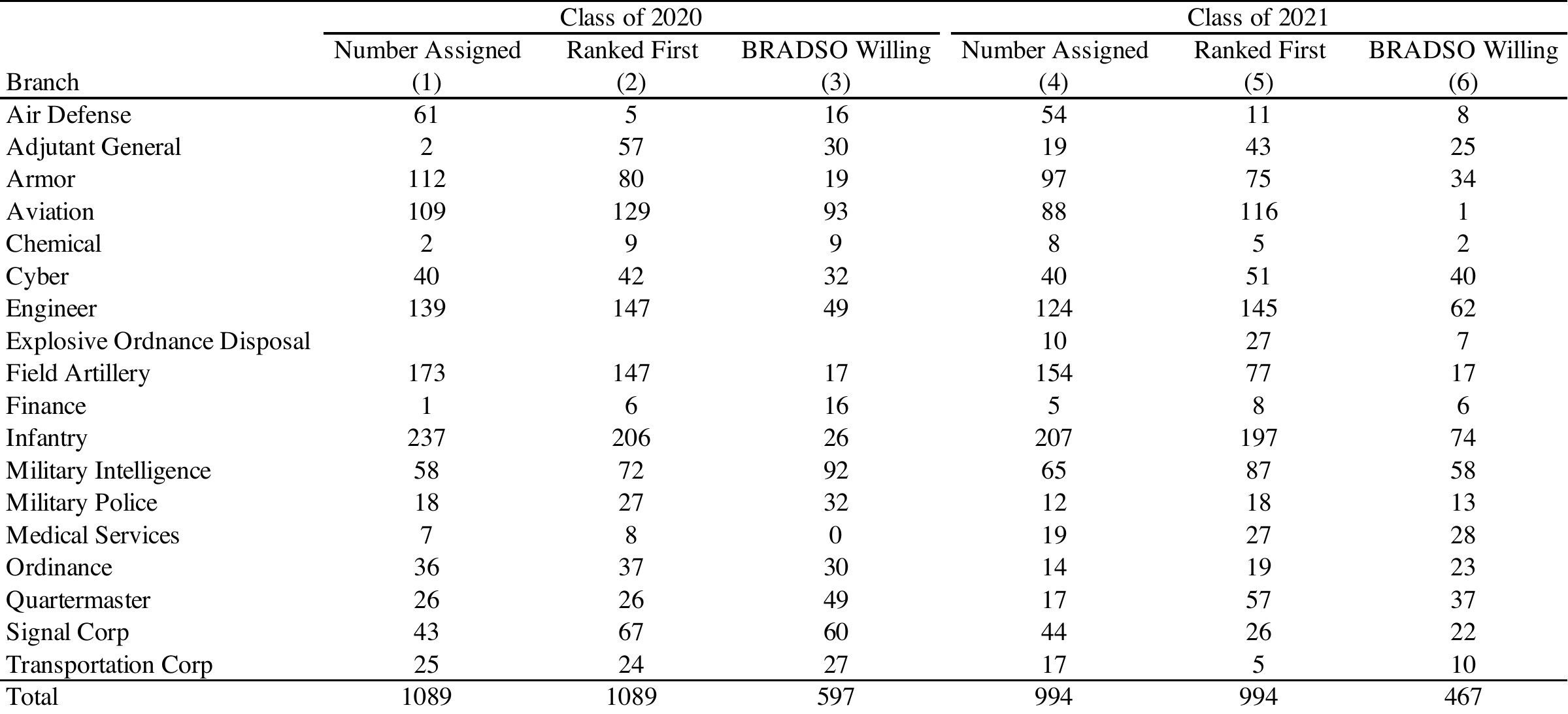}
	\end{center}
    \scriptsize{\textbf{Notes.} This table reports information on branches for the Class of 2020 and 2021.  Number Assigned equals the capacity of the branch.  Ranked First is the number of cadets ranking the branch as their highest rank choice.  BRADSO Willing is the number of cadets who rank a BRADSO contract at the branch anywhere on their rank order list. Explosive Ordnance Disposal was not a branch option for the Class of 2020.}
\end{table}

\begin{table}[htp]
    \begin{center}
    \caption{\textbf{Mechanism Replication Rate}}
    \includegraphics[scale = 0.7]{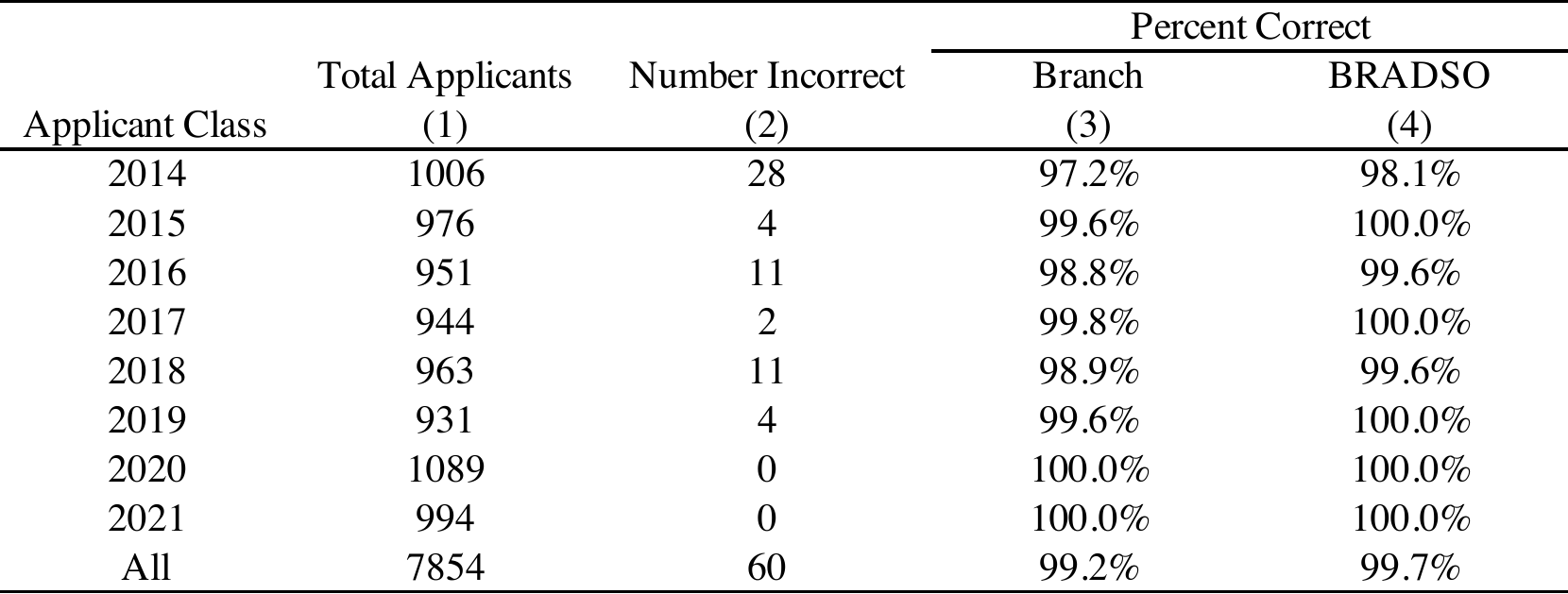}
    \end{center}
   \scriptsize{\textbf{Notes.}  This table reports the replication rate of the USMA assignment mechanism across years. The
    USMA-2006 mechanism is used for the Classes of 2014-2019, USMA-2020 mechanism is used for the Class of 2020, and
    the multi-price Cumulative Offer mechanism is used for the Class of of 2021.  Number incorrect are the number of cadets who obtain a different
    assignment under our replication.   Branch percent correct is the number of branch assignments that we replicate.
    BRADSO percent correct is the number of BRADSO assignments we replicate.
    \label{fig:tableReplicationRate}}
\end{table}

\begin{figure}[htp]
    \begin{center}
    \caption{\textbf{Comparison of Outcomes of the USMA-2006 and USMA-2020 Mechanisms\\}}
       \label{fig:failures}
    \includegraphics[scale=0.75]{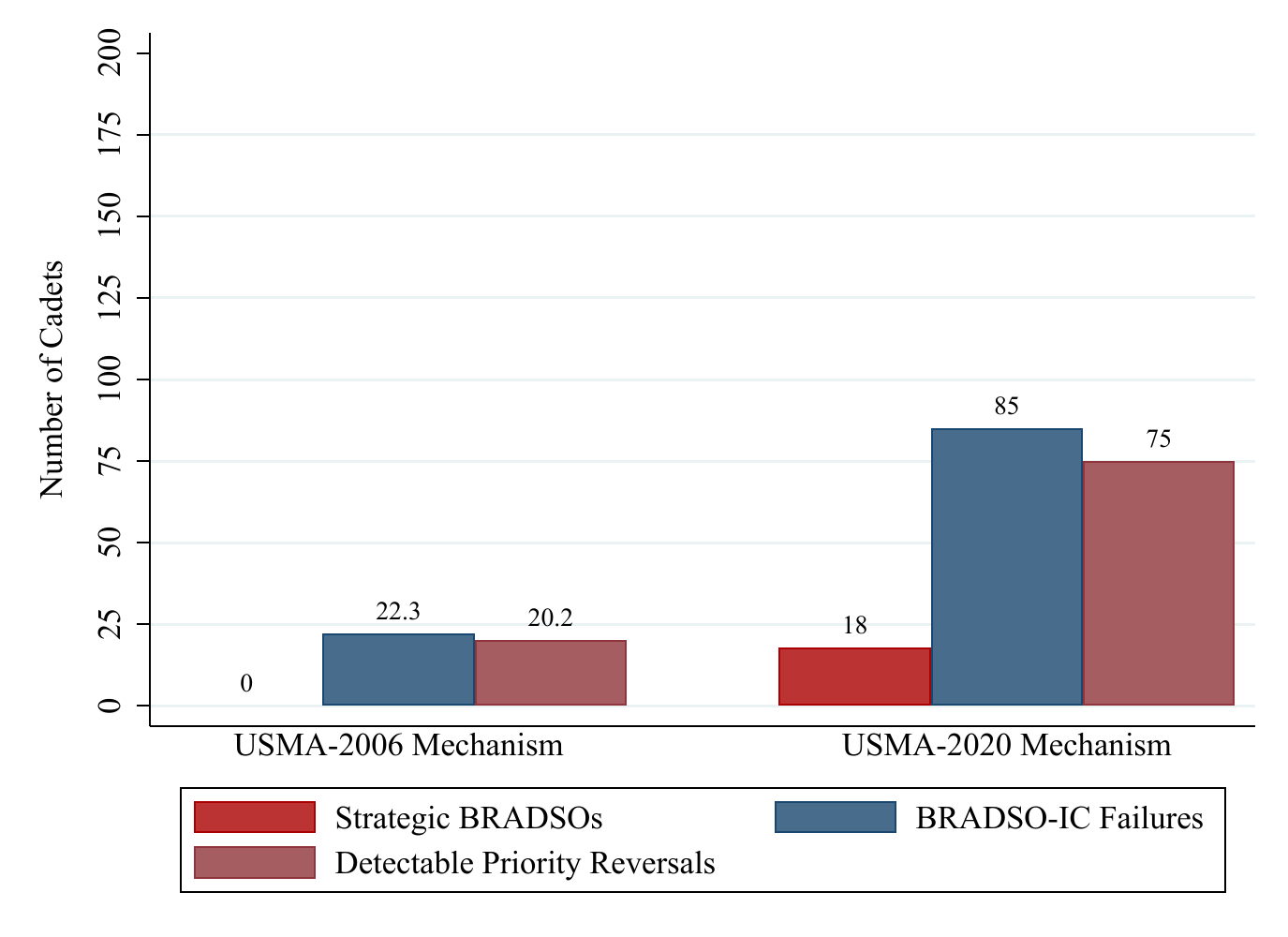}
    \end{center}
\scriptsize{\textbf{Notes.}  This figure reports Strategic BRADSOs, BRADSO-IC Failures, and Detectable Priority Reversals under the USMA-2006  and USMA-2020 Mechanisms.   The first three columns correspond to outcomes under USMA-2006 Mechanism averaged over classes from 2014-2019.  The last three columns correspond to outcomes under USMA-2020 Mechanism for the Class of 2020.} 
\end{figure}

\begin{figure}[htp]
    \begin{center}
    \caption{\textbf{USMA-2006 and USMA-2020 Mechanism Performance under Truthful Strategies Simulated from Preference Data from Class of 2021}\\}
    \label{fig:2021sim}
    \includegraphics[scale = 0.8]{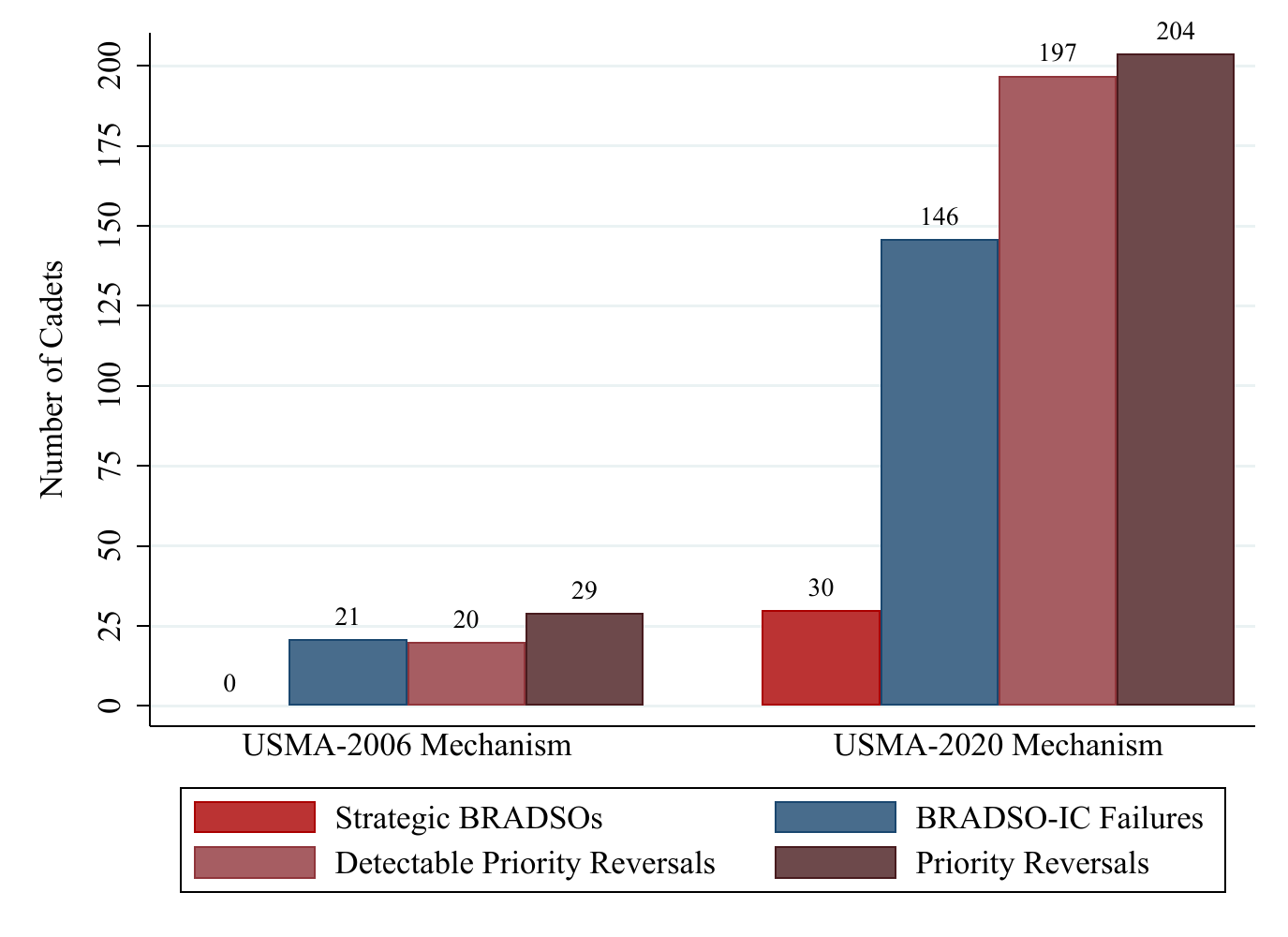}
    \end{center}
\scriptsize{\textbf{Notes.} 
This figure uses data from the Class of 2021 to simulate the outcomes of the mechanisms USMA-2006 and
USMA-2020. We use preferences over branch-price pairs under the MPCO mechanism to
construct truthful strategies for USMA-2006 and USMA-2020 by assuming that willingness to BRADSO at a branch means the cadet's strategy under the USMA-2006 and USMA-2020 mechanisms has her willing to BRADSO. To compute Priority Reversals, we compare a cadet's outcome in either the USMA-2006 or USMA-2020 mechanism to a cadet's preference submitted under the MPCO mechanism. If a cadet prefers a higher ranked choice and has higher priority over a cadet who is assigned that choice, then the cadet is part of a Priority Reversal.}  
\end{figure}

\begin{figure}[htp]
\begin{center}
\caption{\textbf{USMA-2020 Mechanism Performance Under Indicative and Final Strategies}}
\label{fig:usma2020indicative}
    \includegraphics[scale = 0.75]{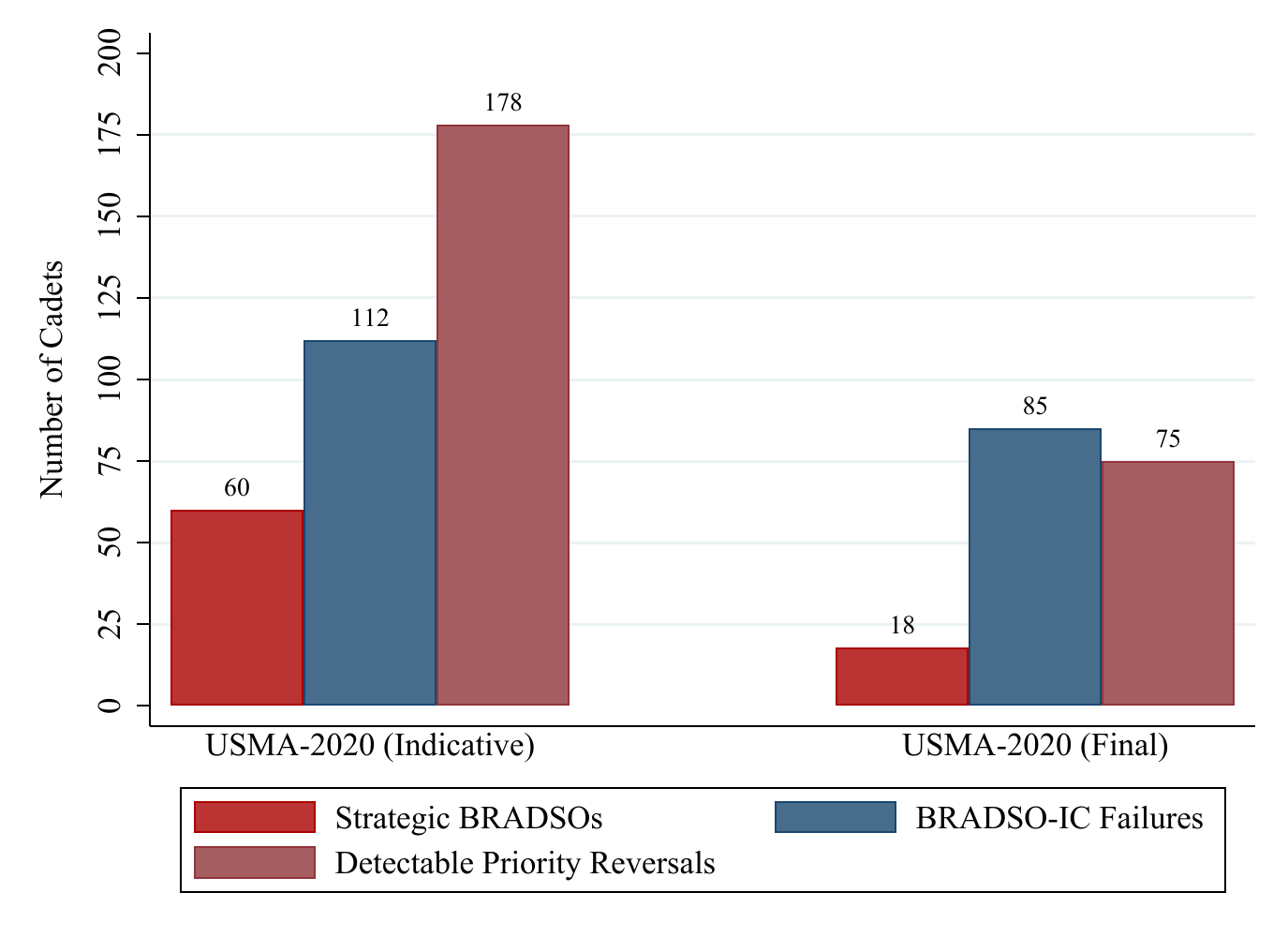}
\end{center}
\scriptsize{\textbf{Notes.} This figure reports on the number of Strategic BRADSOs, BRADSO-IC failures, and Detectable Priority Reversals under indicative strategies submitted
in a dry-run of the USMA-2020 mechanism and final strategies of the USMA-2020 mechanism for the Class of 2020. }
\end{figure}

\begin{figure}[htp]
\begin{center}
\caption{\textbf{USMA-2020 Mechanism Performance under Truthful Strategies Simulated from Indicative and Final Preference Data from Class of 2021}\\}
\label{fig:usma2021indicative}
    \includegraphics[scale = 0.7]{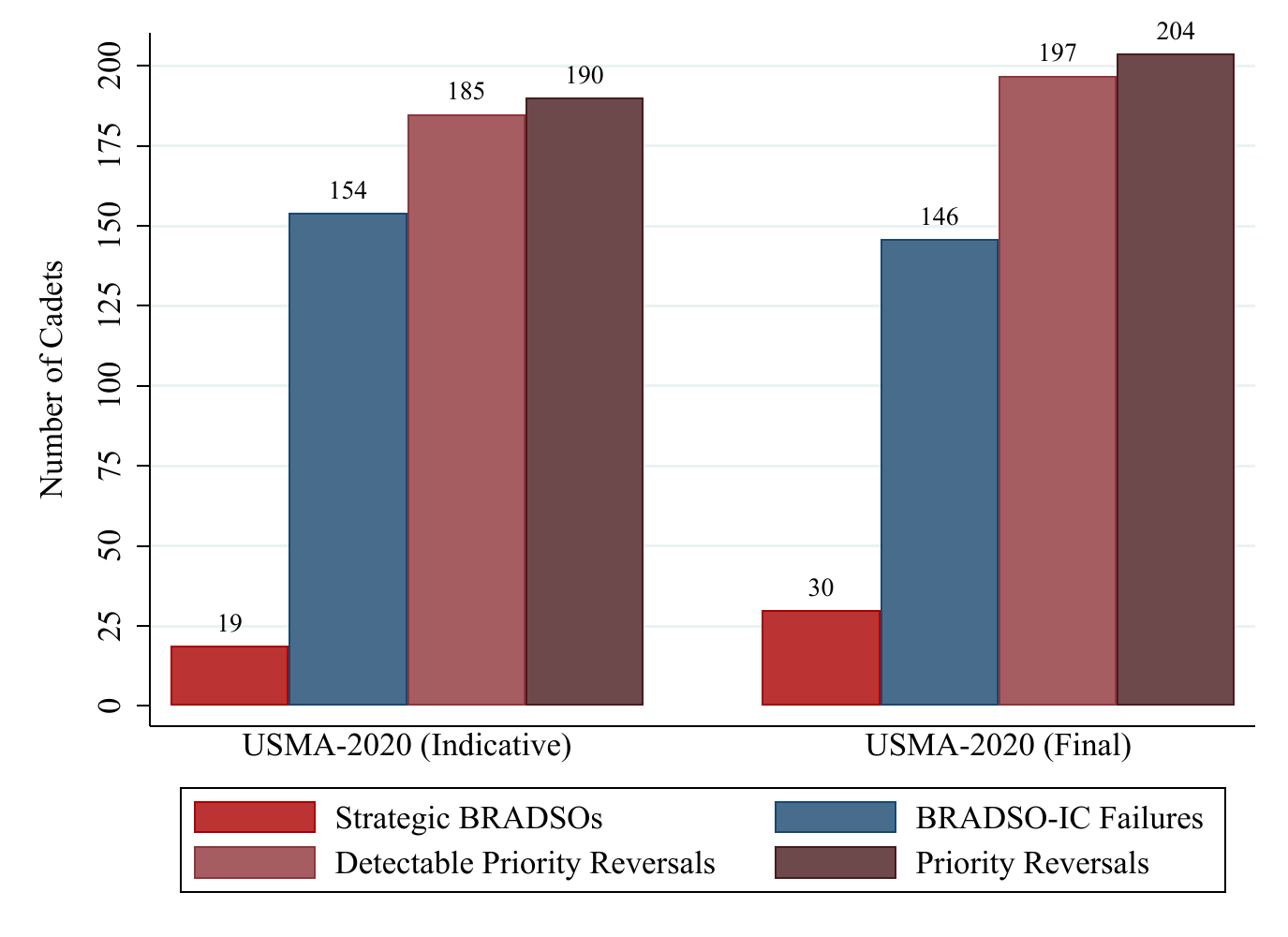}
\end{center}

\scriptsize{\textbf{Notes.} 
USMA used the strategy-proof  MPCO mechanism for the Class of 2021. 
This figure uses data  from the indicative and final rounds from the Class of 2021 on cadet preferences, branch priorities, and branch capacities to simulate the outcome of the 
USMA-2020 mechanism. Since the message space of the mechanism USMA-2020 differs from that of the mechanism MPCO, cadet strategies 
that correspond to truthful branch-preferences and BRADSO willingness are 
are simulated from cadet preferences over branch-price pairs under the  MPCO mechanism. 
Truthful strategies are constructed from Class of 2021 preferences 
by assuming that a preference indicating willingness to BRADSO at a branch means the cadet's strategy under the USMA-2006 and USMA-2020 mechanisms has her willing to BRADSO.  USMA-2020 (Indicative) reports outcomes using strategies constructed from preferences submitted in the dry-run of the MPCO mechanism.
USMA-2020 (Final) reports outcomes using strategies constructed from preferences submitted in the final run of the  MPCO mechanism.}
\end{figure}

\begin{figure}[htp]
    \begin{center}
    \caption{\textbf{BRADSO Ranking Relative to Non-BRADSO Ranking by Class of 2021}\\}
    \label{fig:nonconsecutive_bradso}
    \includegraphics[scale = 0.9]{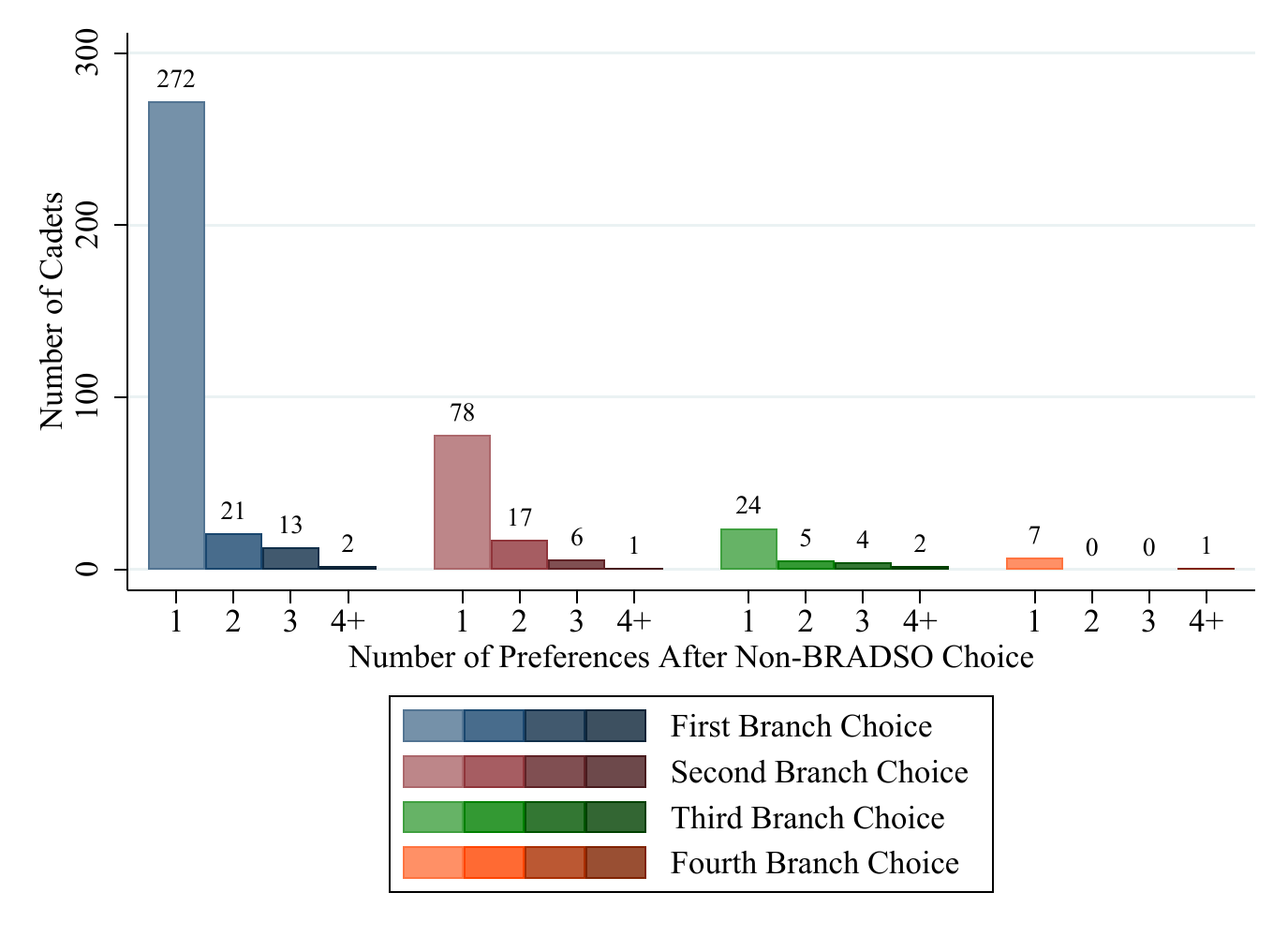}
    \end{center}
\scriptsize{\textbf{Notes.} This figure reports where in the preference list a branch is ranked with BRADSO relative to where it is ranked without BRADSO.  A value of 1 (2 or 3) indicates that the branch is ranked with BRADSO immediately after (two places or three places after, respectively) the branch is ranked at base price.  4+ means that the a branch is ranked with BRADSO four or more choices after the branch is ranked at base price.}
\end{figure}

\subsection{Cadet Survey Questions and Answers}
\label{survey}

\renewcommand\thefigure{\thesection.\arabic{figure}}
\renewcommand\thetable{\thesection.\arabic{table}}
\setcounter{figure}{0}
\setcounter{table}{0}

In September 2019, the Army administered
a survey to West Point cadets in the Class of 2020.  This survey asked two questions related to assignment mechanisms, one on cadet understanding of USMA-2020 and the other on cadet preferences over assignment mechanisms. This section reports the questions and the distribution of survey responses.

\bigskip

\textbf{Question 1.} \textit{What response below best describes your understanding of the impact of volunteering to BRADSO for a branch in this year's branching process?}
\begin{itemize}
     \item[A.] I am more likely to receive the branch, but I am only charged a BRADSO if I would have failed to receive the branch had I not volunteered to BRADSO. (43.3\% of respondents)
     \item[B.] I am charged a BRADSO if I receive the branch, regardless of whether volunteering to BRADSO helped me receive the branch or not. (9.5\% of respondents)
     \item[C.] I am more likely to receive the branch, but I may not be charged a BRADSO if many cadets who receive the same branch not only rank below me but also volunteer to BRADSO. (38.8\% of respondents)
     \item[D.] I am more likely to receive the branch, but I do not know how the Army determines who is charged a BRADSO. (6.7\% of respondents)
     \item[E.] I am NOT more likely to receive the branch even though I volunteered to BRADSO. (1.8 percent of respondents)
\end{itemize}

38.8\% of cadets selected the correct answer (answer C). 43.3\% of cadets believed that the 2020 mechanism would only charge a BRADSO if required to receive the branch (answer A)

\bigskip

\textbf{Question 2}. \textit{A cadet who is charged a BRADSO is required to serve an additional 3 years on Active Duty. Under the current
mechanism, cadets must rank order all 17 branches and indicate if they are willing to BRADSO for each branch choice. For
example:}
\begin{itemize}
     \item \textbf{Current Mechanism Example}:
   \begin{itemize}
       \item  \textbf{1: AV/BRADSO, 2: EN, 3: CY}
   \end{itemize}
\end{itemize}
\bigskip
Under an alternative mechanism, cadets could indicate if they prefer to receive their second branch choice without a
BRADSO charge more than they prefer to receive their first branch choice with a BRADSO charge. For example:
\begin{itemize}
     \item \textbf{Alternative Mechanism Example}:
   \begin{itemize}
       \item  \textbf{1: AV, 2: EN, 3: AV/BRADSO, 4: CY} 
   \end{itemize}
\end{itemize}
\bigskip
When submitting branch preferences, which mechanism would you prefer?

\begin{itemize}
     \item A. Current Mechanism (21.4\% of respondents)
     \item B. Alternative Mechanism (49.7\% of respondents)
     \item C. Indifferent (24.2\% of respondents)
     \item D. Do Not Understand (4.8\% of respondents)
\end{itemize}

\newpage

\section{Potential Applications of Price Responsive Policies in Priority-Based Assignment} \label{sec:potentialapplication}

Our model and main application are inspired by the reform of the U.S. Army's cadet assignment system, but we believe
that price responsive policies could have several other potential applications.  

\subsection{Talent Alignment and Retention in Priority-Based Assignment Markets}

\begin{enumerate}

\item \textit{Diplomat / Foreign Service Officer Placement}

Each year, thousands of applicants compete for diplomatic positions at
more than 285 U.S. embassies and consulates around the world.  
Prioritization is based on scores on the foreign service officer test,
with additional points given for applicants based on veterans or disability status
and foreign language ability \cite[US State][]{fso:19}.   In
this market, a price-responsiveness
policy where willingness to work for an extended tour
in exchange for a priority boost could help manage retention
and talent alignment.

\item \textit{Civil Service Placement}

Governments around the world use centralized systems to place personnel
into positions.  For example,   \cite{khan/khwaja/olken:19}
describe the use of a centralized assignment mechanism
to assign property tax inspectors in Pakistan.  They designed
a scheme where priority was determined by past performance
as an inspector.  In such a scheme, a price-responsiveness
policy, where a willingness to sign an extended service commitment 
generates increased priority in the assignment, could help manage
retention and talent alignment.

\cite{bayer:21} describe the process used to assign police officers
to positions in other districts in Chicago.  The priority
is based on officer seniority.  A challenge in this setting 
is the lack of demand for working in unsafe neighborhoods
and oversubscription in safe neighborhoods.
The officer assignment board may be able to use this oversubscription
to increase retention by awarding desirable positions to officers 
who are willing to extend their time in a posting in exchange for
higher priority.

\item \textit{Centralized Teacher Assignment}

Centralized schemes are used in teacher placement in several countries
including in Czech Republic, France, Germany, Mexico,
Peru, Portugal, Turkey, and Uruguay \citep{combe/terceiux/terrier:22, Combe/Dur/Tercieux/Terrier/Unver:22}.  In these markets,
teachers priority is often based on seniority.  
The central administration aspires to assign teachers respecting
their preferences, while at the same time avoiding a surplus of
inexperienced teachers in disadvantaged areas.   \cite{elacqua:20}
and \cite{bertonia:21} use data to describe Peru's 
national teacher selection process.  In that system,
teachers can rank up to 5 schools and performance
on a standardized test is used for prioritization.  Since there
is oversubscription in advantaged regions of the country,
a price responsiveness policy where lower performing
teachers can buy priority by extending their service commitment
could cause some more experienced teachers to be assigned
to less advantaged regions.

\item \textit{Other Military Sectors: Marines Corps and Air Force}

Centralized placement is also widespread in the military, aside from the 
United States Army.   Graduates of the U.S. Air Force Academy
obtain their career field using a centralized mechanism
where cadets rank fields \citep{armacost/lowe:05}.  The Air Force
judges success of their placement process 
based on retention-related outcomes and an Airman's fit \citep{airforce:21}.  Likewise, the U.S. Marine Corps  struggles with
turnover of marines, and a 2021 manpower report describes
creating a digital talent marketplace to address this retention concern
and balance the needs of units \cite[United States Marine][]{marine:21}.
Both of these markets are situations where the flexibility of a price
responsiveness policy may facilitate a balance between
talent alignment and retention.

\end{enumerate}

\subsection{Priority-Based Assignment with or without Amenities}

Our first examples use a price-responsiveness policy as a tool
to manage retention-related outcomes.  Here we describe two examples where the
mechanism could unbundle the assignment into an assignment
under two terms to manage resource constraints.
First, nearly 15,000 officers and 500 units in the Army participate
in the Army Talent Alignment Process each year
 \citep{army:19}.\footnote{The cadet-branch assignment
process is used to determine a new officer's occupation. The Army Talent Alignment Process is used to match officers to specific jobs at later points in their Army career.}   Starting in 2019,
this system used officer preferences and a version of the deferred acceptance algorithm
for placement into units \citep{davis/greenberg/jones:23,greenberg/crow/wojtaszek:20}.  
In this market, when an officer is assigned outside the U.S., they must reside
in government-controlled military family housing if it is available. However, not all officers
may wish to bring their families abroad and may not require this housing.
Hence, the system could offer job assignment with and without family housing,
with the base price corresponding to housing and the increased price corresponding
to no family housing.   In places where there is scarcity of family housing
options, a price responsiveness policy could allow an officer who is willing to forego family housing
to buy priority for a position over an officer who needs family housing.
The same concept could apply for college admissions, where a student can be assigned
with the right to on-campus housing or without the right to on-campus housing.

Second, consider student assignment at K-12 as in \cite{abdulkadiroglu/sonmez:03}.
In that framework, students are assigned schools, and each position at a school is identical.
However, a school position can be offered to a student under different terms.  For example, 
for kindergarten and pre-kindergarten, a school can sometimes
offer a full-day or half-day option.   These two terms correspond
to the base price and the increased price.  A price responsiveness policy
where an applicant can buy priority if she is willing be assigned a half-day option
 is an instrument that would allow certain lower priority applicants to access a
 sought-after school for a half-day that they could not otherwise access.
 It is possible to envision similar ideas, like offering options for a school with an early start
time or late start time (a common way to manage overcrowding), or offering a school
with meal or without meal service.  A price responsiveness
policy in these cases would allow applicants willing to take the increased cost option
(e.g., starting school early for some or attending school without free breakfast) in exchange
for increased priority.  If these ideas are used within
the context of a centralized mechanism, then our axioms are natural and imply that the MPCO
is the only possible mechanism.

\end{document}